\title{Interactions between Lattice Hadrons
\footnote{Prepared for {\sl World Scientific Publishing Company}
  (International Review of Nuclear Physics, Vol. 9,
  Hadronic Physics from Lattice QCD,
  edited by A. M. Green)}
}
\author{H. Rudolf Fiebig\\{\sl Physics Department, FIU, Miami, Florida 33199, USA}
\vspace{1ex}\\
Harald Markum\\{\sl Atominstitut, Technische Universit{\"a}t Wien, A-1040 Vienna, Austria}}
 
\documentclass[11pt]{article}

\usepackage{amssymb,amsmath}  
\usepackage{amsfonts}         
\usepackage{dcolumn}          
\usepackage{bm}               
\usepackage{graphicx}

\begin{document}
 
\maketitle

\newcommand{\openone}{{\bf 1}}  
\newcommand{\QCD}{$\mbox{QCD}_{3+1}$}
\newcommand{\QED}{$\mbox{QED}_{2+1}$}
\newcommand{\ERI}{effective residual interaction}
\newcommand{\tbl}{\caption}

\hyphenation{me-sons mo-lec-u-les deu-ter-on}

\bibliographystyle{unsrt}

\section{Abstract}
The \ERI\ for a system of hadrons has a long tradition in theoretical
physics. It has been mostly addressed in terms of boson exchange models.
The aim of this review is to describe approaches
based on lattice field theory and numerical simulation. At the
present time this subject matter is in an exploratory stage.
A large array of
problems waits to be tackled, so that known features of hadron-hadron
interactions will eventually be understood in a model-independent way.
The lattice formulation, being capable of
dealing with the nonperturbative regime, describes strong-interaction physics from
first principles, i.e. quantum chromodynamics (QCD). 
Although the physics of hadron-hadron interactions may be intrinsically
complicated, the methods used in lattice simulations are 
simple: For the most part they are based on standard mass calculations.
This chapter addresses commonly used techniques, within \QCD\ and also
simpler lattice models, describes important results, and also gives some insight
into numerical methods for multi-quark systems.

\section{Introductory Overview on Goals, Strategies, and Methods}

\subsection{Modeling Nuclear Forces}

The theory of nuclear forces was pioneered by Yukawa \cite{Yuk35}
in 1935. A particle of intermediate mass, a meson, accounts for the
interaction energy of the proton and neutron. In the framework of a
quantum field theory the meson appears as the quantum of an effective
field that describes the strong interaction \cite{Yuk37}. The approximate
range of the interaction
\begin{equation}
R = \frac{\hbar c}{mc^2}
\label{RY}\end{equation}
is roughly consistent with $ R \approx 2~{\rm fm}$, for a meson mass
in the region of $m \approx 100~{\rm MeV}$.
Those numbers are crude estimates based on the uncertainty relation
(for energy and time) and the contemporary experimental evidence.

An effective neutron-proton interaction can be obtained from solutions
of the Klein-Gordon equation for the meson field $\varphi$
\begin{equation}
(\Box + m^2) \varphi(x) = g{\bar{\psi}(x)}\psi(x) \,,
\label{KG}\end{equation}
with the pion-nucleon coupling $g$.
In the static limit, where the product of nucleon fields $\bar{\psi}\psi$
is replaced with a generic point-like source,
\begin{equation}
(-\Delta + m^2) \varphi(\vec{r}\,) = g\delta(\vec{r}\,) \,,
\label{KGs}\end{equation}
the solution, i.e. the static Green function,
\begin{equation}
\varphi(x) = \frac{g}{4\pi}\frac{e^{-mr}}{r}
\label{YukPot}\end{equation}
is interpreted as a potential energy function, the `Coulomb potential'
of the meson theory. 
(Units are such that $\hbar=c=1$, conversion is
$\hbar c=0.197~\mbox{GeV\,fm}$.)
This line of thought is
in close analogy to electrodynamics where the Coulomb potential
emerges as the solution of a Poisson equation with a static point-like
charge (source).

Since the early work, the meson theory of nuclear forces has received 
much attention and refinements. One of the more significant
insights \cite{Tak51} has been
that the range of the nucleon-nucleon force can be
subdivided into three regions:
\begin{itemize}
\item The long-range region, inter-nucleon distances are above $r\approx 
1.5~\mbox{fm}$.
Here we have attraction. This is the region described by the
Yukawa potential.
\item Intermediate-range attraction, around $r\approx 1.0~\mbox{fm}$.
\item Short-range repulsion, below $r\approx 0.5~\mbox{fm}$.
\end{itemize}
Other salient features are:
\begin{itemize}
\item A tensor force, important for the quadrupole
moment of the deuteron.
\item The presence of a spin-orbit force.
\item Charge independence of the strong interaction.
\end{itemize}
In order to describe the short-distance region, scalar and vector mesons
($\eta, \rho, \omega, \ldots$) are needed in the boson-exchange model.
Remarkably, a $\pi$--$\pi$ correlated state (known as the $\sigma$ meson)
is relied upon to produce intermediate range attraction.
The pseudoscalar meson ($\pi$), which also gives rise to the
long-range (though finite) Yukawa potential, contributes significantly to
the tensor force.

After approximately half a century the meson theory of nuclear forces
is at a mature, sophisticated stage. Exhaustive review articles exist, for
example Refs.~\cite{Holinde:1981se,Lacombe:1981gy,Machleidt:1987hj,Machleidt:1989tm}
may serve the interested reader.
Boson-exchange models of this type presently constitute the
state-of-the-art for effective theories of nuclear forces.

\begin{figure}[htb]
\centerline{\includegraphics[height=84mm]{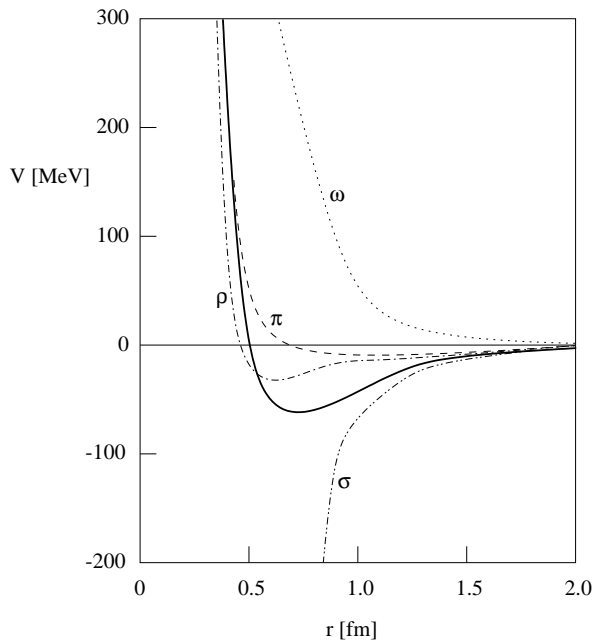}}
\vspace*{8pt}
\caption{Generic form of a central nucleon-nucleon
potential according to boson exchange models and its
contribution from various bosons. 
Schematic drawing after Ref.~\protect\cite{Machleidt:1989tm}.\label{figBXP}}
\end{figure}

Typical for this class of models is an effective Lagrangian, say for
the nucleon-nucleon interaction, such as
\begin{eqnarray}
{\cal L}_{\bar{\psi}\varphi\psi} &=& \sum_{\varphi}\left(\rule{0mm}{5mm}
-g_{\rm P}\overline{\psi}i\gamma_5\psi\varphi^{\rm P}
+g_{\rm S}\overline{\psi}\psi\varphi^{\rm S}\right.\nonumber\\
&&\left.-g_{\rm V}\overline{\psi}\gamma_{\mu}\psi\varphi^{\rm V}_{\mu}
-\frac{f_{\rm V}}{2M}\overline{\psi}\sigma^{\mu\nu}
\psi\partial_{\mu}\varphi^{\rm V}_{\nu}\right) \,,
\label{BXlag}\end{eqnarray}
with pseudoscalar $g_{\rm P}$, scalar $g_{\rm S}$, vector $g_{\rm V}$
and tensor $f_{\rm V}$ couplings for the various bosons $\varphi$, and $M$ is the
nucleon mass \cite{Machleidt:1989tm}. Potentials are obtained from tree-level diagrams
to the nucleon-nucleon propagator. Those are most naturally written in momentum
space.
The simplest contribution derived from (\ref{BXlag}) is
\begin{equation}
V(\vec{k}) = -\frac{g_{\rm P}^2}{12M} \left[
\vec{\sigma}_1\cdot\vec{\sigma}_2
-\vec{\sigma}_1\cdot\vec{\sigma}_2\frac{m_{\rm P}^2}{\vec{k}^2+m_{\rm P}^2}
+\frac{S_{12}(\vec{k})}{\vec{k}^2+m_{\rm P}^2} \right]
\vec{\tau}_1\cdot\vec{\tau}_2 \ldots \,,
\label{BXpot}\end{equation}
where $m_{\rm P}$ is the pseudoscalar mass.
This already is the non-relativistic reduction where the three terms
correspond to a (coordinate-space) $\delta$-function, a Yukawa,
and a tensor potential, the latter containing
\begin{equation}
S_{12} = 3(\vec{\sigma}_1\cdot\vec{k})(\vec{\sigma}_2\cdot\vec{k})
-\vec{\sigma}_1\cdot\vec{\sigma}_2 \vec{k}^2 \,.
\label{S12}\end{equation}
The momentum transfer is $\vec{k}=\vec{q}\,'-\vec{q}$ and $\vec{\sigma}_1$,
$\vec{\sigma}_2$ are the nucleon spins. In addition, for each type of coupling
momentum cutoff form factors are included with the coupling constants
$g_{\rm P},g_{\rm S},g_{\rm V}$ and $f_{\rm V}$, for example
\begin{equation}
g_{\rm P}^2 \rightarrow g_{\rm P}^2\
\left( \frac{\Lambda_{\rm P}^2-m_{\rm P}^2}
{\Lambda_{\rm P}^2+\vec{k}^2} \right)^{n_{\rm P}} \,.
\label{formfac}\end{equation}
The inverse of a momentum cutoff parameter $\Lambda_{\rm P}$ is a
length, and as such indicates the region of validity of the model.
Values of $\approx 1~{\rm GeV}$, or equivalently $\approx 0.2~{\rm fm}$
are typical.
Other refinements of the model include isobar ($\Delta$) degrees of 
freedom, and the extension to the strange hadron
sector \cite{Lohse:1990ew,Janssen:1995wn}.
Typical potentials are illustrated in Fig.~\ref{figBXP}.

\begin{figure}[htb]
\centerline{\includegraphics[width=56mm]{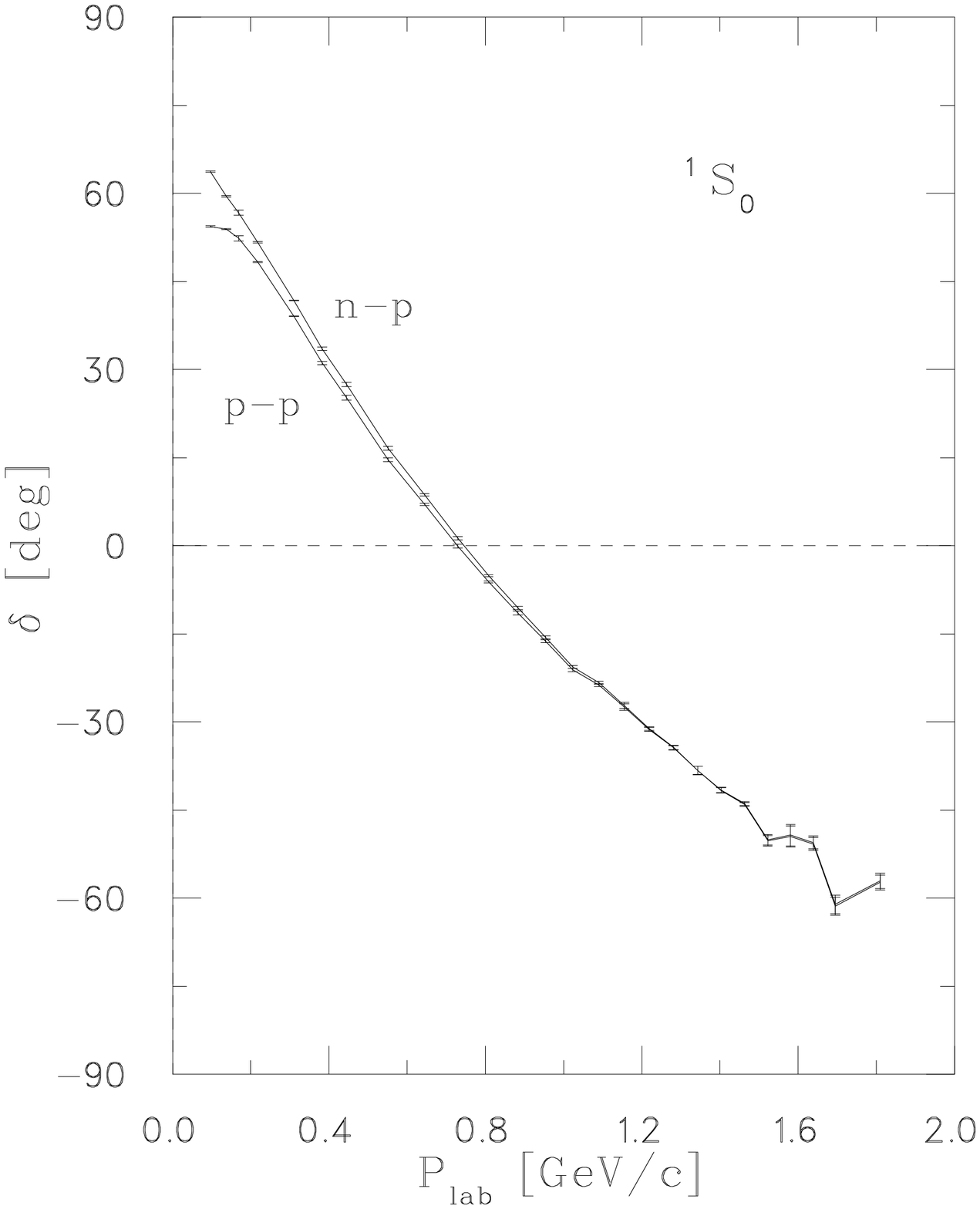}
            \includegraphics[width=56mm]{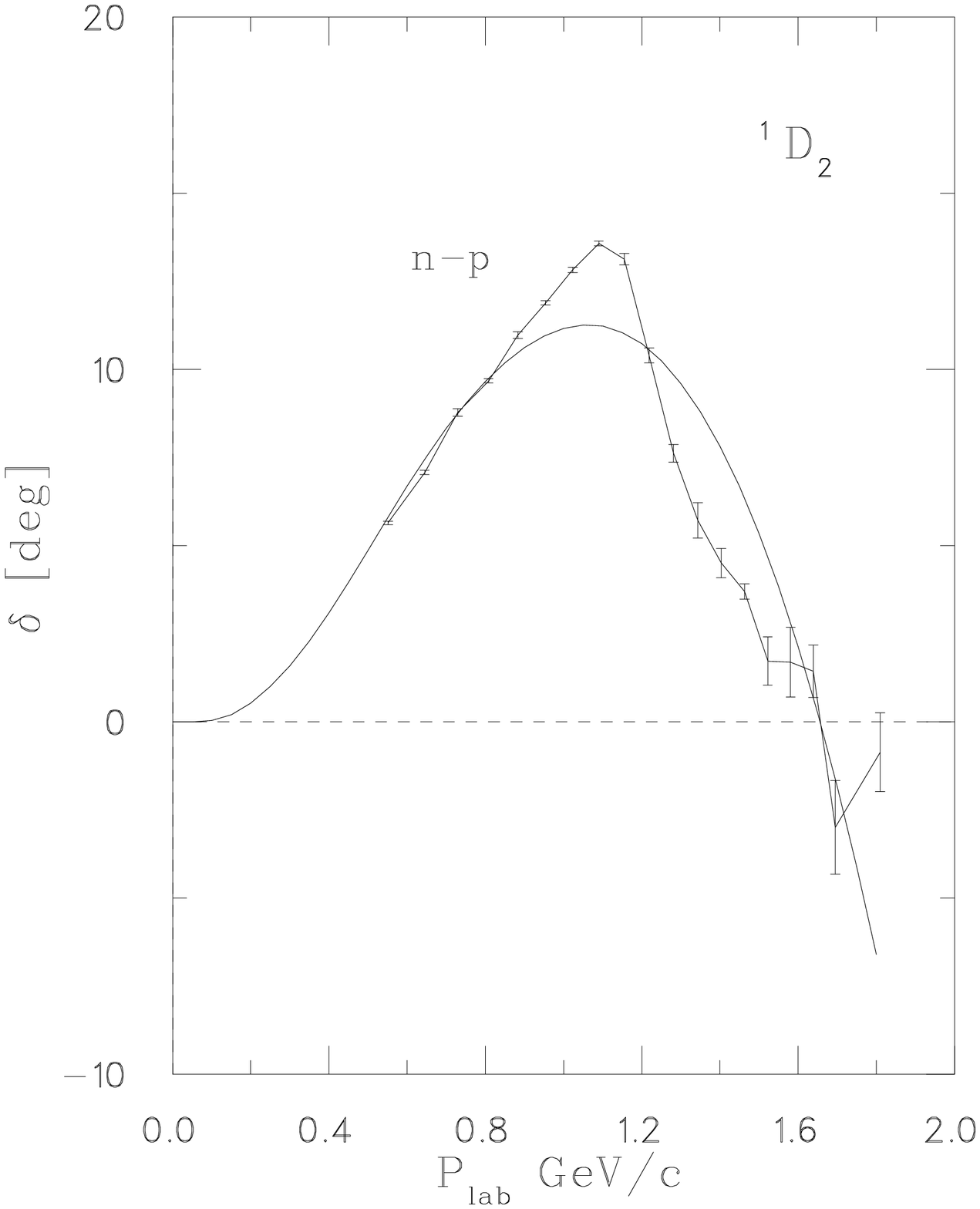}}
\vspace*{8pt}
\caption{Nucleon-nucleon scattering phase shift for channels  $^1S_0$ and
$^1D_2$ compiled from Ref.~\protect\cite{SAID}. In the  $^1D_2$ channel the
smooth curve is a `Bonn-potential' result.\label{figNNsca}}
\end{figure}

The philosophy of boson exchange models is to
take the phenomenology of low-energy excitations of the hadronic
vacuum as a starting point. The experimentally known spectrum of
mesons and hadrons is used to reconstruct the forces between those
particles. The result is an effective theory with adjustable
parameters and predictive power
in a reasonably well-defined realm of validity.
An example is shown in Fig.~\ref{figNNsca}.
Naturally, the
question of how hadronic interaction results from the underlying
microscopic structure of hadrons and the, as we now know, complicated
structure of the hadronic vacuum, is beyond the scope, and aim,
of the model.

In the present context we wish to look at hadronic interactions
from the perspective of lattice QCD.
Since the physics of hadronic interactions is contained in the
low-energy, non-perturbative, regime of QCD the lattice formulation
is well suited for this task. 

\subsection{The Lattice QCD Perspective}

From the lattice QCD point of view the three-fold subdivision of the inter-nucleon
interaction region, see Fig.~\ref{figBXP}, appears in a new light:
\begin{itemize}
\item The long-range region is expected to be dominated by excitations
of the QCD vacuum that are recovered only in the chiral limit of
the lattice simulation. If Wilson fermions are used this region can
be monitored by extrapolating to the zero-mass limit of the pseudoscalar
meson, which is the Goldstone boson of QCD. The mechanism for pion
exchange is available on the lattice because correlated quark-antiquark
pairs are dynamically provided by the lattice action. This is true
even if the simulation is limited to the quenched approximation.
\item In the intermediate-distance region various excitations of
the QCD vacuum of somewhat higher mass will take over. Those may be
pure hadronic states, gluonic excitations, or combinations of such.
In contrast to boson exchange models the precise nature of those need not
be known as input. Again those would be accounted for by the lattice action.
At $r\approx 1.0~\mbox{fm}$ this is really a transition region between
an appropriate description in terms of effective hadronic fields and
the asymptotically free regime of the interior quark-gluon cores of
the hadrons. The validity of traditional boson exchange models seems
somewhat stretched here. With the currently available lattice constants
($\approx 0.1$ -- $0.4~\mbox{fm}$) the intermediate-distance region appears
appropriate for lattice QCD.
\item For small distances the compositeness of the hadrons, as finite-size
quark-gluon systems, will play the dominant role. Not surprisingly, quark
models have been used to study the nucleon-nucleon interaction.  
The energies involved are naturally larger, for example $\approx 2~{\rm GeV}$,
and thus point to lattice constants of about $\approx 0.1~{\rm fm}$,   
being not out of reach of contemporary lattice simulations.
Lattice QCD provides dynamically
for the compositeness of the hadrons. Internal hadronic structure is, in
turn, crucial for the features of the residual interaction.
\end{itemize}
Thus lattice QCD is in a position to combine the classic 
nucleon-nucleon interaction regions in a single unified approach. Moreover, the
strange hadron sector, or effects of strangeness in the light-quark
sector are straightforwardly incorporated.

\subsection{Short and Long Term Goals for Lattice QCD}

\begin{figure}[htb]
\centerline{\includegraphics[height=64mm]{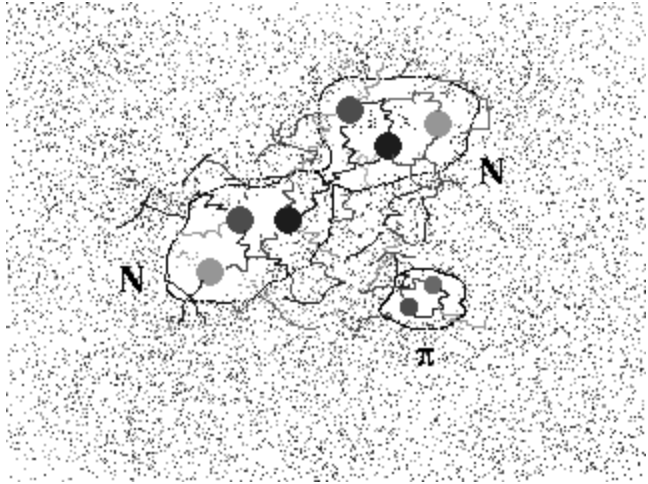}}
\vspace*{8pt}
\caption{Cartoon of a strongly interacting hadron-hadron
system from the QCD perspective.\label{figQCDhh}}
\end{figure}

While boson exchange models have a long tradition, attempts to study residual
interactions between systems of hadrons within lattice QCD are only in an 
experimental stage.
The physics of hadronic interactions in terms of first principles
is rather involved, see Fig.~\ref{figQCDhh}.
Conceptual, technical,
and computational questions are very much in the foreground at this time.
The tentative nature of this situation will be apparent throughout this chapter.
 
Nevertheless, there is potential for new physics, or at least for understanding
known phenomena in a new light. In particular, probing
hadronic matter at somewhat higher energies, say in the 4 -- 8~GeV region
with electron beams, has become a new focus in experimental nuclear physics. 
The interest here lies in the boundary between particle 
(high energy) and nuclear (low energy) physics which we will refer to as
hadronic physics.
Our short-term working list is as follows: 
\begin{itemize}
\item Discuss various basic concepts and numerical methods using simple lattice 
models.
\item Study simple systems, mostly meson-meson, and extract
non-rel\-a\-tiv\-is\-tic potentials. 
\item Explore alternative approaches to the problem, like momentum
space and coordinate space formulations.
\end{itemize}
As a long-term perspective we have in mind:
\begin{itemize}
\item Study realistic meson-meson, meson-baryon, and baryon-baryon systems.
For example, the pion-nucleon system features the prominent delta resonance,
a hallmark of strong-interaction physics.
\item Residual interactions in the strange-hadron sector.
\item Develop advanced methods, possibly the direct extraction of scattering
amplitudes from the lattice.
\end{itemize}

\subsection{Probing the Lattice}

A hadron-hadron force may be viewed as a hyperfine interaction in the
hadron-hadron system. In one way or another, we must therefore compare
the mass spectrum of the interacting hadron-hadron system to the spectrum
of the non-interacting hadron-hadron system.

About two decades ago it was realized that finite volume effects may in
principle be exploited to achieve this end \cite{Hamber:1983vu}.
The energy levels of a two-hadron system, say
two mesons with mass $m$ each, are subject to finite-size corrections.
Those decrease with powers of $1/L$ as the  lattice size $L$ increases.
The formal study of finite-size effects in this context is due to
L\"{u}scher \cite{Luscher:1986dn,Luscher:1986pf}. In Ref.~\cite{Luscher:1986pf}
the following equation for the ground-state two-body energy shift
\begin{equation}
W_0-2m=-\frac{4\pi a_0}{mL^3}\left\{ 1+c_1\frac{a_0}{L}+c_2\frac{a_0^2}{L^2}
\right\} + o(L^{-6})
\label{ScatLen}\end{equation}
is derived. It is valid for s-wave elastic scattering, below the particle
production threshold.
Above $c_1=-2.837297$, $c_2=6.375183$, and $a_0$ is the s-wave scattering length.
This result has been the basis for a number of 
studies \cite{Guagnelli:1990jb,Sharpe:1991nv,Sharpe:1992pp,Gupta:1993rn,Kuramashi:1993ka,Kuramashi:1993yu,Fukugita:1994na,Fukugita:1995ve,Kuramashi:1996sc,Aoki:1999pt,Aoki:2001hc,Aok02hc,Aoki:2002in}.
The extraction of scattering information beyond the static limit,
like phase shifts at nonzero relative energy, is formally more
involved \cite{Luscher:1990ck,Luscher:1991ux}.

\begin{figure}[htb]
\centerline{\includegraphics[height=84mm]{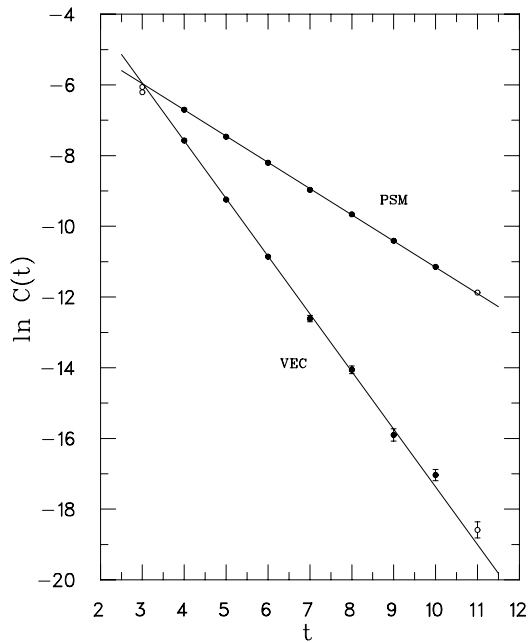}}
\vspace*{8pt}
\caption{Typical time correlation functions (eigenvalues) extracted from a
lattice simulation. The examples are for a pseudoscalar meson (PSM) and a vector
meson (VEC) using \QCD\ and Wilson fermions.\label{figCt}}
\end{figure}

In general it will be necessary to probe the lattice (vacuum) with
one- and two-hadron operators. Let $\phi_{\vec{p}}(t)$ be a composite
field, say, constructed from quark--antiquark (meson) or from three--quark
(baryon) fields. Let $\vec{p}$ be the hadron momentum, for simplicity we 
disregard additional quantum numbers. 
The time correlation matrix built from one-hadron operators
\begin{equation}
C^{(2)}_{\vec{p}\,\vec{q}}(t,t_0)=
\langle\phi^{\dagger}_{\vec{p}}(t)\,\phi_{\vec{q}}(t_0)\rangle-
\langle\phi^{\dagger}_{\vec{p}}(t)\rangle
\langle\phi^{\phantom{\dagger}}_{\vec{q}}(t_0)\rangle
\label{C2}\end{equation}
possesses eigenvalues $\lambda^{(2)}_{m_1}(t,t_0)$ which behave exponentially,
\begin{equation}
\lambda^{(2)}_{m_1}(t,t_0)\propto e^{-w_{m_1}(t-t_0)} \,.
\label{C2eval}\end{equation}
The time correlation matrix is just the euclidean version
of a propagator, in our example a 2-point function $C^{(2)}$ on the hadronic
level (a 4- or 6-point function on the quark level). Thus from (\ref{C2eval})
the energies $w_{m_1}$ of the one-meson system can be extracted. Typically, 
this is done numerically. A practical example is shown in Fig.~\ref{figCt}.
The interpretation of the $w_{m_1}$
is that those are energies of the hadron moving through the lattice.
In principle, the hadron could be in an internally
excited state.  

The fields $\Phi_{\vec{p}}(t)=\phi_{-\vec{p}}(t)\,\phi_{+\vec{p}}(t)$
can be used for probing two-hadron states.
These are operators with total momentum $\vec{P}=0$. Note that $\vec{p}$
in this example means the relative momentum of the two-hadron system. 
We construct a 4-point correlation matrix
\begin{equation}
C^{(4)}_{\vec{p}\,\vec{q}}(t,t_0)=
\langle\Phi^{\dagger}_{\vec{p}}(t)\,\Phi_{\vec{q}}(t_0)\rangle-
\langle\Phi^{\dagger}_{\vec{p}}(t)\rangle
\langle\Phi^{\phantom{\dagger}}_{\vec{q}}(t_0)\rangle \,,
\label{C4}\end{equation}
and, correspondingly, the eigenvalues
\begin{equation}
\lambda^{(4)}_n(t,t_0)\propto e^{-W_n(t-t_0)}
\label{C4eval}\end{equation}
yield the energy levels $W_n$ of the two-hadron system.
Most importantly, these levels somehow contain the desired residual interaction.

A sensible definition of the latter, of course, also involves the eigenvectors
of $C^{(2)}$ and $C^{(4)}$, however, comparison of the level schemes
\begin{equation}
\{\overline{W}_{m=(m_1,m_2)} \stackrel{\rm def}{=} w_{m_1}+w_{m_2} \}
\quad\mbox{to}\quad \{W_n\}
\label{lev0}\end{equation}
already gives important insight into the nature of the hadron-hadron force.

Probing the lattice with momentum-type operators is only one
of several alternatives. The use of coordinate space operators is another popular
choice \cite{Markum:1985es,Ric89,Richards:1990xf,Rabitsch:1993up,Mihaly:1997ue,Michael:1999nq,Fiebig:1999hs}.

\subsection{Finite-size Methods}

The spectrum $\{W_n\}$ of the interacting two-body system may be utilized
in a variety of ways. L\"{u}scher has shown how to extract scattering
phase shifts directly from the spectrum \cite{Luscher:1991ux}. The idea is to match the
poles of (interacting) Green functions that live on the finite-sized
periodic lattice to the poles of (free) Green functions in infinite volume.
Since the lattice Green functions comprise the interaction, comparison in 
the asymptotic (free) region yields the phase shifts. Using methods similar
to the `Jost' formalism of scattering theory \cite{Tay72} an equation
of the following form
\begin{equation}
\det \left[ e^{2i\delta}-(M+i)^{-1}(M-i)\right] =0
\label{LueDet}\end{equation}
is derived. Here $M=M(q^2)$ is an analytically constructed energy-dependent
matrix ($W^2=m^2+q^2$).
It reflects features of the `natural'
(free) spectrum of the periodic lattice. The other
matrix $e^{2i\delta}$ in (\ref{LueDet})
contains a set of scattering phase shifts $\delta_\ell(q^2)$. The lattice
simulation gives a discrete spectrum $\{W_n\}$. A certain energy $W_n$, for a
fixed $n$, is then used to calculate $M(q_n^2)$ and (\ref{LueDet}) is solved
for the corresponding $\delta_\ell(q_n^2)$. This yields a set of phase shifts,
as many as the size of $M$, at each energy $W_n$. The physical
$\delta_\ell(q_n^2)$, at a discrete $q_n^2$, is contained in this set.

\begin{figure}[htb]
\centerline{\includegraphics[height=84mm]{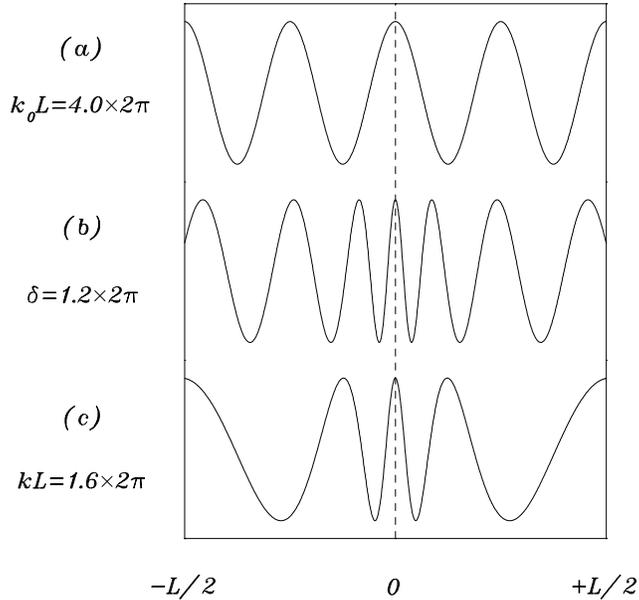}}
\vspace*{8pt}
\caption{Generic solutions $\psi(x)$ of a Schr{\"o}dinger equation: (a) no
interaction, (b) an attractive potential around $x\approx 0$, and (c) with
decreased wave number $k$ to restore the original phase conditions.\label{figWAVE}}
\end{figure}

A simple illustration of the idea behind this formalism \cite{Luscher:1990ck} is the
following:
Consider a solution $\psi (x)$ of a Schr\"odinger equation in the
noninteracting case with a phase $kL = 2\pi\,n$. It satisfies $\psi (x) = \psi (x\!+\!L)$,
see Fig.~\ref{figWAVE}(a).  If the interaction is turned on, say an attractive
potential with range $|x|\le 0.2L$, the periodic
boundary condition would be violated, see Fig.~\ref{figWAVE}(b),
unless compensated for by a
simultaneous change in the kinematical phase $kL$.  In other words, the change
in $kL$ necessary to restore the original periodic boundary
conditions, see Fig.~\ref{figWAVE}(c), is a measure for the phase shift.

The kinematical phase $kL$ is changed by varying $k$ or $L$. Varying $L$ is
reminiscent of a finite-size effect, see (\ref{ScatLen}). Finding the change
in $k$ at fixed $L$ is equivalent to finding all (discrete) two-body levels
$\{W_n\}$ for the interacting case.  
This method, which already appears in Ref.~\cite{Luscher:1986pf}, has been used by
L\"uscher and Wolff \cite{Luscher:1990ck} to obtain scattering phase shifts in an O(3)
nonlinear sigma model on a 1+1 dimensional lattice.

\subsection{Residual Interaction}

Scattering phase shifts are very close to observable quantities (cross
sections).  This is an appealing feature of L\"{u}scher's method, 
as the above procedure came to be known. 
On the other hand an \ERI\ between hadrons
is more in line with the historic development 
and has a heuristic value which should not be underestimated. We will therefore
place considerable emphasis on this aspect. In this approach the
information contained in the 2- and 4-point correlation matrices, see
(\ref{C2}) and (\ref{C4}), will be utilized in a different way.

In principle an \ERI\ is subject to definition.
Towards this end we will first split a free, or non-interacting, part
off the 4-point correlation matrix (\ref{C4})
\begin{equation}
C^{(4)}(t,t_0)=\overline{C}(t,t_0)+C_I(t,t_0)\,,
\label{C4bar}\end{equation}
where $\overline{C}$ is built from 2-point correlators $C^{(2)}$ and
describes two non-interacting hadrons on the lattice. This can be 
accomplished by means of diagrammatic considerations.
In terms of (\ref{C4bar}) we then define an \ERI\ as
\begin{equation}
H_I = -\frac{\partial}{\partial t}\ln \left(
\overline{C}^{\,-1/2} C_I\, \overline{C}^{\,-1/2} \right) \,.
\label{HI}\end{equation}
In principle the interaction Hamiltonian (potential) is non-local.
Our focus is to compute $H_I$ from a lattice simulation and eventually
calculate the corresponding phase shifts.

The above method relies on lattice sizes large enough to accommodate two
hadrons. Assuming a manageable lattice $L^3\times T$ with, say, $L=10$
and physical extent of about $4.0$~fm this translates into a lattice
constant of $a=0.4$~fm. As experience with hadron spectroscopy has shown
the standard Wilson plaquette action \cite{Wilson:1974sk} requires
$a\approx 0.1-0.2$~fm in order to work reasonably
close to the continuum limit. This does not sound very encouraging
for putting multi-hadron systems on the lattice. 

\subsection{Use of Improved Actions}

Fortunately, considerable progress in recent years has provided us with
so-called `improved' lattice actions \cite{Lepage:1993xa}. Here, in the simplest
form, one supplements the Wilson action, built from elementary $1\times 1$
plaquettes (pl), with larger $1\times 2$ rectangles (rt)
\begin{equation}
S_{G}[U] = \beta\left(\rule{0pt}{9pt}S_{pl}[U] + c_{rt}S_{rt}[U]\right) \,.
\label{ImpAct}\end{equation}
The Wilson part of the action (pl) approaches the classical (tree-level)
limit like $o(a^4)$ as $a\rightarrow 0$. If $c_{rt}=-1/20$ the (rt) term in 
(\ref{ImpAct}) improves this behavior to order $o(a^5)$. We refer to such an
action as $o(a)$ tree-level improved. Better yet, one can also account for
a large part of the quantum fluctuations of the gluon field through
renormalizing the link variables $U_{\mu}(x)$. Of course, this would destroy the
tree-level cancellations between the (pl) and (rt) terms in (\ref{ImpAct}).
Within mean-field theory it can be `repaired' by changing the factor $c_{rt}$ 
to
\begin{equation}
c_{rt}=-\frac{1}{20 u^2}\,,
\label{Crt}\end{equation}
where $u$, known as the tadpole factor, is the fourth root of the average
(4-link) plaquette
\begin{equation}
u=\langle\frac13\mbox{Re}\,\mbox{Tr}\,U_{pl}\rangle^{1/4} \,.
\label{tadpole}\end{equation}
The tadpole factor can easily be computed self-consistently within a
running lattice simulation.
There are also radiative corrections which can be incorporated into
the lattice action to some extent \cite{Symanzik:1983dc,Symanzik:1983gh,Luscher:1985zq}.

The benefit of improvement is that the physical continuum limit, 
$a\rightarrow 0$, is approached `faster' compared to the Wilson action.
The `same physics' is described with improved actions
at larger lattice constants $a$. For example, a Wilson action at $a\approx
0.1$~fm may give a similar mass spectrum as an improved action at $a\approx
0.3$~fm. For our present purpose of putting two hadrons on the lattice,
this means that, stretching things a bit
and choosing $a\approx 0.4$~fm, we would only need an $L^3\times T$ lattice
with about $L=10$ for a lattice of 4.0~fm across. This should give us
a first window on two-hadron systems on the lattice.

Improved actions have been a big innovation for lattice field theory \cite{Lepage:1996ph}.
There are versions of improved actions including fermions \cite{Sheikholeslami:1985ij},
and work pushing the improvement to higher order, for example
$o(a^2)$ improved actions \cite{Eguchi:1984xr,Alford:1995ui,Alford:1995hw}.
-- In Appendix~\ref{appB} we have collected some topics that are relevant to this chapter.
-- Strong-interaction physics, in particular the boundary between
particle and nuclear physics,
is now open to studies within quantum chromodynamics.

\section{A Simple U(1) Lattice Model in 2+1 Dimensions}\label{secU1}

In order to discuss some practical work we look at a simple but non-trivial
model. We consider U(1) gauge theory on an $L^d\times T$ lattice, 
in $d+1$ dimensions, with $d=2$.
The intent is to study the interaction between two mesons living in a plane.
For the sake of simplicity we
work with the (unimproved) Wilson action and use staggered fermions. 

\subsection{Lattice Action}

For a U(1) gauge group the link variables are
\begin{equation}
U_{\mu}(x) = e^{i\Theta_{\mu}(x)} \in \mbox{U(1)}  \,,
\label{U1link}\end{equation}
with $\mu=1\ldots d+1$, $x=(\vec{x},t)$, and $\Theta_{\mu}(x)\in [-\pi,\pi)$.
The plaquette (pl) variables are
\begin{eqnarray}
U_{pl}(\mu,\nu;x) &=& U_{\mu}(x) U_{\nu}(x+\hat{\mu}) 
U^{\dagger}_{\mu}(x+\hat{\nu}) U^{\dagger}_{\nu}(x) \nonumber\\
 &=& \exp\left(\rule{0pt}{9pt}i\Theta_{pl}(\mu,\nu;x)\right)\,,
\label{Upl}\end{eqnarray}
where
\begin{equation}
\Theta_{pl}(\mu,\nu;x) = \Theta_{\mu}(x)+\Theta_{\nu}(x+\hat{\mu})
-\Theta_{\mu}(x+\hat{\nu})-\Theta_{\nu}(x)
\label{Tpl}\end{equation}
is the oriented plaquette angle.
We write $\hat{\mu}$ for a vector of length $a$ in direction $\mu$.
The (compact form of the) Wilson gauge field action is
\begin{equation}
S_G = \beta \sum_{x} \sum_{\mu < \nu} \mbox{Re}
\left(\rule{0pt}{9pt}1-U_{pl}(\mu,\nu;x)\right)
 = \beta \sum_{x} \sum_{\mu < \nu}  \left(\rule{0pt}{9pt}
1-\cos \Theta_{pl}(\mu,\nu;x)\right) \,.
\label{WilU1}\end{equation}

For simplicity, fermions are treated in the staggered, or Kogut-Susskind, scheme where
Dirac indices have been `spin-diagonalized' away \cite{Kogut:1975ag,Susskind:1977jm,Kawamoto:1981hw}.
Thus we work with one-component Grassmann fields
$\chi_{f}(x)$ defined on the lattice sites
$x=(\vec{x},t)$, with (external) flavor index
$f=u,d,s$. There is just one color index, $C=1$, which we omit.
The fermionic action can be written as
\begin{equation}
S_F = \sum_{x,y} \sum_{f} \bar\chi_f(x) \: G^{-1}(x,y)[U]\,\chi_f(y) \,,
\label{StaFerAct}\end{equation}
with the flavor-independent fermion matrix
\begin{equation}
G^{-1}(x,y) =\frac12 \sum_\mu \eta_\mu(x) \left[ U_\mu(x)
\delta_{x+\hat{\mu},y} - U^\dagger_\mu(y) \delta_{x,y+\hat{\mu}} \right] +
m_F\,\delta_{x,y} \,,
\label{FerMat}\end{equation}
including a mass term. Here
$\eta_\mu(x)=(-1)^{x_1+\ldots x_{\mu-1}}, \mu=2\ldots d+1, \eta_1(x)=1$
is the staggered phase.

This lattice field model is confining in finite volume \cite{Coddington:1986jk}, and exhibits
chiral symmetry breaking \cite{Dagotto:1989id} and monopole condensation \cite{Fiebig:1990uh},
for appropriately chosen inverse gauge coupling $\beta$.
Its basis is the euclidean generating functional integral
\begin{equation}
Z[\bar{J},J] = \int [dU][d\chi][d\bar{\chi}]
e^{-S_G[U]-S_F[U,\chi,\bar{\chi}]+\bar{J}\chi+\bar{\chi}J} \,,
\label{Zint}\end{equation}
with the fermionic sources $\bar{J}\chi=\sum_{x}\bar{J}(x)\chi(x)$
and $\bar{\chi}J=\sum_{x}\bar{\chi}(x)J(x)$.
We will use gauge configurations in the quenched approximation.

\subsection{Meson Fields}\label{secMesF}

The simplest hadrons we can work with are pseudoscalar mesons. In the
staggered scheme suitable one-meson operators are
\begin{equation}
\phi_{\vec{p}}(t)=L^{-2}\sum_{\vec{x}}\,e^{i\vec{p}\cdot\vec{x}}
\bar{\chi}_{d}(\vec{x},t)\,\chi_{u}(\vec{x},t) \,.
\label{onemes}\end{equation}
The $\vec{x}$ sum extends over the spatial sites of the lattice. The fixed
flavor assignment $u$,$d$ leading to a $\pi^+$ will
simplify computations. For $d=2$, a planar lattice, the momentum parameters are
\begin{equation}
\vec{p}=\frac{2\pi}{L}(k_1,k_2)\,, \quad \mbox{where} \quad
k_{1,2}=-(\frac{L}{2}-1)\ldots\frac{L}{2}\,, \quad \mbox{even $L$}\,,
\label{momp}\end{equation}
if $L$ is odd we have $k_i=-(L-1)/2\ldots(L-1)/2$.
The operators (\ref{onemes}) probe the lattice (vacuum) for
composite meson states with momentum $\vec{p}$.
Products of those operators probe for multi-meson excitations. For example
\begin{equation}
\Phi_{\vec{p}}(t)=\phi_{-\vec{p}}(t)\,\phi_{+\vec{p}}(t)
\label{twomes}\end{equation}
describes two-meson fields with total momentum $\vec{P}=0$ and relative
momentum $\vec{p}$. The reduced mass is $m/2$.

\subsection{Correlation Matrices}\label{secCor}

The time-correlation matrix for the one-meson system is 
\begin{equation} 
C^{(2)}_{\vec{p}\,\vec{q}}(t,t_0)=
\langle\phi^{\dagger}_{\vec{p}}(t)\,\phi_{\vec{q}}(t_0)\rangle-
\langle\phi^{\dagger}_{\vec{p}}(t)\rangle
\langle\phi^{\phantom{\dagger}}_{\vec{q}}(t_0)\rangle \,.
\label{eq41}\end{equation}
It can be worked out with Wick's theorem in terms of contractions between 
the Grassmann fields. In our simple example we use, see Appendix~\ref{appA},
\begin{eqnarray} 
\ldots\stackrel{n}{\chi}_{f}(x) \stackrel{n}{\bar{\chi}}_{f'}(x')
\ldots &=& \ldots\delta_{ff'} {G}(x,x')\ldots \label{eq420}\\
\ldots \stackrel{n}{\bar{\chi}}_{f}(x) \stackrel{n}{\chi}_{f'}(x')\ldots &=&
\ldots \delta_{ff'} {G}^{\ast}(x,x')\ldots \,, \label{eq42}
\end{eqnarray}
where ${n}$ indicates the partners of the contraction, and $G$ is the
inverse fermion matrix, cf. (\ref{FerMat}).
This may be used to work out the correlation matrices. For example, consider
\begin{equation}
\langle\phi^{\dagger}_{\vec{p}}(t)\,\phi_{\vec{q}}(t_0)\rangle =
 L^{-4}\sum_{\vec{x}}\sum_{\vec{y}}\,
e^{-i\vec{p}\cdot\vec{x}} e^{i\vec{q}\cdot\vec{y}} 
\langle\bar{\chi}_{u}(\vec{x},t)\chi_{d}(\vec{x},t)\,
\bar{\chi}_{d}(\vec{y},t_0)\chi_{u}(\vec{y},t_0)\rangle \,.
\label{chi4}\end{equation}
Because of the flavor assignment in (\ref{onemes}) the separable term
in (\ref{eq41}) is zero, and there is only one group of contractions
in the first term
\begin{equation}
C^{(2)} \sim\;\;\, \stackrel{2}{\bar{\chi}}_{u} \,
\stackrel{1}{\chi}_{d} \,
\stackrel{1}{\bar{\chi}}_{d} \,
\stackrel{2}{\chi}_{u} \,,
\label{con2112}\end{equation}
with
\begin{eqnarray}
\ldots \stackrel{1}{\chi}_{d}(\vec{x},t)
\stackrel{1}{\bar{\chi}}_{d}(\vec{y},t_0)\ldots &=&
\ldots G(\vec{x}t,\vec{y}t_0)\ldots \\
\ldots \stackrel{2}{\bar{\chi}}_{u}(\vec{x},t)
\stackrel{2}{\chi}_{u}(\vec{y},t_0)\ldots &=&
\ldots G^{\ast}(\vec{x}t,\vec{y}t_0)\ldots\,.
\label{con1122}\end{eqnarray}
Thus we obtain
\begin{equation} 
C^{(2)}_{\vec{p}\,\vec{q}}(t,t_0)=
L^{-4}\sum_{\vec{x}}\sum_{\vec{y}}
e^{-i\vec{p}\cdot\vec{x}+i\vec{q}\cdot\vec{y}}
\langle|G(\vec{x}t,\vec{y}t_0)|^2\rangle \,.
\label{eq43}\end{equation}
Assuming translational invariance
\begin{equation}
\langle|G(\vec{x},t;\vec{y},t_0)|^2\rangle=
\langle|G(\vec{x}+\vec{a},t;\vec{y}+\vec{a},t_0)|^2\rangle
\label{transinv}\end{equation}
renders the 2-point correlator diagonal
\begin{equation} 
C^{(2)}_{\vec{p}\,\vec{q}}(t,t_0)=
\delta_{\vec{p}\,\vec{q}}\;L^{-2}\sum_{\vec{x}}
e^{-i\vec{p}\cdot\vec{x}}
\langle|G(\vec{x}t,\vec{x}_0t_0)|^2\rangle  \,\, e^{i\vec{p}\cdot\vec{x}_0} \,.
\label{eq44}\end{equation}
The point $\vec{x}_0$ is arbitrary, $C^{(2)}$ is 
independent of $\vec{x}_0$, and the phase factor is irrelevant. 

\begin{figure}[htb]
\centerline{\includegraphics[width=112mm]{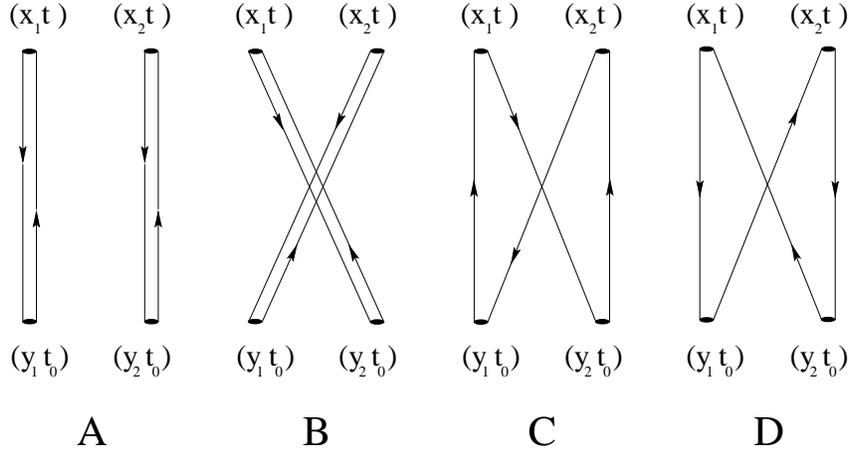}}
\vspace*{8pt}
\caption{Diagrammatic classification of the correlation matrix $C^{(4)}$
according to (\protect\ref{eq46}).\label{figABCD}}
\end{figure}

The 4-point correlator describes the propagation of two
interacting mesons on the lattice
\begin{equation} 
C^{(4)}_{\vec{p}\,\vec{q}}(t,t_0)=
\langle\Phi^{\dagger}_{\vec{p}}(t)\,\Phi_{\vec{q}}(t_0)\rangle-
\langle\Phi^{\dagger}_{\vec{p}}(t)\rangle
\langle\Phi^{\phantom{\dagger}}_{\vec{q}}(t_0)\rangle \,.
\label{eq45}\end{equation}
Here $\vec{p}$ and $\vec{q}$ are the relative momenta in the 
meson-meson system.  In terms of the fermion propagator this correlator reads
\begin{eqnarray}
C^{(4)}_{\vec{p}\,\vec{q}} (t,t_0) &&= L^{-8} \: \sum_{\vec x_1} \: 
\sum_{\vec x_2} \: \sum_{\vec y_1} \: \sum_{\vec y_2} \: {\rm e}^{i\vec p 
\cdot(\vec x_2 -
\vec x_1) + i \vec q \cdot (\vec y_2 - \vec y_1)} \label{eq46}\\
&& \mbox{} \left\langle \left( \: G^\ast(\vec x_2t, \vec y_2t_0) \quad
G(\vec x_2t, \vec y_2t_0) \quad G^\ast(\vec x_1t, \vec y_1t_0) \quad
G(\vec x_1t, \vec y_1t_0) \right. \right. \nonumber \\
&& \mbox{} + G^\ast(\vec x_1t, \vec y_2t_0) \quad
G(\vec x_1t, \vec y_2t_0) \quad G^\ast(\vec x_2t, \vec y_1t_0) \quad
G(\vec x_2t, \vec y_1t_0) \nonumber \\
&& \mbox{} - G^\ast(\vec x_2t, \vec y_1 t_0) \quad
G(\vec x_2t, \vec y_2 t_0) \quad G^\ast(\vec x_1t, \vec y_2 t_0) \quad
G(\vec x_1t, \vec y_1 t_0) \nonumber \\
&& \mbox{} - \left. \left. G(\vec x_2t, \vec y_1 t_0) \quad
G^\ast(\vec x_2t, \vec y_2 t_0) \quad G(\vec x_1t, \vec y_2 t_0) \quad
G^\ast(\vec x_1t, \vec y_1 t_0)  \right) \right\rangle \nonumber \,.
\end{eqnarray}
For an SU(N) gauge group, $G\rightarrow G_{CC'}$, there would be
sums over four sets of color indices corresponding to the
$\vec{x}_1,\vec{x}_2,\vec{y}_1,\vec{y}_2$ assignment in (\ref{eq46}).
A diagrammatic classification of the various terms which
contribute to $C^{(4)}$ proves useful.  Such a classification arises
naturally from working out the contractions between the quark fields in
(\ref{eq45}) and (\ref{eq41}).  Let us write
\begin{eqnarray}
C^{(4)} &=& C^{(4A)} + C^{(4B)} - C^{(4C)} - C^{(4D)} \label{eq47} \\
        &=&
\mbox{ \begin{picture}(18,17)
       \put(2,-6){\line(0,1){15}}
       \put(4,-6){\line(0,1){15}}
       \put(11,-6){\line(0,1){15}}
       \put(13,-6){\line(0,1){15}}
       \put(2,-6){\line(1,0){2}}
       \put(2,9){\line(1,0){2}}
       \put(11,-6){\line(1,0){2}}
       \put(11,9){\line(1,0){2}}
       \end{picture} }
+
\mbox{ \begin{picture}(18,17)
       \put(2,-6){\line(3,5){9}}
       \put(4,-6){\line(3,5){9}}
       \put(11,-6){\line(-3,5){9}}
       \put(13,-6){\line(-3,5){9}}
       \put(2,-6){\line(1,0){2}}
       \put(2,9){\line(1,0){2}}
       \put(11,-6){\line(1,0){2}}
       \put(11,9){\line(1,0){2}}
       \end{picture} }
-
\mbox{ \begin{picture}(18,17)
       \put(2,-6){\line(0,1){15}}
       \put(3,-6){\line(3,5){9}}
       \put(12,-6){\line(-3,5){9}}
       \put(13,-6){\line(0,1){15}}
       \put(2,4){\vector(0,1){0}}
       \put(2,-6){\line(1,0){1}}
       \put(2,9){\line(1,0){1}}
       \put(12,-6){\line(1,0){1}}
       \put(12,9){\line(1,0){1}}
       \end{picture} }
-
\mbox{ \begin{picture}(18,17)
       \put(2,-6){\line(0,1){15}}
       \put(3,-6){\line(3,5){9}}
       \put(12,-6){\line(-3,5){9}}
       \put(13,-6){\line(0,1){15}}
       \put(2,-1){\vector(0,-1){0}}
       \put(2,-6){\line(1,0){1}}
       \put(2,9){\line(1,0){1}}
       \put(12,-6){\line(1,0){1}}
       \put(12,9){\line(1,0){1}}
       \end{picture} } 
\label{eq48}\end{eqnarray}
for the four terms as they occur in (\ref{eq46}).  Each diagram line
corresponds to a fermion propagator, see Fig.~\ref{figABCD}.
Specifically, using the notation introduced in (\ref{eq420}) and (\ref{eq42}),
we have
\begin{equation}
C^{(4A)}=\langle \stackrel{43\phantom{9}}{\phi^{\dagger}_{+\vec{p}}}
            \stackrel{21\phantom{99}}{\phi^{\dagger}_{-\vec{p}}}
        \stackrel{12\phantom{99}}{\phi^{\phantom{\dagger}}_{-\vec{q}}}
    \stackrel{34\phantom{9}}{\phi^{\phantom{\dagger}}_{+\vec{q}}} \rangle
        =\langle \stackrel{43\phantom{9}}{\phi^{\dagger}_{+\vec{p}}}
             \stackrel{34\phantom{9}}{\phi^{\phantom{\dagger}}_{+\vec{q}}}
        \stackrel{21\phantom{99}}{\phi^{\dagger}_{-\vec{p}}}
    \stackrel{12\phantom{99}}{\phi^{\phantom{\dagger}}_{-\vec{q}}} \rangle
\label{eq49}\end{equation}
\begin{equation} 
C^{(4B)}=\langle \stackrel{21\phantom{9}}{\phi^{\dagger}_{+\vec{p}}}
            \stackrel{43\phantom{99}}{\phi^{\dagger}_{-\vec{p}}}
        \stackrel{12\phantom{99}}{\phi^{\phantom{\dagger}}_{-\vec{q}}}
    \stackrel{34\phantom{9}}{\phi^{\phantom{\dagger}}_{+\vec{q}}} \rangle
        =\langle \stackrel{43\phantom{99}}{\phi^{\dagger}_{-\vec{p}}}
             \stackrel{34\phantom{9}}{\phi^{\phantom{\dagger}}_{+\vec{q}}}
         \stackrel{21\phantom{9}}{\phi^{\dagger}_{+\vec{p}}}
     \stackrel{12\phantom{99}}{\phi^{\phantom{\dagger}}_{-\vec{q}}} \rangle \,,
\label{eq410}\end{equation}
where the ${n}={1}\ldots {4}$ identify the partners
$\chi$ and $\bar{\chi}$ of a contraction.
In diagrams A and B no quark exchange between the mesons occurs. Diagram B
describes the exchange of (composite) mesons as a whole.
On the other hand diagrams C and D both contain quark exchange
between the mesons
\begin{equation} 
C^{(4C)}=\langle \stackrel{23\phantom{9}}{\phi^{\dagger}_{+\vec{p}}}
             \stackrel{41\phantom{99}}{\phi^{\dagger}_{-\vec{p}}}
         \stackrel{12\phantom{99}}{\phi^{\phantom{\dagger}}_{-\vec{q}}}
     \stackrel{34\phantom{9}}{\phi^{\phantom{\dagger}}_{+\vec{q}}} \rangle
\label{eq411}\end{equation}
\begin{equation} 
C^{(4D)}=\langle \stackrel{41\phantom{9}}{\phi^{\dagger}_{+\vec{p}}}
             \stackrel{23\phantom{99}}{\phi^{\dagger}_{-\vec{p}}}
         \stackrel{12\phantom{99}}{\phi^{\phantom{\dagger}}_{-\vec{q}}}
     \stackrel{34\phantom{9}}{\phi^{\phantom{\dagger}}_{+\vec{q}}} \rangle \,.
\label{eq412}\end{equation}
Thus these diagrams must be considered as sources of effective interaction.

The propagation of free, noninteracting, mesons is contained in
$C^{(4A)}+C^{(4B)}$. However, gluonic correlations between the mesons are
also present because the gauge configuration average $\langle\ldots\rangle$
is taken over the product of all four fields.
(This is also true for $C^{(4C)}$ and $C^{(4D)}$.)
Gluonic correlations of course contribute to the \ERI. Hence
it is essential to isolate the uncorrelated
part contained in $C^{(4A)}+C^{(4B)}$.

\subsection{Correlation Matrix for Noninteracting Mesons}\label{secCor0}

\begin{figure}[htb]
\centerline{\includegraphics[height=82mm]{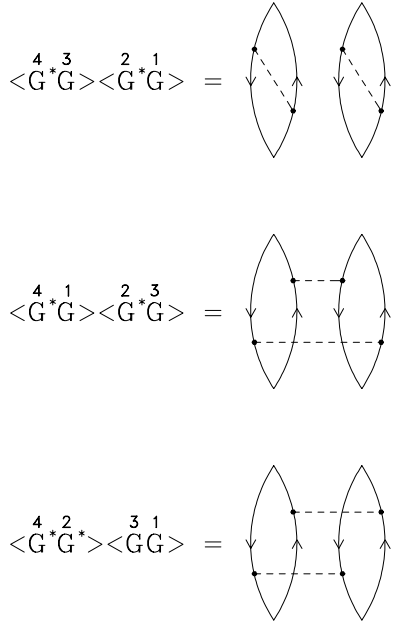}}
\vspace*{8pt}
\caption{Illustration  of the cumulant expansion of $C^{(4)}$. The
dashed lines represent gluonic correlations.\label{figCUMUL}}
\end{figure}

Let us examine $C^{(4A)}$.  
Expressing the contractions in (\ref{eq49}) in terms of quark propagators 
with the notation introduced in (\ref{eq420}) and (\ref{eq42}) we have,
in generic notation
\begin{equation} 
C^{(4A)} \sim \langle
\stackrel{4}{G^{\ast}} \stackrel{3}{G^{\phantom{\ast}}}\!
\stackrel{2}{G^{\ast}} \stackrel{1}{G^{\phantom{\ast}}}\! \rangle \,,
\label{eq51}\end{equation}
where the $\sim$ indicates the Fourier sums, etc, which carry over from
(\ref{onemes}), see (\ref{eq46}). 
The gauge configuration average in (\ref{eq51}) may be analyzed 
by means of the cumulant (or cluster) expansion theorem\footnote{The cluster
expansion theorem involves the definition
of the cumulant as a generalization of the standard deviation. For two
random variables $X_1$ and $X_2$, for example, we have $\langle X_1 X_2
\rangle = \langle X_1 \rangle \langle X_2 \rangle + \langle\!\langle X_1 X_2
\rangle\!\rangle$. If $X_1=X_2$ the square of the standard deviation
is $\sigma^2=\langle\!\langle X_1 X_2 \rangle\!\rangle$ \cite{MaS85}.}.
Taking advantage of $\langle G\rangle=0$ we have
\begin{eqnarray} 
\lefteqn{\langle \stackrel{4}{G^{\ast}} \stackrel{3}{G^{\phantom{\ast}}}
\stackrel{2}{G^{\ast}} \stackrel{1}{G^{\phantom{\ast}}} \rangle=}&& \label{eq52}\\
& & \langle \stackrel{4}{G^{\ast}} \stackrel{3}{G^{\phantom{\ast}}} \rangle 
\langle \stackrel{2}{G^{\ast}} \stackrel{1}{G^{\phantom{\ast}}} \rangle +
\langle \stackrel{4}{G^{\ast}} \stackrel{1}{G^{\phantom{\ast}}} \rangle 
\langle \stackrel{2}{G^{\ast}} \stackrel{3}{G^{\phantom{\ast}}} \rangle +
\langle \stackrel{4}{G^{\ast}} \stackrel{2}{G^{\ast}} 
\rangle \langle  \stackrel{3}{G^{\phantom{\ast}}}
\stackrel{1}{G^{\phantom{\ast}}} \rangle +
\langle\!\langle \stackrel{4}{G^{\ast}} \stackrel{3}{G^{\phantom{\ast}}}
\stackrel{2}{G^{\ast}} \stackrel{1}{G^{\phantom{\ast}}} \rangle\!\rangle \,.\nonumber
\end{eqnarray}
The last term defines the cumulant.
The first three terms on the right-hand side of (\ref{eq52})
are illustrated in Fig.~\ref{figCUMUL}.
The dashed lines indicate that the propagators are correlated through gluons.
Evidently only the first one of the three separable terms in (\ref{eq52})
represents free, uncorrelated, mesons.
All other terms in (\ref{eq52}) are sources of residual effective
interaction between the mesons.  We therefore define
\begin{equation} 
\overline{C}^{(4A)}=
\langle \stackrel{43\phantom{9}}{\phi^{\dagger}_{+\vec{p}}}
        \stackrel{34\phantom{9}}{\phi^{\phantom{\dagger}}_{+\vec{q}}} \rangle
\langle \stackrel{21\phantom{99}}{\phi^{\dagger}_{-\vec{p}}}
        \stackrel{12\phantom{99}}{\phi^{\phantom{\dagger}}_{-\vec{q}}} \rangle
\sim
\langle \stackrel{4}{G^{\ast}} \stackrel{3}{G^{\phantom{\ast}}} \rangle 
\langle \stackrel{2}{G^{\ast}} \stackrel{1}{G^{\phantom{\ast}}} \rangle \,.
\label{eq53}\end{equation}
A similar analysis of $C^{(4B)}$ leads to
\begin{equation} 
\overline{C}^{(4B)}=
\langle \stackrel{43\phantom{99}}{\phi^{\dagger}_{-\vec{p}}}
        \stackrel{34\phantom{9}}{\phi^{\phantom{\dagger}}_{+\vec{q}}} \rangle
\langle \stackrel{21\phantom{9}}{\phi^{\dagger}_{+\vec{p}}}
       \stackrel{12\phantom{99}}{\phi^{\phantom{\dagger}}_{-\vec{q}}}\rangle\,.
\phantom{ \sim
\langle \stackrel{4}{G^{\ast}} \stackrel{3}{G^{\phantom{\ast}}} \rangle 
\langle \stackrel{2}{G^{\ast}} \stackrel{1}{G^{\phantom{\ast}}} \rangle }
\label{eq54}\end{equation}
The sum of those is the free meson-meson time correlation matrix
\begin{equation} 
\overline{C}^{(4)}_{\vec{p}\,\vec{q}}(t,t_0) =
\overline{C}^{(4A)}_{\vec{p}\,\vec{q}}(t,t_0) +
\overline{C}^{(4B)}_{\vec{p}\,\vec{q}}(t,t_0) \,.
\label{eq55}\end{equation}
$\overline{C}^{(4)}$ is an additive part of the full 4-point correlator,
\begin{equation} 
C^{(4)}=\overline{C}^{(4)}+C^{(4)}_I \,.
\label{eq56}\end{equation}
The remainder
\begin{equation} 
C^{(4)}_I = \left(\rule{0pt}{0pt} C^{(4A)}+C^{(4B)}-\overline{C}^{(4)}\right)
-C^{(4C)}-C^{(4D)}
\label{Cint}\end{equation}
contains all sources of the residual interaction, 
be it from gluonic correlations, or quark exchange (or quark-antiquark loops,
if the simulation is unquenched) or any other effects.
The free correlator $\overline{C}^{(4)}$ describes two noninteracting 
identical lattice mesons. Their internal
structure and masses consistently arise from the dynamics 
determined by the lattice field model and its numerical implementation.
As already mentioned the
separable term in (\ref{eq41}) is identical zero because of the $u,d$ flavor 
assignment to the Grassmann fields of the one-meson operators (\ref{onemes}).  

$\overline{C}^{(4)}$ may be expressed in terms of the 2-point
correlator $C^{(2)}$. Using (\ref{eq41}) and (\ref{eq53}),(\ref{eq54}) gives
\begin{equation} 
\overline{C}^{(4)}_{\vec{p}\,\vec{q}}=
C^{(2)}_{\vec{p},\vec{q}}\,C^{(2)}_{-\vec{p},-\vec{q}}+ 
C^{(2)}_{-\vec{p},\vec{q}}\,C^{(2)}_{\vec{p},-\vec{q}} \,. 
\label{eq57}\end{equation}
$C^{(2)}$ is diagonal in the momentum indices. Writing
\begin{equation} 
C^{(2)}_{\vec{p}\,\vec{q}}(t,t_0)=
\delta_{\vec{p}\,\vec{q}}\,c_{\vec{p}}(t,t_0)\,e^{i\vec{p}\cdot\vec{x}_0} \,,
\label{eq58}\end{equation}
where $c_{\vec{p}}(t,t_0)$ is given by (\ref{eq44})
and has the obvious property
\begin{equation} 
c_{-\vec{p}}(t,t_0)=c^{\ast}_{\vec{p}}(t,t_0) \, ,
\label{eq59}\end{equation}
we obtain
\begin{equation} 
\overline{C}^{(4)}_{\vec{p}\,\vec{q}}(t,t_0)=
\left( \delta_{\vec{p},\vec{q}}+\delta_{-\vec{p},\vec{q}} \right)
\left| c_{\vec{p}}(t,t_0) \right|^2 \,.
\label{eq510}\end{equation}
To be consistent with the computation of 
$C^{(4)}_{\vec{p}\,\vec{q}}(t,t_0)$, using (\ref{eq46}),
$c_{\vec{p}}(t,t_0)$ should be computed from (\ref{eq43}).  
It is interesting to note that the property
\begin{equation} 
\overline{C}^{(4)}_{\vec{p},\vec{q}}=\overline{C}^{(4)}_{-\vec{p},\vec{q}}=
\overline{C}^{(4)}_{\vec{p},-\vec{q}}=\overline{C}^{(4)}_{-\vec{p},-\vec{q}} \,,
\label{eq511}\end{equation}
which is evident from (\ref{eq57}) and is explicit in (\ref{eq46}), 
also holds for $C^{(4)}$. This property 
reflects Bose symmetry with respect to 
the {\em composite} mesons. Permutation of the mesons, one with momentum 
$+\vec{p}$ the other with momentum $-\vec{p}$, results in the substitution 
$\vec{p}\rightarrow -\vec{p}$. Clearly, combining diagrams
$\overline{C}^{(4A)}$ and $\overline{C}^{(4B)}$ into $\overline{C}^{(4)}$
is crucial for the symmetry (\ref{eq511}) to hold.

\subsection{Computation with Random Sources}\label{secRanS}

The computation of  $C^{(4)}$ from the expression (\ref{eq46}) is a 
formidable numerical task. First, propagator matrix elements are needed
for arbitrary $\vec{y}$, as opposed to only one column of $G$ like in
(\ref{eq44}). Typically $G$ is computed by solving a linear equation like
$G^{-1}X^{(R)}=R$ where $R$ is a given source vector; the solution vector 
$X^{(R)}$ then represents `some column' of $G$. For arbitrary $\vec{y}$
such a strategy translates into $L^d$ linear equations, all for a fixed
time slice $t_0$ and not even counting color indices (or Dirac indices
in case of Wilson fermions).

Furthermore, the sum over lattice sites in (\ref{eq46}) contains $L^{4d}$
terms. Even for $d=2$ and our simple U(1) gauge model those are
$\approx 10^8$ terms. In a realistic calculation, for $d=3$ and SU(3),
these would be $L^{4d}3^4\approx 10^{14}$ for a meson-meson system,
and even larger if baryons are involved.
It is clear that an efficient numerical technique is called for to deal with
computing quark propagators and correlator matrices.

A practical way is the use of random
estimators \cite{Batrouni:1985jn,Duane:1986iw,Scalettar:1986uy,Gottlieb:1987eg}.
Let $R(x)$ be a vector of independent complex random deviates with
\begin{equation}
\langle R(x)R^{\ast}(y) \rule{0pt}{12pt}\rangle_R = \delta_{x,y}\,,
\label{RanDel}\end{equation}
where $ \langle\ldots\rangle_R$ denotes the random R-average. Gaussian deviates
come to mind, but $Z^2$ noise also works and may sometimes be more
desirable \cite{Dong:1994pk}. The idea is to solve
\begin{equation}
\sum_y G^{-1}(x,y)\,X^{(R)}(y) = R(x)
\label{GXR}\end{equation}
for $X^{(R)}$, recalling that the fermion matrix $G^{-1}$ is known,
then the statistical average
\begin{equation}
\langle X^{(R)}(y)R^{\ast}(z)\rule{0pt}{12pt}\rangle_R = G(y,z)
\label{Gave}\end{equation}
is an estimator for the inverse $G$ of $G^{-1}$.

Since random sources $R$ with various characteristics may be chosen the
technique is very versatile. Let us expand our notation, where
\begin{equation}
R^{(r;\alpha)}(x)\,, \quad \mbox{with} \quad r=1\ldots N_R\,,
\label{Rralpha}\end{equation}
is a finite sample of $N_R$ vectors drawn from the distribution. The label
$\alpha$ distinguishes between various types of sources, for example we
take $\alpha=t_0$ to mean	
\begin{equation}
R^{(r;t_0)}(\vec{x},t) =  R^{(r;t_0)}(\vec{x}) \delta_{t}^{t_0} \,.
\label{Rt0}\end{equation}
Here, random sources are placed on one time slice only. A more complicated
example is $\alpha=(t_0,\vec{p}\,)$ with
\begin{equation}
R^{(r;t_0,\vec{p})}(\vec{x},t) =  L^{-2}\,e^{i\vec{p}\cdot\vec{x}}
R^{(r;t_0,\vec{p})}(\vec{x}) \delta_{t}^{t_0} \,.
\label{Rt0p}\end{equation}
These are Fourier-modified (boosted) sources on time slice $t_0$.

Given a finite sample (\ref{Rralpha}), an estimator for the statistical
average of (\ref{RanDel}) is
\begin{equation}
\frac{1}{N_R}\sum_{r=1}^{N_R}  R^{(r;\alpha)}(x) {R^{(r;\beta)}}^{\ast}(y)
\simeq \sum_{\langle r \rangle}  R^{(r;\alpha)}(x) {R^{(r;\beta)}}^{\ast}(y) =
\delta_{x,y}\delta^{\alpha,\beta} \,.
\label{RanDelN}\end{equation}
The above, intuitive, notation $\sum_{\langle r \rangle}$ for the statistical
R-average
$\langle\ldots\rangle_R$ is convenient in expressions of correlation functions.
The second $\delta$ on the r.h.s. in (\ref{RanDelN}) comes from drawing
stochastically independent samples for each type of source.

For definiteness we continue with sources of type (\ref{Rt0}) set at a fixed
time slice $t_0$. Then the solutions $X$ of the linear equations (\ref{GXR})
\begin{equation}
\sum_{\vec{y}\,y_4}\,G^{-1}(\vec{x}x_4,\vec{y}y_4)\,X^{(r;t_0)}(\vec{y}y_4)
=R^{(r;t_0)}(\vec{x})\delta_{x_4}^{t_0}
\label{GXRt0}\end{equation}
can be used to estimate columns of the propagator $G$. Applying $G$ to
both sides of (\ref{GXRt0}) gives
\begin{equation}
X^{(r;t_0)}(\vec{z}z_4)
=\sum_{\vec{x}} G(\vec{z}z_4,\vec{x}t_0) R^{(r;t_0)}(\vec{x}) \,.
\label{XGRt0}\end{equation}
Multiplication with ${R^{(r;t_0)}}^{\ast}(\vec{y})$,
taking the R-average and using (\ref{RanDelN}) yields
\begin{equation}
G(\vec{z}z_4,\vec{y}t_0) =
\sum_{\langle r\rangle} X^{(r;t_0)}(\vec{z}z_4){R^{(r;t_0)}}^{\ast}(\vec{y})\,.
\label{GRX}\end{equation}
In this form the random estimator can be used directly in the expression
(\ref{eq43}) and (\ref{eq46}) for $C^{(2)}$ and $C^{(4)}$, respectively.

Another, potentially useful, relation follows from (\ref{XGRt0})
\begin{equation}
\sum_{\langle r\rangle}X^{(r;t_0)}(\vec{z}z_4){X^{(r;t_0)}}^{\ast}(\vec{y}y_4)=
\sum_{\vec{x}}\,G(\vec{z}z_4,\vec{x}t_0)\,G(\vec{y}y_4,\vec{x}t_0)^{\ast}\,.
\label{XXGG}\end{equation}
A more general version of (\ref{XXGG}) can be derived
when Fourier-modified sources (\ref{Rt0p}) are used. We have
\begin{equation}
\sum_{\langle r\rangle} X^{(r;t_0,\vec{p})}(\vec{z}z_4)
{X^{(r;t_0,\vec{q})}}^{\ast}(\vec{y}y_4)\!=\!
L^{-4}\!\sum_{\vec{x}} G(\vec{z}z_4,\vec{x}t_0)\,
e^{i(\vec{p}-\vec{q})\cdot\vec{x}}\,
G(\vec{y}y_4,\vec{x}t_0)^{\ast}.
\label{XXGpqG}\end{equation}
Direct application of (\ref{XXGpqG}) gives the 2-point correlator
(\ref{eq43}) in the form
\begin{equation} 
C^{(2)}_{\vec{p}\vec{q}}(t,t_0) = \langle 
L^{-2}\sum_{\vec{x}}\,e^{-i\vec{p}\cdot\vec{x}}
\sum_{\langle r\rangle}\, X^{(r;t_0,\vec{q})}(\vec{x}t)\,
{X^{(r;t_0,\vec{0})}}^{\ast}(\vec{x}t) \rangle \,.
\label{eq72}\end{equation}
This equation is particularly useful for testing diagonality, see (\ref{eq44}). 

In the case of the 4-point correlator two independent sets of random sources
are necessary. The four contributions $C^{(4A)}$
through $C^{(4D)}$, see (\ref{eq47}), may be computed separately, for example
\begin{eqnarray}
C^{(4A)}_{\vec p \vec q} (t,t_0) &=& 
\langle \sum_{\langle r_1\rangle} \, \sum_{\vec{x}_1} \,
{\rm e}^{-i\vec{p}\cdot\vec{x}_1}\;
X^{(r_1;t_0,\vec{0})}(\vec{x}_1t)\,{X^{(r_1;t_0,\vec{q})}}^{\ast}(\vec{x}_1t) 
\nonumber \\ &\phantom{xx}& 
\sum_{\langle r_2\rangle}\,\sum_{\vec{x}_2}\,{\rm e}^{+i\vec{p}\cdot\vec{x}_2}\;
X^{(r_2;t_0,\vec{q})}(\vec{x}_2t)\,{X^{(r_2;t_0,\vec{0})}}^{\ast}(\vec{x}_2t) 
\rangle \,.
\label{eq30w}\end{eqnarray}
There are similar expressions for $C^{(4B)}$ and $C^{(4C)}$, $C^{(4D)}$.
The noninteracting correlator $\overline{C}^{(4A)}$ is computed via
(\ref{eq72}) and the use of (\ref{eq57}).

Having selected an even number $N_R$ of random sources, in the above case
a subset of $N_R/2$ of those can be used in $\sum_{\langle r_1\rangle}$ and
the other half in $\sum_{\langle r_2\rangle}$. There are
$\binom{N_R}{\frac12 N_R}$
possibilities to choose such a subdivision (e.g. 70 for $N_R=8$ and 12870 for
$N_R=16$).

\section{Effective Residual Interaction}\label{secERI}

Potentials are not unique in the sense that a class of potentials can
produce the same phase shifts (phase-equivalent). Therefore an \ERI\
extracted from the lattice is subject to definition. In this section we
study the perturbative expansion of the time correlation matrix for
an interacting elementary Bose field on a euclidean lattice.
Our intention is to find guidance towards a sensible definition for the 
lattice simulation.

\subsection{Perturbative Definition}\label{secPDH}

Let $\hat{\cal L}_0=\hat{\cal L}_0(\hat{\phi},\partial\hat{\phi})$ be the free 
Lagrangian for an elementary Bose field $\hat{\phi}(x)$ defined on the sites 
$x=(\vec{x},t)$ of the lattice.  It is understood that $\hat{\phi}$ is subject 
to the usual canonical quantization through commutators.  Let 
$\hat{\cal L}=\hat{\cal L}_0+\hat{\cal L}_I$ such that 
$\hat{\cal L}_I=\hat{\cal L}_I(\hat{\phi})$ is a (small) interaction.  In the 
usual way $\hat{\cal L}$ gives rise to a Hamiltonian
\begin{equation} 
\hat{H}=\hat{H}_0+\hat{H}_I\,,
\label{eqA1}\end{equation}
where $\hat{H}_0$ is the free part and $\hat{H}_I$ a perturbative interaction.  
In view of (\ref{onemes}) and (\ref{twomes}) define
\begin{equation} 
\hat{\phi}_{\vec{p}}(t)=
L^{-2}\sum_{\vec{x}}\,e^{i\vec{p}\cdot\vec{x}}\hat{\phi}(\vec{x},t)
\label{eqA2}\end{equation}
and
\begin{equation} 
\hat{\Phi}_{\vec{p}}(t)=\hat{\phi}_{-\vec{p}}(t)\,\hat{\phi}_{+\vec{p}}(t) \,.
\label{eqA3}\end{equation}
In the correlation matrix
\begin{equation} 
\hat{C}^{(4)}_{\vec{p}\,\vec{q}}(t,t_c)=
\langle 0|\hat{\Phi}^{\dagger}_{\vec{p}}(t)\,
\hat{\Phi}_{\vec{q}}(t_c)|0\rangle
\label{eqA4}\end{equation}
the separable term of (\ref{eq45}) has been dropped since it is zero for 
the flavored quark fields, like in Sec.~\ref{secCor}.
The (nondegenerate) vacuum state
$|0\rangle$ satisfies $\hat{H}|0\rangle=W_0|0\rangle$.  We will assume that 
its energy is zero, $W_0=0$.  Thus the time dependence of the correlator 
(\ref{eqA4}) may be made explicit
\begin{equation} 
\hat{C}^{(4)}_{\vec{p}\,\vec{q}}(t,t_c)=
\langle 0|\hat{\Phi}^{\dagger}_{\vec{p}}(t_c)\,
e^{-\hat{H}(t-t_c)}\,\hat{\Phi}_{\vec{q}}(t_c)|0\rangle \,.
\label{eqA5}\end{equation}
Switching to the interaction picture, we define
\begin{equation} 
\hat{H}_I(t)=e^{\hat{H}_0(t-t_c)}\,\hat{H}_I\,e^{-\hat{H}_0(t-t_c)}
\label{eqA6}\end{equation}
and the (euclidean) time evolution operator\footnote{This is in analogy
to the transfer matrix formalism.}
\begin{equation} 
\hat{U}(t,t_c)=e^{\hat{H}_0(t-t_c)}\,e^{-\hat{H}(t-t_c)} \,. 
\label{eqA7}\end{equation}
The perturbative expansion of the latter
\begin{equation} 
\hat{U}(t,t_c)=\sum_{N=0}^{\infty}\,\frac{(-1)^N}{N!}\,
\int_{t_c}^t\!dt_1\ldots\int_{t_c}^t\!dt_N\,
\mbox{T}[\hat{H}_I(t_1)\ldots\hat{H}_I(t_N)]
\label{eqA8}\end{equation}
then induces a perturbative expansion of the correlator
\begin{eqnarray}
\hat{C}^{(4)}_{\vec{p}\,\vec{q}}(t,t_c)&=&
\langle 0|\hat{\Phi}^{\dagger}_{\vec{p}}(t_c)\,
e^{-\hat{H}_0(t-t_c)}\,\hat{U}(t,t_c)\,\hat{\Phi}_{\vec{q}}(t_c)|0\rangle
\label{eqA9} \\ 
&=& \sum_{N=0}^{\infty}\,\hat{C}^{(4;N)}_{\vec{p}\,\vec{q}}(t,t_c) \,.
\label{eqA10}\end{eqnarray}
The zero-order term $\hat{C}^{(4;N=0)}$ evidently describes 
noninteracting mesons.\vspace{2ex}
\\
\noindent \underline{Order $N=0$}\vspace{2ex}
\\
Let $|n\nu\rangle$ be a complete orthogonal set of eigenstates of $\hat{H}_0$
\begin{equation} 
\hat{H}_0 |n\nu\rangle = W^{(0)}_n |n\nu\rangle\,,
\label{eqA11}\end{equation}
where $W^{(0)}_n$ are free two-meson energies on the lattice, and $\nu$ is a
degeneracy index.  Lattice effects set aside, those energies should be close 
to $2\sqrt{m^2+p^2}$ with $m$ being the rest mass of one meson and
$p=|\vec{p}\,|$ its momentum.  Also, define the relative meson-meson 
momentum-space wave functions $\psi^{(0)}_{n\nu}(\vec{p}\,)$ through
\begin{equation} 
c^{(0)}_{n\nu}\psi^{(0)}_{n\nu}(\vec{p}\,) =
\langle n\nu|\hat{\Phi}_{\vec{p}}(t_c)|0\rangle^{\ast}\,,
\label{eqA12}\end{equation}
where $c^{(0)}_{n\nu}$ are normalization factors. The order $N=0$ correlator 
then is
\begin{equation} 
\hat{C}^{(4;N=0)}_{\vec{p}\,\vec{q}}(t,t_c) = 
\sum_{n\nu} |c^{(0)}_{n\nu}|^2 e^{-W^{(0)}_n(t-t_c)} 
\psi^{(0)}_{n\nu}(\vec{p}\,) \psi^{(0)\ast}_{n\nu}(\vec{q}\,) \,.
\label{eqA13}\end{equation}
With properly chosen normalization factors $c^{(0)}_{n\nu}$ we expect 
orthonormality and completeness
\begin{equation} 
\sum_{\vec{p}}\,\psi^{(0)\ast}_{n\nu}(\vec{p}\,) \psi^{(0)}_{m\mu}(\vec{p}\,) = 
\delta_{n m} \delta_{\nu\mu}
\label{eqA14}\end{equation}
\begin{equation} 
\sum_{n\nu}\,\psi^{(0)}_{n\nu}(\vec{p}\,) \psi^{(0)\ast}_{n\nu}(\vec{q}\,) = 
\delta_{\vec{p}\vec{q}} \,.
\label{eqA15}\end{equation}
For a free elementary Bose field this is almost a trivial point since
the $\psi^{(0)}_{n\nu}$ are merely  plane (lattice) waves.  A glance at 
(\ref{eqA13}) shows that, technically, those could be obtained as (normalized)
eigenvectors from diagonalizing the correlation matrix $\hat{C}^{(4;N=0)}$, 
where $|c^{(0)}_{n\nu}|^2$ are the eigenvalues, at $t=t_c$. For the 
case considered here all eigenvalues of $\hat{C}^{(4;N=0)}$ will be nonzero.

In a lattice model, however, where the role of 
$\hat{\Phi}_{\vec{p}}(t_c)$ is assumed to be a composite operator made from 
fermion fields, see (\ref{onemes}) and (\ref{twomes}), it can not be a priori 
excluded that an operator matrix element, of the type as it occurs in 
(\ref{eqA12}), is identically zero for all $\vec{p}$. In this case the free 
correlator $\overline{C}^{(4)}$ would have an eigenvalue
zero (for all $t$).  Likewise, if, for some reason, the set of hadron
operators used to construct the correlation matrices on the lattice is
linearly dependent, one should expect $\overline{C}^{(4)}$ to have an 
eigenvalue zero.\vspace{2ex}
\\
\noindent \underline{Order $N=1$}\vspace{2ex}
\\
From (\ref{eqA8})--(\ref{eqA10}) we obtain
\begin{equation} 
\hat{C}^{(4;N=1)}_{\vec{p}\,\vec{q}}(t,t_c) =
-\langle 0|\hat{\Phi}^{\dagger}_{\vec{p}}(t_c)\,
e^{-\hat{H}_0(t-t_c)}\,\int_{t_c}^{t}\!dt_1\hat{H}_I(t_1)\,
\hat{\Phi}_{\vec{q}}(t_c)|0\rangle \,.
\label{eqA16}\end{equation}
Upon inserting the complete set $|n\nu\rangle$ on both sides of 
$\hat{H}_I(t_1)$  and using (\ref{eqA6}) the $t_1$ integral over exponentials 
can be carried out explicitly.  The result is
\begin{eqnarray} 
\hat{C}^{(4;N=1)}_{\vec{p}\,\vec{q}}(t,t_c) &=& 
-\sum_{n\nu}\,\sum_{m\mu}\,\psi^{(0)}_{n\nu}(\vec{p}\,) 
\psi^{(0)\ast}_{m\mu}(\vec{q}\,) \label{eqA17}\\ & &
\langle n\nu|\hat{H}_I|m\mu\rangle c^{(0)\ast}_{n\nu} c^{(0)}_{m\mu}
\exp\left[-\frac{W^{(0)}_n+W^{(0)}_m}{2}(t-t_c)\right] \nonumber \\ & &
\left\{ (t-t_c)\delta_{nm} + 
\frac{\sinh\left[\frac{W^{(0)}_n-W^{(0)}_m}{2}(t-t_c)\right]}
{\frac{W^{(0)}_n-W^{(0)}_m}{2}}(1-\delta_{nm})
\right\} \,. \nonumber
\end{eqnarray}
Without loss of generality the normalization constants $c^{(0)}_{n\nu}$ may be 
chosen real and positive, with the phase factors being absorbed into 
$\psi^{(0)}_{n\nu}(\vec{p}\,)$, as is evident from (\ref{eqA12}). A glance 
at (\ref{eqA13}) shows that the two normalization factors and the exponential 
in (\ref{eqA17}) may be removed by multiplying the correlation matrix 
$\hat{C}^{(4;N=1)}$ from both sides with the inverse square root of
$\hat{C}^{(4;N=0)}$. Hence the matrix elements of
\begin{equation} 
\hat{\mathfrak C}^{(4;N=1)}(t,t_c) = {\hat{C}^{(4;N=0)}(t,t_c)}^{-1/2}\,
\hat{C}^{(4;N=1)}(t,t_c)\,{\hat{C}^{(4;N=0)}(t,t_c)}^{-1/2} 
\label{eqA18}\end{equation}
in the basis $\psi^{(0)}_{n\nu}(\vec{p}\,)$ are products of 
$\langle n\nu|\hat{H}_I|m\mu\rangle$ and the expression inside 
$\left\{\ldots\right\}$ of (\ref{eqA17}).  The $t$ derivative of the latter is 
equal to one at $t=t_c$.  Thus we have
\begin{equation} 
\left[ \frac{\partial \hat{\mathfrak C}^{(4;N=1)}_{\vec{p}\,\vec{q}}(t,t_c)}
{\partial t} \right]_{t=t_c} =
-\sum_{n\nu}\,\sum_{m\mu}\,\psi^{(0)}_{n\nu}(\vec{p}\,) 
\langle n\nu|\hat{H}_I|m\mu\rangle \psi^{(0)\ast}_{m\mu}(\vec{q}\,) \,.
\label{eqA19}\end{equation}
Using (\ref{eqA14}) and (\ref{eqA15}), this translates into an explicit 
equation for $\hat{H}_I$ independent of the basis
\begin{equation} 
\hat{H}_I = - \left[ \frac{\partial \hat{\mathfrak C}^{(4;N=1)}_{\vec{p}\,\vec{q}}
(t,t_c)}{\partial t} \right]_{t=t_c} \,,
\label{eqA20}\end{equation}
which is valid to order $N=1$.
Finally, we may replace $\hat{C}^{(4;N=1)}$ in 
(\ref{eqA18}) with the full correlation matrix $\hat{C}^{(4)}$.  The 
corresponding expression (\ref{eqA20}) for $\hat{H}_I$ will still be valid up 
to order $N=1$ in perturbation theory.  Thus, summarizing, define the effective 
correlator
\begin{equation} 
\hat{\mathfrak C}^{(4)}(t,t_c) = {\hat{C}^{(4;N=0)}(t,t_c)}^{-1/2}\,
\hat{C}^{(4)}(t,t_c)\,{\hat{C}^{(4;N=0)}(t,t_c)}^{-1/2} 
\label{eqA21}\end{equation}
understood as a matrix product, then the meson-meson interaction $\hat{H}_I$
satisfies
\begin{equation} 
\hat{\mathfrak C}^{(4)}(t,t_c) = e^{-\hat{H}_I(t-t_c)}
\label{eqA22}\end{equation}
up to (at least) order $N=1$ in perturbation theory.

\begin{figure}[htb]
\centerline{\includegraphics[width=102mm]{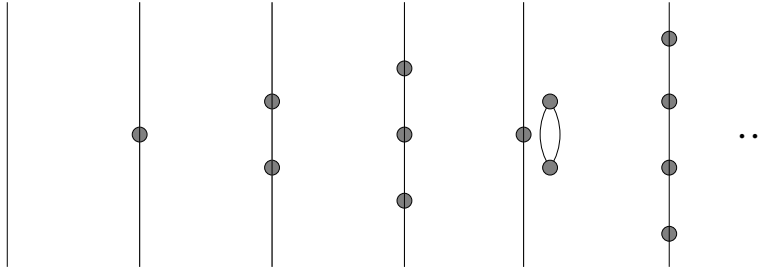}}
\vspace*{8pt}
\caption{Low-order diagrams for the elementary meson field correlator.
The circles represent $ \hat{H}_I$.\label{figHIpert}}
\end{figure}

The utility of these results in the framework of a lattice simulation
lies in the analogy which can be drawn between $\hat{C}^{(4;N=0)}$ and the free
correlator $\overline{C}^{(4)}$, and between $\hat{C}^{(4)}$ and the full
correlator $C^{(4)}$.  The analogue of (\ref{eqA22}) may then be considered
as the definition of an effective interaction. \vspace{2ex}
\\
\noindent \underline{Order $N=2$}\vspace{2ex}
\\
A calculation similar to the above reveals that (\ref{eqA22}) is valid
to order $N=2$. However, for $N>2$ the situation is complicated by the
presence of disconnected diagrams, see Fig.~\ref{figHIpert}. The diagramatic
series of connected insertions of $\hat{H}_I$ still gives rise to the
exponential form (\ref{eqA22}).
We conclude that, within the region of validity of (\ref{eqA22}), the operator
\begin{equation} 
\hat{H}_I = - \frac{\partial\ln\hat{\mathfrak C}^{(4)}(t,t_c)}{\partial t}
\label{hatERI}\end{equation}
is independent of $t$ and $t_c$.

\subsection{Effective Interaction for Composite Operators}

The above result is now easily transcribed to a system of composite hadrons.
In this case the effective correlator (\ref{eqA21}) becomes
\begin{eqnarray} 
{\mathfrak C}^{(4)}(t,t_0)
&=& {\overline{C}^{(4)}(t,t_0)}^{-1/2}\,
{C}^{(4)}(t,t_0)\,\,{\overline{C}^{(4)}(t,t_0)}^{-1/2}\nonumber\\
&=& {\overline{C}^{(4)}(t,t_0)}^{-1/2}\, 
{C}_I^{(4)}(t,t_0)\,\,{\overline{C}^{(4)}(t,t_0)}^{-1/2}+{\openone}\label{Ceff1}\\
&\stackrel{\rm def}{=}&{\mathfrak C}_I^{(4)}(t,t_0)+{\openone}\,,\nonumber
\label{Ceff2}\end{eqnarray}
where (\ref{eq56}) has been used. This then defines the \ERI
\begin{equation} 
{\cal H}_I\stackrel{\rm def}{=}-\lim_{t\rightarrow\infty}\frac{\partial
\ln{\mathfrak C}^{(4)} (t,t_0)}{\partial t} \,.
\label{ERI}\end{equation}
In the lattice simulation the eigenvalues of 
${\mathfrak C}_I^{(4)} (t,t_0)$ at asymptotic times (within lattice limits) select
the low-energy part
of the excitation spectrum of the two-body system. This will help to suppress
intrinsic excitations of the composite particles and thus yields an \ERI\
between the mesons (or hadrons) in their respective ground states. 
How well the excitation of intrinsic degrees of freedom is suppressed
depends on the lattice size and on the design of the meson operators.

\subsection{Lattice Symmetries}\label{secLatSym}

Lattice symmetries should be utilized to reduce the size
of the correlation matrices $C^{(2)}$ and $C^{(4)}$. For an $L^d\times T$
lattice the action is (usually) invariant under the group $O(d,\mathbb Z)$
of discrete transformations of the cubic sublattice. The representation
theory of these groups is standard \cite{Ham64}. For example, if $d=2$, there
are five irreducible representations $\Gamma=A_1,A_2,B_1,B_2,E$, using a
common nomenclature, with dimensionalities $N_{\Gamma}=1,1,1,1,2$ respectively.
For $d=3$ there are, again in common nomenclature, 
$\Gamma=A_1^{\pm},A_2^{\pm},E^{+},T_1^{\pm},T_2^{\pm}$ with respective
dimensionalities $N_{\Gamma}=1,1,2,3,3$ .

Given a fixed, discrete, lattice momentum $\vec{p}\,'$ the application of
group transformations $g\in O(d,{\mathbb Z})$ generates a set of momenta
$\vec{p}={\cal O}_g\,\vec{p}\,'$ that all have the same length
$p=|\vec{p}\,|$. These transformations define a representation of 
$O(d,{\mathbb Z})$ with basis vectors $|\vec{p}>$ which is in general
reducible. Let
\begin{equation}
|\vec{p}> = \sum_{\Gamma} \sum_{\epsilon} |(\Gamma,p)\epsilon> 
<(\Gamma,p)\epsilon|\vec{p}> \,,
\label{pRep}\end{equation}
where $|(\Gamma,p)\epsilon>$ denote a set of basis vectors,
$\epsilon=1\ldots$, of the subspace that belongs to $\Gamma$. 
Those can be constructed with group
theoretical techniques \cite{Ham64}. A simple example, which will
suffice for our present purposes, is given by choosing $\vec{p}=(p,0)$,
assuming $d=2$ and $pL/2\pi\in{\mathbb N}$, and the trivial one-dimensional 
representation $\Gamma=A_1$,
thus
\begin{equation}
|(A_1,p)1> =
\frac{1}{2}\left( |(+p,0)> +|(0,+p)> +|(-p,0)> +|(0,-p)> \right) \,.
\label{A1p}\end{equation}

Since both the full and the free correlators $C^{(4)}$ and $\overline{C}^{(4)}$,
respectively, commute with all group operations $g\in O(d,{\mathbb Z})$
there exist reduced matrices, say $C^{(4;\Gamma)}$ and
$\overline{C}^{(4;\Gamma)}$, within each irreducible representation $\Gamma$
such that (Schur's lemma)
\begin{equation}
<(\Gamma,p)\epsilon|C^{(4)}(t,t_0)|(\Gamma',q)\epsilon'> = 
\delta_{\Gamma\Gamma'}\delta_{\epsilon\epsilon'}
C^{(4;\Gamma)}_{p\,q}(t,t_0)
\label{C4Gam}\end{equation}
and similarly for $\overline{C}^{(4;\Gamma)}$.  Above we 
identify $C_{\vec{p}\,\vec{q}}$ with $<\vec{p}\,|C|\vec{q}>$.

Similar group theoretical considerations are of course also applicable
on a coordinate space $|\vec{r}>$ lattice.

\begin{figure}[htb]
\centerline{\includegraphics[height=82mm]{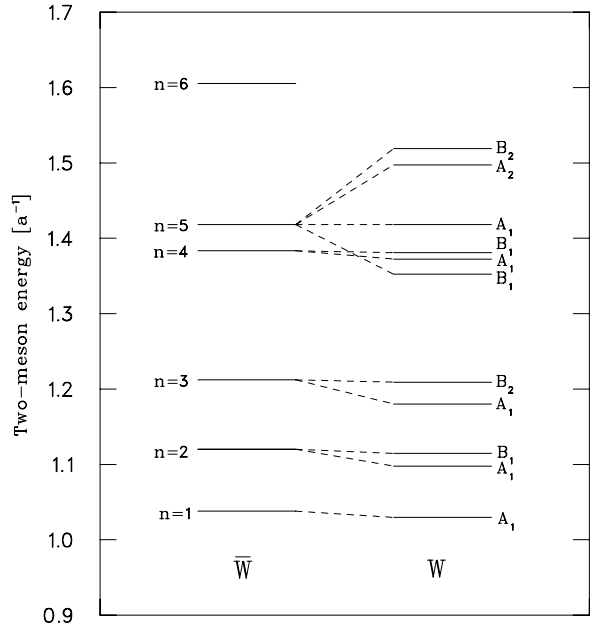}}
\vspace*{8pt}
\caption{Hyperfine splitting due to the residual interaction for the \QED\
model \protect\cite{Canosa:1997xr}. The labels for $\{ W\}$ refer to irreducible
representations of the lattice symmetry group $O(2,{\mathbb Z})$.\label{figHYPERF}}
\end{figure}

An example from the \QED\ model is shown in Fig.~\ref{figHYPERF}.
The set $\{ \overline{W}\}$ of energy levels, with labels $n=1\ldots 6$, belongs
to the free system. Those were extracted from the asymptotic time
behavior of $\overline{C}^{(4)}_{\vec{p}\,\vec{q}}(t,t_0)$, see (\ref{eq510}), 
and are degenerate (in general) for momenta of the same magnitude but different
directions. The degeneracy is lifted by the residual interaction, the set
$\{ {W}\}$ being extracted from  $C^{(4;\Gamma)}_{p\,q}(t,t_0)$.
Since the angular momentum content of the various representations
$\Gamma=A_1,A_2,B_1,B_2$ is different ($A_1$ mostly contains $\ell=0$,
$B_1$ and $B_2$ contain $\ell=2$, etc) the hyperfine splitting can be
attributed to a spin-orbit like force.

\subsection{Truncated Momentum Basis\label{secTmb}}

It is obvious from (\ref{eq510})  and (\ref{eq511}) that the reduced matrix
elements of the free correlator have the form
\begin{equation} 
\overline{C}^{(4;\Gamma)}_{p\,q}(t,t_0) = \delta_{p\,q}\,
\left| \bar{c}^{(\Gamma)}_{p}(t,t_0) \right| ^2 \, .
\label{eq73}\end{equation}
The functions $\bar{c}^{(\Gamma)}_{p}(t,t_0)$ are linear combinations of
2-point correlator elements $C^{(2)}_{\vec{p}\,\vec{p}\,'}$ with momenta
$\vec{p}$ and $\vec{p}\,'$ that differ only by their directions.

The eigenvalues of $\overline{C}^{(4;\Gamma)}(t,t_0)$
describe the time evolution of the eigenstates
of a free meson-meson Hamiltonian, say ${\cal H}_0$. Since ${\cal H}_0$
relates to a composite system, it is useful to think of its degrees of freedom
in terms of the relative motion of the two mesons (as clusters) and the
intrinsic motion of the individual mesons. Assume, for the purpose of
discussion, that the internal excitation energies are large (hard modes)
compared to the relative excitation energies (soft modes).
In principle, the separation of soft and hard modes may be enforced by choosing
a big enough $L^d\times T$ lattice:
\begin{itemize}
\item Large $L$ will yield a dense momentum spectrum $p\propto L^{-1}$.
\item Large $T$ will permit suppression of intrinsic excitations, at large $t$.
\end{itemize}
In practice there are of course numerical limitations, however, one may
check how well those conditions are satisfied.
We assume that $\overline{C}^{(4;\Gamma)}$
and $C^{(4;\Gamma)}$ are available on a truncated set of lattice momenta
and both have size $N^{(\Gamma)}$.
Naturally this will allow us to study features of the \ERI\ in the long and
intermediate range.

The advantage of working in momentum space is that $\overline{C}^{(4;\Gamma)}$
is already diagonal, see (\ref{eq73}). Thus the effective 
correlator (\ref{eqA21}),(\ref{Ceff1}) simply becomes
\begin{equation} 
{\mathfrak C}^{(4;\Gamma)}_{p\,q}(t,t_0) =
\frac{C^{(4;\Gamma)}_{p\,q}(t,t_0)}
{ \left| \bar{c}^{(\Gamma)}_{p}(t,t_0) \right|
  \left| \bar{c}^{(\Gamma)}_{q}(t,t_0) \right| }\,.
\label{eq75}\end{equation}
We envision numerical diagonalization, say 
\begin{equation}
{\mathfrak C}^{(4;\Gamma)}_{p\,q}(t,t_0) =
\sum_{n=1}^{N^{(\Gamma)}} \, v^{(\Gamma)}_n(p)
\lambda^{(\Gamma)}_n(t,t_0) \, v^{(\Gamma)}_n(q) \,.
\label{C4diag}\end{equation}
The eigenvectors $v^{(\Gamma)}_n$ are time independent \cite{Luscher:1990ck}, whereas
the eigenvalues behave exponentially for asymptotic ($\simeq$) times,
\begin{equation}
\lambda^{(\Gamma)}_n(t,t_0) \simeq
a^{(\Gamma)}_{n}\exp(-w^{(\Gamma)}_{n}(t-t_0)) \,,
\label{Wt}\end{equation}
or, if periodic boundary conditions across the time extent of the lattice
are imposed,
\begin{equation}
\lambda^{(\Gamma)}_n(t,t_0) \simeq
c^{(\Gamma)}_{n}\cosh(w^{(\Gamma)}_{n}(t-t_c)) \,.
\label{Wtcosh}\end{equation}
Now, extraction of the \ERI\ as defined in (\ref{ERI}) becomes a trivial matter.
Provided that the $v^{(\Gamma)}_n$ are orthonormal we have
\begin{equation}
<p|{\cal H}_I^{(\Gamma)}|q> =
\sum_{n=1}^{N^{(\Gamma)}} \, v^{(\Gamma)}_n(p)
\, w^{(\Gamma)}_n \, v^{(\Gamma)}_n(q) \,.
\label{HIpq}\end{equation}
These are the desired matrix elements of the \ERI\ in the truncated momentum
basis.  They may be used in a variety of ways. For example, in the basis
$|\vec{p}>$ of lattice momenta we have
\begin{equation}
<\vec{p}\,|{\cal H}_I|\vec{q}> = \sum_{\Gamma} \sum_{\epsilon} 
<\vec{p}\,|(\Gamma,p)\epsilon> 
<p\,|{\cal H}_I^{(\Gamma)}|q>
<(\Gamma,q)\epsilon|\vec{q}> \,.
\label{HIpRep}\end{equation}
This would require computation of the reduced matrix elements for all
irreducible representations
$\Gamma$ of the cubic lattice symmetry group. However, depending
on the system studied, it may well be that $A_1$ dominates the series.

The coordinate space matrix elements of the \ERI\ are obtained from
(\ref{HIpRep}) by (discrete lattice) Fourier transformation
\begin{equation}
<\vec{r}\,|{\cal H}_I|\vec{s}> = 
L^{-d}\sum_{\vec{p}}\sum_{\vec{q}}
e^{ i\vec{p}\cdot(\vec{r}-\vec{s})}
e^{-i\vec{q}\cdot(\vec{r}+\vec{s})}
<\vec{p}-\vec{q}\,|{\cal H}_I|\vec{p}+\vec{q}> \,.
\label{HIrRep}\end{equation}
It is useful to write the matrix element of ${\cal H}_I$ as a sum of 
$<-\vec{q}\,|{\cal H}_I|+\vec{q}>$ and a remainder. A reference system
where the relative momenta before and after a scattering event are $\pm\vec{q}$,
respectively, is known as the Breit frame \cite{Fes92}.
Then, the sum over $\vec{p}$ in (\ref{HIrRep}) gives rise to 
$\delta_{\vec{r}\,\vec{s}}$ and thus to a local potential ${\cal V}(\vec{r}\,)$
for the Breit-frame contribution, and a genuinely nonlocal potential
${\cal W}(\vec{r},\vec{s}\,)$ for the remainder
\begin{equation}
<\vec{r}\,|{\cal H}_I|\vec{s}> = 
\delta_{\vec{r}\vec{s}}\, {\cal V}(\vec{r}\,) + {\cal W}(\vec{r},\vec{s}\,) \,.
\label{HIVW}\end{equation}
The potentials are
\begin{equation}
{\cal V}(\vec{r}\,) =
\sum_{\vec{q}} e^{-i2\vec{q}\cdot\vec{r}}
<-\vec{q}\,|{\cal H}_I|+\vec{q}>
\label{Vpot}\end{equation}
\begin{eqnarray}
{\cal W}(\vec{r},\vec{s}\,) &=&
L^{-d}\sum_{\vec{p}}\sum_{\vec{q}}
e^{ i\vec{p}\cdot(\vec{r}-\vec{s})}
e^{-i\vec{q}\cdot(\vec{r}+\vec{s})} \nonumber\\
&&\left( <\vec{p}-\vec{q}\,|{\cal H}_I|\vec{p}+\vec{q}>
-<-\vec{q}\,|{\cal H}_I|+\vec{q}> \right) \,.
\label{Wpot}\end{eqnarray}

\begin{figure}[htb]
\centerline{\includegraphics[width=56mm]{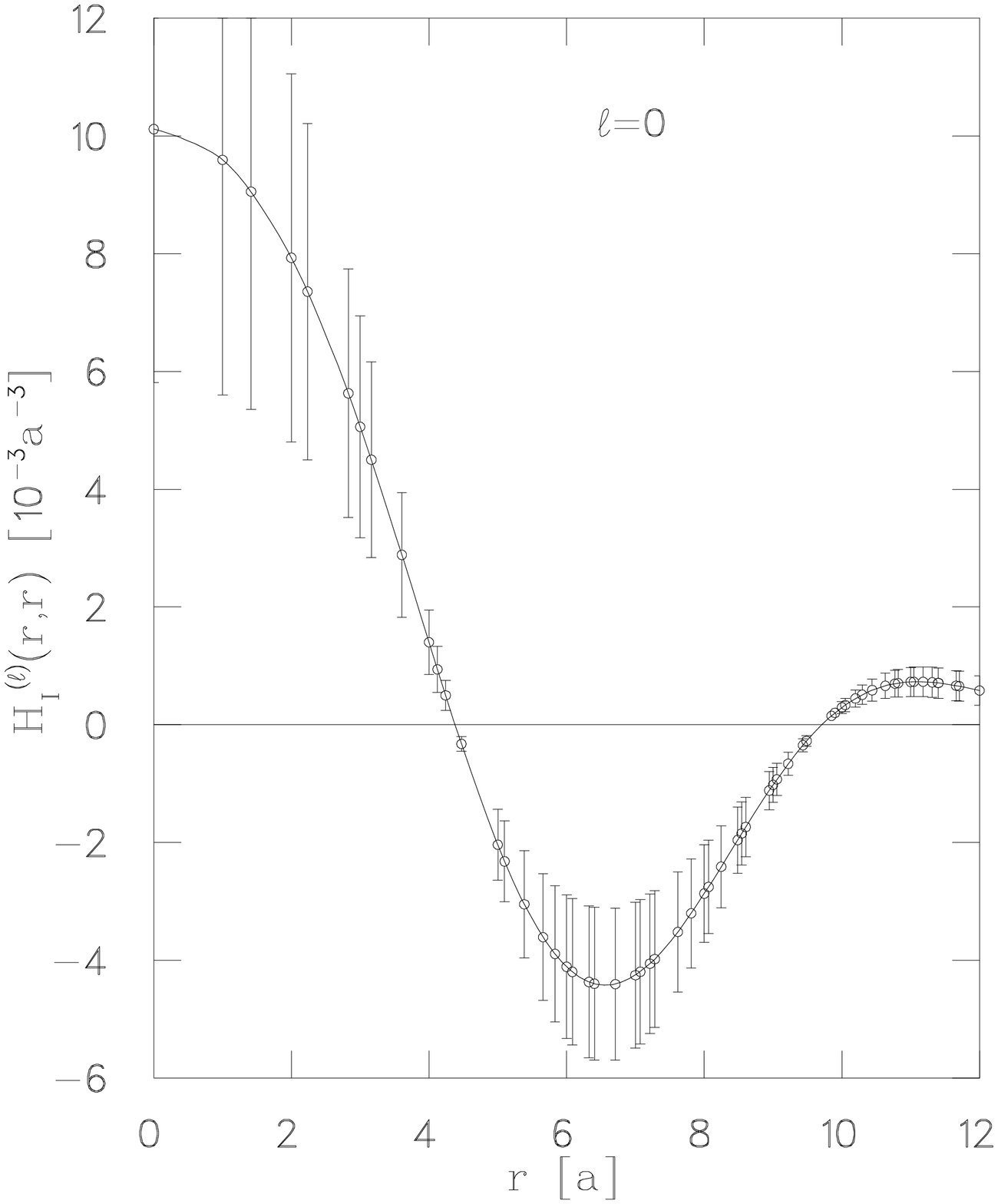}
            \includegraphics[width=56mm]{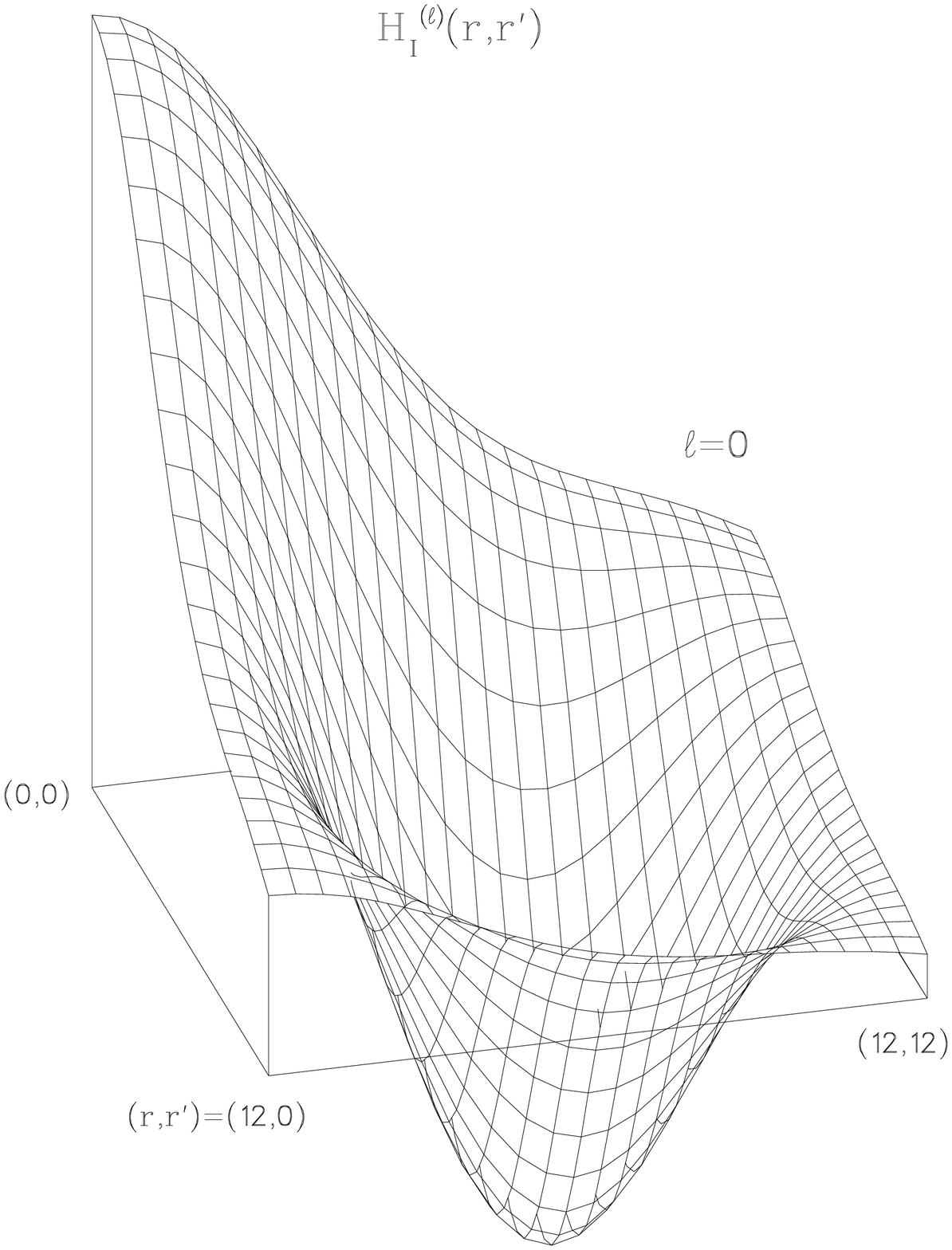}}
\vspace*{8pt}
\caption{Example of a meson-meson potential derived within the
momentum-space approach. Local and nonlocal potentials are shown.
The model is \QED\ in 2+1 dimensions \protect\cite{Canosa:1997xr}\label{figEVT30_L0}.}
\end{figure}

\begin{figure}[htb]
\centerline{\includegraphics[height=82mm]{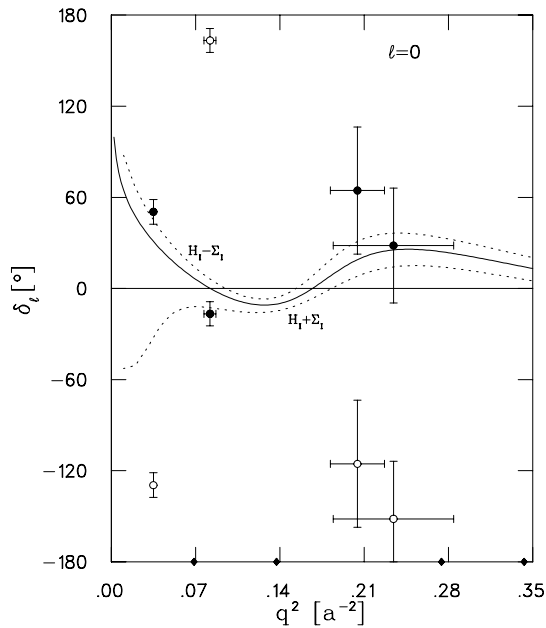}}
\vspace*{8pt}
\caption{Scattering phase shifts, for the $s$-wave, obtained from the effective
interaction depicted in Fig.~\protect\ref{figEVT30_L0}.
The four data points (filled
circles) were computed in Ref.~\protect\cite{Fiebig:1994qi}, using L\"{u}scher's method.
The same set of phase shifts are also shown modulo $\mbox{180}^\circ$ (open
circles). The diamonds on the abscissa correspond to the values of
$p^2=(2\pi/L)^2k^2$ for an $L=24$ lattice.\label{figJOST2D_pha}}
\end{figure}

Results from an actual simulation are displayed in 
Fig.~\ref{figEVT30_L0}. The (lattice) Fourier transform, see (\ref{HIrRep}),
of (\ref{HIpRep}) is shown. An angular momentum projection was employed. Most
of the $\ell=0$ partial wave is contained in the $A_1$ sector.
The corresponding scattering phase shifts, obtained from a
Schr\"{o}dinger equation, are shown in Fig.~\ref{figJOST2D_pha}.

\subsection{Adiabatic Approximation}\label{secAdi}

Probing the lattice with a set of coordinate space operators of the type
\begin{equation}
\phi_{\vec{x}}(t)=\sum_{\vec{p}}\,e^{-i\vec{p}\cdot\vec{x}}
\phi_{\vec{p}}(t) =
\bar{\chi}_{d}(\vec{x},t)\,\chi_{u}(\vec{x},t)\,,
\label{onemesx}\end{equation}
where we have used (\ref{onemes}), gives us an alternative window on the
\ERI. The Fourier transform of the two-meson operator  (\ref{twomes}) becomes
\begin{equation}
\Phi_{\vec{r}}(t)= 
\sum_{\vec{p}}\,e^{-i\vec{p}\cdot\vec{r}} \Phi_{\vec{p}}(t) =
\sum_{\vec{x}}\sum_{\vec{y}} \delta_{\vec{r},\vec{x}-\vec{y}}
\,\phi_{\vec{y}}(t)\,\phi_{\vec{x}}(t) \,.
\label{Phir}\end{equation}
It corresponds to two composite mesons with relative distance $\vec{r}$ and
total momentum zero.

Following Secs.~\ref{secCor} and \ref{secCor0} we consider the free
meson-meson correlator
\begin{equation}
\overline{C}^{(4)}_{\vec{r}\,\vec{s}}(t,t_0) =
\sum_{\vec{p}}\sum_{\vec{q}}\,
e^{i\vec{p}\cdot\vec{r}} e^{-i\vec{q}\cdot\vec{s}}
\,\,\overline{C}^{(4)}_{\vec{p}\,\vec{q}}(t,t_0) \,.
\label{Cbarx}\end{equation}
It may also be written, using (\ref{eq510}), as
\begin{equation}
\overline{C}^{(4)}_{\vec{r}\,\vec{s}}(t,t_0) =
\sum_{\vec{p}}
\left( e^{i\vec{p}\cdot(\vec{r}
-\vec{s})}+e^{i\vec{p}\cdot(\vec{r}+\vec{s})}\right)
|c_{\vec{p}}(t,t_0)|^2 \,.
\label{barCrs}\end{equation}
Here $c_{\vec{p}}(t,t_0)$ is the correlation function of a single-meson
interpolating operator with momentum $\vec{p}$, see (\ref{eq58}). Its
(exponential) time behavior is determined by the total energy $E(p)=
\sqrt{m^2+p^2}$, assuming continuum dispersion for the moment. In the
non-relativistic limit, $E(p)\rightarrow m+p^2/2m$ as $m\rightarrow\infty$, 
the correlator thus becomes independent of $\vec{p}$ and may be replaced
with $c_{\vec{p}=0}(t,t_0)$ in (\ref{barCrs}). This leads to the 
approximation
\begin{equation}
\overline{C}^{(4)}_{\vec{r}\,\vec{s}}(t,t_0) \approx
L^2(\delta_{\vec{r},\vec{s}}+\delta_{\vec{r},-\vec{s}})
|c_{\vec{p}=0}(t,t_0)|^2 \,.
\label{barCdeltars}\end{equation}
In coordinate space the free correlator is diagonal in the 
$m\rightarrow\infty$ limit, whereas in momentum space diagonality is exact.

In terms of the fermion propagator we have, see (\ref{eq58}) and
(\ref{eq43}),(\ref{eq44}),
\begin{equation} 
c_{\vec{p}=0}(t,t_0)=
\langle L^{-4}\sum_{\vec{x}}\sum_{\vec{y}}
|G(\vec{x}t,\vec{y}t_0)|^2\rangle =
\langle L^{-2}\sum_{\vec{x}}
|G(\vec{x}t,\vec{x}_0t_0)|^2\rangle \,.
\label{cp0}\end{equation}
This is just the time correlation function for a single meson at rest.

Using lattice symmetry, see Secs.~\ref{secLatSym} and \ref{secTmb}, the reduced
correlator matrix has the form
\begin{equation}
\overline{C}^{(4;\Gamma)}_{r\,s}(t,t_0) \approx
\delta_{r,s} |\overline{c}^{(\Gamma)}(t,t_0)|^2 \,,
\label{barCdeltarsGam}\end{equation}
where the asymptotic time behavior is given by
\begin{equation}
\overline{c}^{(\Gamma)}(t,t_0) \simeq
\bar{a}^{(\Gamma)}\,e^{-m(t-t_0)}\,,
\label{barcexp}\end{equation}
at least for $\Gamma=A_1$. The full, interacting, correlator matrix is
built from (\ref{Phir})
\begin{equation} 
C^{(4)}_{\vec{r}\,\vec{s}}(t,t_0)=
\langle\Phi^{\dagger}_{\vec{r}}(t)\,\Phi_{\vec{s}}(t_0)\rangle-
\langle\Phi^{\dagger}_{\vec{r}}(t)\rangle
\langle\Phi^{\phantom{\dagger}}_{\vec{s}}(t_0)\rangle \,.
\label{C4rs}\end{equation}
Again, the separable term vanishes in our example (flavor assignment).
Expressing $C^{(4)}$,
via Wick's theorem, in terms of the fermion propagator and the diagrammatic
classification proceeds just like in Secs.~\ref{secCor} and \ref{secCor0}.
Alternatively, we may simply Fourier transform (\ref{eq46}),
\begin{eqnarray}
C^{(4)}_{\vec{r}\,\vec{s}} (t,t_0) &=& L^{-4} \: \sum_{\vec x_1} \: 
\sum_{\vec x_2} \: \sum_{\vec y_1} \: \sum_{\vec y_2} \:
\delta_{\vec{r},\vec x_1 - \vec x_2} \delta_{\vec{s},\vec y_2 - \vec y_1}\\
&& \mbox{} \left\langle \left( \: G^\ast(\vec x_2t, \vec y_2t_0) \,
G(\vec x_2t, \vec y_2t_0) \, G^\ast(\vec x_1t, \vec y_1t_0) \,
G(\vec x_1t, \vec y_1t_0) \right. \right. \nonumber \\
&& \mbox{} + G^\ast(\vec x_1t, \vec y_2t_0) \,
G(\vec x_1t, \vec y_2t_0) \, G^\ast(\vec x_2t, \vec y_1t_0) \,
G(\vec x_2t, \vec y_1t_0) \nonumber \\
&& \mbox{} - G^\ast(\vec x_2t, \vec y_1 t_0) \,
G(\vec x_2t, \vec y_2 t_0) \, G^\ast(\vec x_1t, \vec y_2 t_0) \,
G(\vec x_1t, \vec y_1 t_0) \nonumber \\
&& \mbox{} - \left. \left. G(\vec x_2t, \vec y_1 t_0) \,
G^\ast(\vec x_2t, \vec y_2 t_0) \, G(\vec x_1t, \vec y_2 t_0) \,
G^\ast(\vec x_1t, \vec y_1 t_0)  \right) \right\rangle \,.\nonumber
\label{eq46rs}\end{eqnarray}
In keeping with the $m\rightarrow\infty$ limit it is reasonable to assume
that the relative distance between the (heavy) mesons will not change
much during the propagation from $t_0$ to $t$.
After all, the meson propagation is described by (\ref{cp0}).
This leads us to neglect
the off-diagonal elements of $C^{(4)}$. More precisely, since it costs
no energy to change the system according to a transformation that belongs
to the lattice symmetry group $O(d,{\mathbb Z})$, the correct approximation is
\begin{equation}
C^{(4;\Gamma)}_{r\,s}(t,t_0) \approx \delta_{r\,s}\,
C^{(4;\Gamma)}_{r\,r}(t,t_0) \,.
\label{Cdiag}\end{equation}
Now, the eigenvalues of $C^{(4;\Gamma)}(t,t_0)$ are just the diagonal
elements. Those behave exponentially for asymptotic times.
\begin{equation}
C^{(4;\Gamma)}_{r\,r}(t,t_0) \simeq
a^{(\Gamma)}_r\,e^{-W^{(\Gamma)}_r(t-t_0)}\,.
\label{Crrt}\end{equation}
The effective correlator (\ref{eq75}) is also diagonal in this approximation
\begin{equation}
{\mathfrak C}^{(4;\Gamma)}_{r\,s}(t,t_0) \approx \delta_{r\,s}\,
{\mathfrak C}^{(4;\Gamma)}_{r\,r}(t,t_0) \,,
\label{calCdiag}\end{equation}
with
\begin{equation}
{\mathfrak C}^{(4;\Gamma)}_{r\,r}(t,t_0) \simeq 
\frac{a^{(\Gamma)}_r}{|\bar{a}^{(\Gamma)}|^2}
e^{-\left[ W^{(\Gamma)}_r-2m\right](t-t_0)}\,.
\label{calCt}\end{equation}
According to our definition (\ref{ERI}) the \ERI\ is
\begin{equation} 
<r|{\cal H}_I^{(\Gamma)}|s> \approx  \delta_{r\,s} {\cal V}^{(\Gamma)}(r)\,,
\quad\mbox{with}\quad
{\cal V}^{(\Gamma)}(r)= W^{(\Gamma)}_r -2m \,.
\label{HIVWrs}\end{equation}

The above line of arguments is somewhat similar in spirit to the
Born-Oppenheimer approximation often used with systems of two atoms or
mo\-lecules \cite{Mes66b}. The time scale for the dynamics of the
interaction is much faster than the time scale for the dynamics of the
motion for the atoms. Borrowing a thermodynamics term, we speak of an
adiabatic approximation. In our case we simply compute the total
energy of the two-meson system at fixed (static) relative distance $r$
and interpret the difference to the ground state energy as a
potential \cite{Fiebig:1996mx}.

\begin{figure}[htb]
\centerline{\includegraphics[width=56mm]{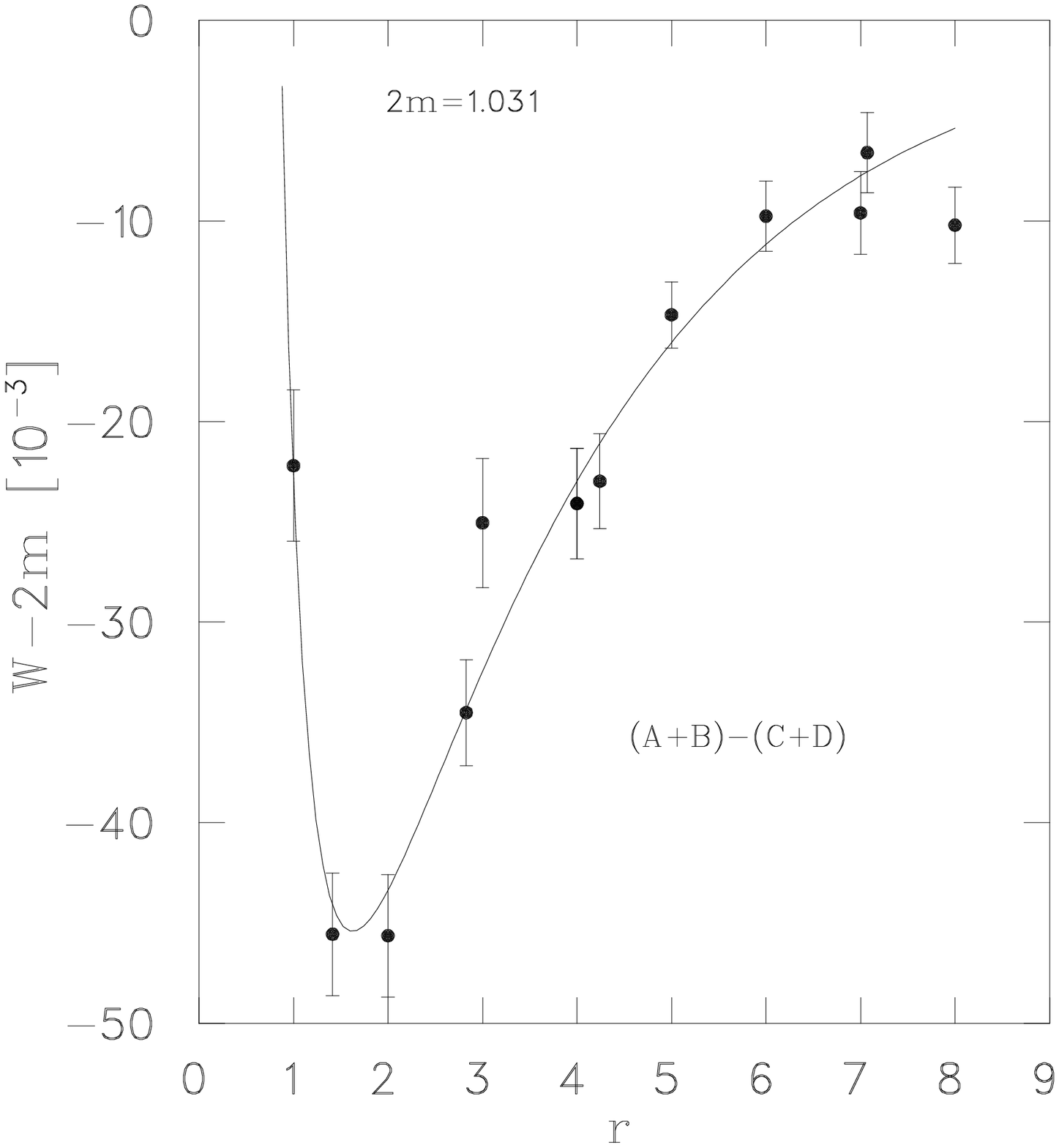}
            \includegraphics[width=56mm]{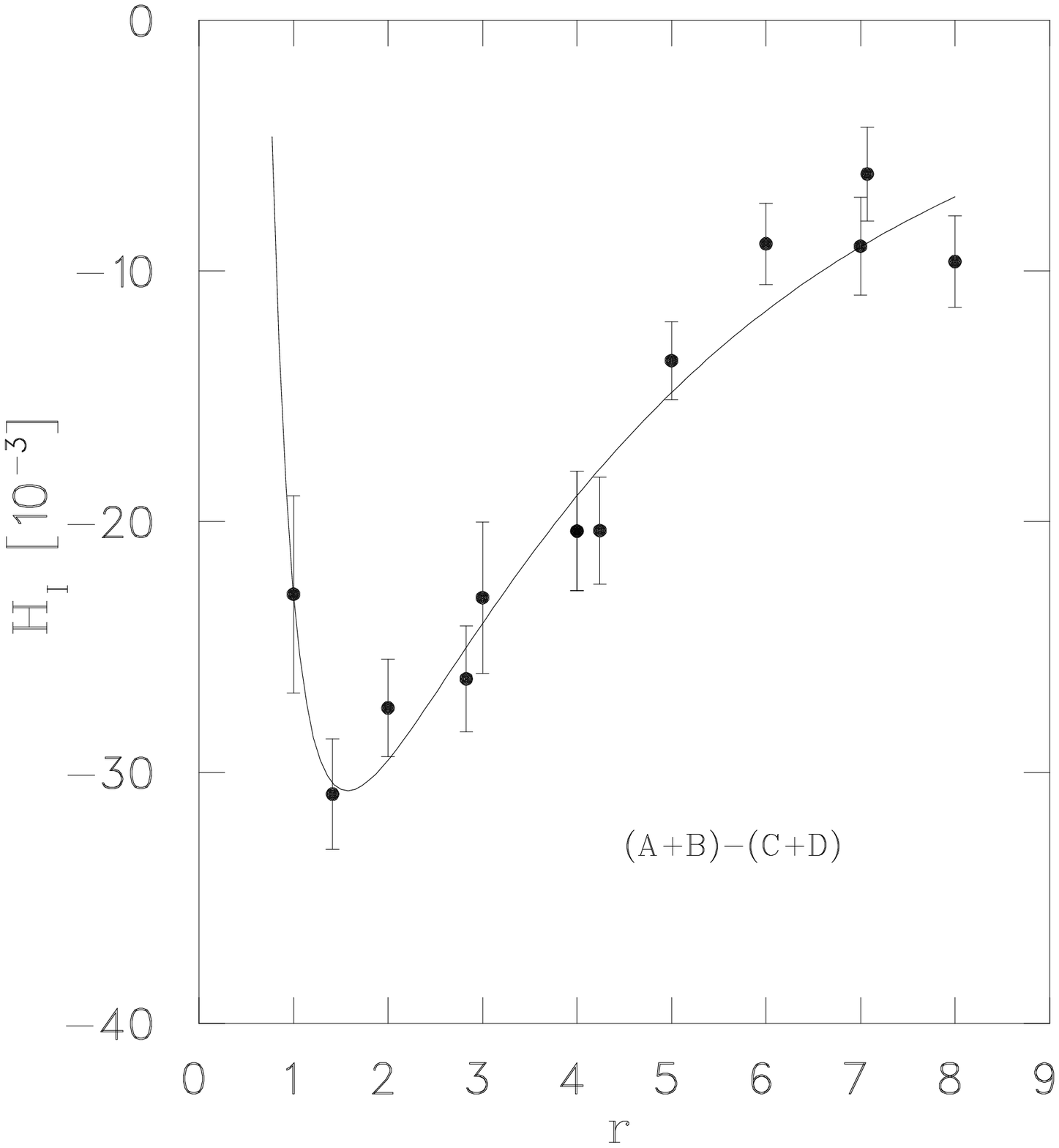}}
\vspace*{8pt}
\caption{Static meson-meson potential for the \QED\ model according to
\protect(\ref{HIVWrs}), on the left. On the right a somewhat different
approximation according to (\protect\ref{Var}) serves to check the
perturbative definition of ${\cal H}_I$.\label{figMAS2}}
\end{figure}

An example from numerical work on the \QED\ model \cite{Fiebig:1996xi} is shown in
Fig.~\ref{figMAS2}. The left part displays the static potential
${\cal V}^{(\Gamma)}$ according to (\ref{HIVWrs}). The distance
$\vec{r}=0$ is forbidden by the Pauli exclusion principle. This is the reason
for the increase (repulsive core) of the potential as $r\rightarrow 0$.

The right part of Fig.~\ref{figMAS2} shows a potential computed slightly
differently, for the sake of comparison. The perturbative version of
(\ref{calCt}) leads to
\begin{equation}
{\cal V}^{(\Gamma)}(r) \approx \frac{1}{t-t_0} \left[ 1 -
\frac{{\mathfrak C}^{(4;\Gamma)}_{r\,r}(t,t_0)} {a^{(\Gamma)}_r
|\bar{a}^{(\Gamma)}|^{-2}} \right] \,.
\label{Var}\end{equation}
In this way the $r$-dependence of $a^{(\Gamma)}_r$ enters the analysis.
The results in Fig.~\ref{figMAS2} are similar.
They shed some light on the validity of the perturbative definition
of ${\cal H}_I$ put forward in Sec.~\ref{secPDH}.

Of course, only the local part of the \ERI\ is accessible in this way.
Note, however, that a local potential
\begin{equation}
{\cal V}(\vec{r}\,) = \sum_{\Gamma} \sum_{\epsilon} 
<\vec{r}\,|(\Gamma,r)\epsilon> 
{\cal V}^{(\Gamma)}(r)
<(\Gamma,r)\epsilon|\vec{r}> \,,
\end{equation}
compared to (\ref{HIpRep}), may not be the same
as the approximate local potential of (\ref{Vpot}) even if
the untruncated correlation matrices are used. The reason is that 
diagonal elements of the non-local part of the \ERI\ may contribute to
the adiabatic local potential.

The adiabatic approximation hinges on the assumption of heavy
($m\rightarrow\infty$) partners. This should be reasonable if at least
one of the quarks in each hadron is heavy, say an s-quark. 
However, in the chiral limit ($m_{\pi}\rightarrow 0$), using Wilson
fermions, the adiabatic approximation is expected to fail if pseudoscalar
mesons are involved. In this case momentum space methods like in
Sec.~\ref{secTmb} are preferable.
Finally, the quality of the adiabatic approximation can of course be tested
numerically by computing off-diagonal elements of the correlation matrices.

\subsection{Analysis on a Periodic Lattice}\label{secPeriLat}

Periodic boundary conditions across the space extent of the lattice are a
common choice. The potentials ${\cal V}(\vec{r}\,)$ and
${\cal W}(\vec{r},\vec{s}\,)$, see (\ref{Vpot}) and (\ref{Wpot}), extracted
from the lattice simulation, in one or the other way, will reflect those
conditions. In particular, the maximal usable relative distance $r$ is $L/2$.

\begin{figure}[htb]
\centerline{\includegraphics[height=74mm]{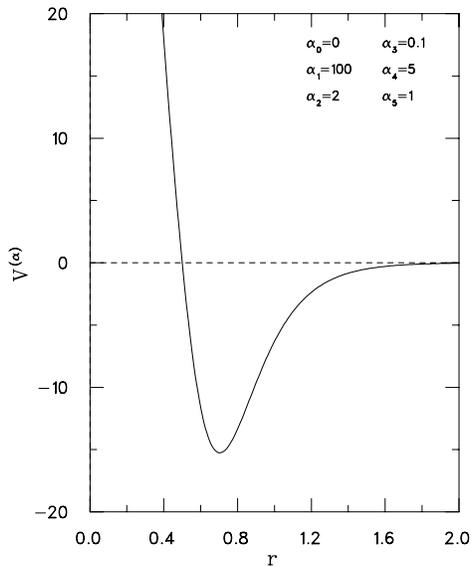}}
\vspace*{8pt}
\caption{Example of $V^{(\alpha)}$ as defined in (\protect\ref{Valpha}). The
parameters $\alpha$ are arbitrarily chosen, for the purpuse of
illustration.\label{figVlocFit}}
\end{figure}

Suppose it is desired to make a fit to the lattice local potential with a
class of functions $V^{(\alpha)}(\vec{r}\,)$ depending on a set of parameters
$\alpha$. An example is
\begin{equation}
V^{(\alpha)}(\vec{r}\,) = \alpha_1\frac{1-\alpha_2 r^{\alpha_5}}
{1+\alpha_3 r^{\alpha_5+1}e^{\alpha_4 r}} + \alpha_0 \,.
\label{Valpha}\end{equation}
This class (see Fig.~\ref{figVlocFit}) has enough flexibility
to match heuristic features of the hadronic interaction,
like short-range repulsion or attraction as the case may be, and a
long range Yukawa form
\begin{equation}
V^{(\alpha)}(\vec{r}\,) \rightarrow - \frac{\alpha_1\alpha_2}{\alpha_3}
\frac{e^{-\alpha_4 r}}{r} + \alpha_0
\quad\mbox{as}\quad r\rightarrow\infty \,.
\label{Vyukawa}\end{equation}
The periodic extension, defined as
\begin{equation}
V^{(\alpha)}_L(\vec{r}\,) = \sum_{\vec{n}}
V^{(\alpha)}(\vec{r}-\vec{n}L) \,,
\label{ValphaL}\end{equation}
where $\vec{n}$ are vectors with $d$ whole number components, can then be
used to fit the lattice potentials with
\begin{equation}
\chi^2(\alpha) = \frac{1}{N}\sum_{\vec{r}}
|{\cal V}(\vec{r}\,)-V^{(\alpha)}_L(\vec{r}\,)|^2
\frac{1}{\sigma^2(\vec{r}\,)} \,.
\label{chi2alpha}\end{equation}
Here the sum runs over the set of $N$ lattice distances $\vec{r}$ for which
${\cal V}(\vec{r}\,)$ has been computed.

The idea behind using a periodic extension, like (\ref{ValphaL}), is to
account for the fact that the lattice hadrons are interacting
with their replicas in adjacent copies of the periodic lattice. 
With $\chi^2$ minimized, we would take $V^{(\alpha)}(\vec{r}\,)$ and 
$W^{(\beta)}(\vec{r},\vec{s}\,)$ obtained from a similar fit,
as the results of the simulation.

\section{Current State in QCD in 3+1 Dimensions}

At the time of this writing, lattice work on the subject of hadronic
interaction is in an exploratory phase.
Nevertheless, it is useful to convey what has been
done through some selected examples. We will not include in this section
results other than \QCD, for example $O(4)$ symmetric field theories
and four-fermion models \cite{Zimmermann:1993kx,Gockeler:1994rx,Gockeler:1996mu},
models exploring static quark geometry\footnote{The contribution by A. M. Green
is devoted to this subject.} \cite{Green:1993ag,Green:1993yw}, studies with an
SU(2) gauge group \cite{Stewart:1998hk}, and work using the hamiltonian lattice
formulation \cite{Chaara:1994ar}.

\subsection{Scattering Lengths for $\pi$ and N Systems}\label{secScatLen}

To date the most elaborate application within \QCD\ has been via L\"{u}\-scher's
formula (\ref{ScatLen}) by the CP-PACS collaboration aiming at scattering
lengths for the $\pi$--$\pi$, $\pi$--N, K--N, $\bar{\rm K}$--N, and N--N
systems \cite{Kuramashi:1993yu,Kuramashi:1993ka,Fukugita:1994na,Fukugita:1995ve,Kuramashi:1996sc,Aoki:1999pt,Aoki:2001hc,Aok02hc,Aoki:2002in}.
The collaboration has applied considerable resources to this problem,
recently using lattices as large as $L^3\times T=32^3\times 60$
\cite{Aoki:2001hc}.
However, for most of their earlier explorations \cite{Fukugita:1995ve}
lattices typically of size $12^3\times 20$ with lattice lattice constants in the range
of $a\approx 0.15$ -- $0.30$~fm were used.

Extracting the s-wave scattering length $a_0$ from L\"{u}scher's formula
\begin{equation}
W_{h_1 h_2}-(m_{h_1}+m_{h_2}) =
-\frac{2\pi(m_{h_1}+m_{h_2})a_0}{m_{h_1}m_{h_2}L^3}
[1+c_1\frac{a_0}{L}+c_2(\frac{a_0}{L})^2+ o(L^{-3})]\,,
\label{ScatLen1}\end{equation}
where $h_1$ and $h_2$ denote hadrons 1 and 2, see (\ref{ScatLen}),
requires a standard mass calculation on the lattice.
One therefore needs to compute correlation functions of suitable operators,
say $\phi_{h_1}(t)$ and $\phi_{h_2}(t)$.
For illustration, some examples using Wilson fermions are
\begin{eqnarray}
\phi_{\pi^+}(t) &=& 
 -\sum_{\vec{x}}\bar{\psi}_{dA}(\vec{x},t) \gamma_5 {\psi}_{uA}(\vec{x},t)\\
\phi_{\pi^0}(t) &=& \frac{1}{\sqrt{2}} \sum_{\vec{x}}
\left(\bar{\psi}_{uA}(\vec{x},t) \gamma_5 {\psi}_{uA}(\vec{x},t)
 -\bar{\psi}_{dA}(\vec{x},t) \gamma_5 {\psi}_{dA}(\vec{x},t) \right)\rule{6ex}{0mm} \\
\phi_{K^+}(t) &=& 
 \sum_{\vec{x}}\bar{\psi}_{sA}(\vec{x},t) \gamma_5 {\psi}_{uA}(\vec{x},t) \,,
\end{eqnarray}
etc, where $A=1,2,3$ represents SU(3) color. Only zero-momentum interpolating
fields are needed here. Kogut-Susskind,
or staggered, fermions \cite{Kogut:1975ag,Susskind:1977jm,Kawamoto:1981hw} were also
used by the CP-PACS collaboration. The construction of operators with specific
hadron quantum numbers is technically more complicated, however \cite{Kilcup:1987dg}.
The authors of Ref.~\cite{Fukugita:1995ve} chose to construct
correlation functions where sources and sinks between $h_1$ and $h_2$
are one time slice apart. For the one-hadron systems this means
\begin{eqnarray}
C_{h_1}(t,0) &=& \langle \phi_{h_1}(t) \phi^{\dagger}_{h_1}(0) \rangle
-\langle \phi_{h_1}(t) \rangle\langle \phi^{\dagger}_{h_1}(0) \rangle \\
&\simeq& \bar{a}_{h_1} e^{-m_{h_1}t} \nonumber\\
C_{h_2}(t+1,1) &=& \langle \phi_{h_2}(t+1) \phi^{\dagger}_{h_2}(1) \rangle
-\langle \phi_{h_2}(t+1) \rangle\langle \phi^{\dagger}_{h_2}(1) \rangle \\
&\simeq& \bar{a}_{h_2} e^{-m_{h_2}t}\,,\nonumber
\end{eqnarray}
where $\simeq$ indicates the asymptotic behavior.
The two-hadron interpolating fields are constructed from linear combinations
of products of one-hadron operators, respecting the one-time-slice offset.
Denoting those by $\Phi_{h_1 h_2}(t,t+1)$ illustrational examples
in the $\pi$--$\pi$ system with isospin $I$ are
\begin{eqnarray}
\Phi_{\pi\pi}^{(I=0,I_3=0)}(t,t+1)&=&
\frac{1}{\sqrt{3}} \left( \phi_{\pi^+}(t)\phi_{\pi^-}(t+1)\right. \\
&& \left. -\phi_{\pi^0}(t)\phi_{\pi^0}(t+1)
+\phi_{\pi^-}(t)\phi_{\pi^+}(t+1) \right)\nonumber \\
\Phi_{\pi\pi}^{(I=2,I_3=2)}(t,t+1) &=&
\phi_{\pi^+}(t)\phi_{\pi^+}(t+1) \,.
\end{eqnarray}
In the s-wave $I=1$ is not allowed (Bose symmetry).
The corresponding two-hadron system correlator is
\begin{eqnarray}
\lefteqn{C_{h_1 h_2}(t+1,t,1,0) =}&& \\
&& \langle \Phi_{h_1 h_2}(t,t+1)\Phi^{\dagger}_{h_1 h_2}(0,1)\rangle
-\langle \Phi_{h_1 h_2}(t,t+1)\rangle
\langle\Phi^{\dagger}_{h_1 h_2}(0,1)\rangle \,.\nonumber
\end{eqnarray}
The measurement of the s-wave scattering length $a_0$ via L\"{u}scher's formula
(\ref{ScatLen1}) requires the energy difference
$W_{h_1 h_2}-(m_{h_1}+m_{h_2})$, which can be extracted from
the asymptotic time behavior of the ratio of correlation functions 
\begin{equation}
R(t)\stackrel{\rm def}{=}\frac{C_{h_1 h_2}(t+1,t,1,0)}{C_{h_1}(t,0)C_{h_2}(t+1,1)} \simeq
\frac{a_{h_1 h_2}}{\bar{a}_{h_1}\bar{a}_{h_2}}
e^{-( W_{h_1 h_2}-(m_{h_1}+m_{h_2}))t} \,.
\label{RCCCC}\end{equation}

Baryon interpolating field operators are more complicated. For the
proton ($N^+$) and neutron ($N^0$) standard choices for
zero-momentum operators are
\begin{eqnarray}
\phi_{N^+}(t) &=& \sum_{\vec{x}}\epsilon_{ABC}\left[
\left(\bar{\psi}_{uA}^T(\vec{x},t)\tilde{\cal C}\psi_{dB}(\vec{x},t)\right)
\psi_{uB}(\vec{x},t)\right.\nonumber \\
&&\phantom{xxxxxxxx}\left.-\left(\bar{\psi}_{dA}^T(\vec{x},t)\tilde{\cal C}\psi_{uB}(\vec{x},t)\right)
\psi_{uB}(\vec{x},t) \right] \\
\phi_{N^0}(t) &=& \sum_{\vec{x}}\epsilon_{ABC}\left[
\left(\bar{\psi}_{uA}^T(\vec{x},t)\tilde{\cal C}\psi_{dB}(\vec{x},t)\right)
\psi_{dB}(\vec{x},t)\right.\nonumber \\
&&\phantom{xxxxxxxx}\left.-\left(\bar{\psi}_{dA}^T(\vec{x},t)\tilde{\cal C}\psi_{uB}(\vec{x},t)\right)
\psi_{dB}(\vec{x},t) \right]\,,
\label{Cpn}\end{eqnarray}
where $\tilde{\cal C}$ means charge conjugation and $A,B,C$ are color indices.
Operators for the ${}^3S_1$ and ${}^1S_0$ channels of the N--N system
are constructed in Ref.~\cite{Fukugita:1995ve}
\begin{eqnarray}
\Phi_{{}^3S_1}(t,t+1) &=& \frac{1}{\sqrt{2}}\left(\phi_{N^+}(t)\phi_{N^0}(t+1)
-\phi_{N^0}(t)\phi_{N^+}(t+1)\right)\rule{6ex}{0mm} \\
\Phi_{{}^1S_0}(t,t+1) &=& \phi_{N^+}(t)\phi_{N^+}(t+1) \,.
\end{eqnarray}
Again, ratios like (\ref{RCCCC}) allow the extraction of the desired mass shifts.

\begin{figure}[htb]
\centerline{\includegraphics[width=56mm]{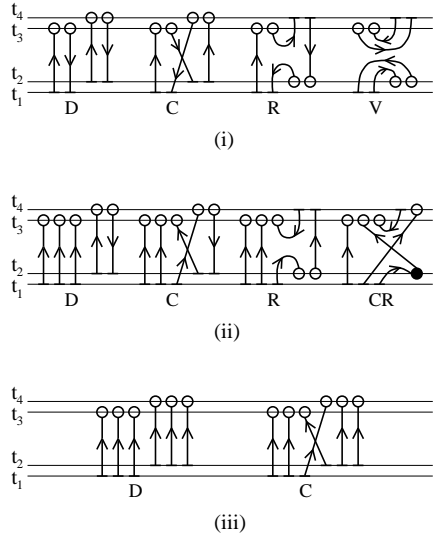}}
\vspace*{8pt}
\caption{Diagrams contributing to hadron-hadron 4-point functions,
from Ref.~\protect\cite{Fukugita:1995ve}.
Arrowed lines represent quark or antiquark propagators. Wall sources are
depicted by short bars, and circles represent sinks for hadron operators.
The diagrams in (i), (ii), and (iii) are for the
$\pi$--$\pi$, $\pi$--N, and N--N systems, respectively.
The decompositions (\protect\ref{RI0}) and (\protect\ref{RI2}), for
example, refer to the labels D,C,R,V of diagrams (i).\label{fig123}}
\end{figure}

Computation of the above time correlation functions requires
matrix elements of the quark propagator $G=D^{-1}$, where $D(x,y)$ denotes the
Dirac, or fermion, matrix and $x=(\vec{x},t)$, are space-time lattice sites.
Since a complete solution of
\begin{equation}
\sum_{x{''}}D(x{'},x{''})\,G(x{''},x)=\openone\,\delta_{x{'},x}\,,
\label{DGdelta}\end{equation}
where $\openone$ means the unit matrix in color-Dirac space,
is not feasible in practice source techniques are typically used to
obtain at least some columns of $G$. The CP-PACS collaboration has employed
so called wall sources. These are defined by placing a point source of
value $\openone$ on each spatial site of a certain time slice $t$, specifically
\begin{equation}
\sum_{\vec{x}{''}t''}D(\vec{x}{'}t',\vec{x}{''}t'')\,G_t(\vec{x}{''}t'')=
\openone\,\sum_{\vec{x}}\delta_{(\vec{x}{'}t'),(\vec{x}t)}=
\openone\,\delta_{t't}\,.
\label{wallsrc}\end{equation}
Note that the uniform nature of the wall source implies that the solution
vector $G_t$ must be independent of $\vec{x}{'}$.
Indeed, employing (\ref{DGdelta}) it is obvious that
\begin{equation}
G_t(\vec{x}{''}t'')=\sum_{\vec{x}}\,G(\vec{x}{''}t'',\vec{x}t)\,.
\label{Gt}\end{equation}
Loosely speaking, since the spatial source is structureless
quark propagation to $x{''}$ proceeds
in `the same way' regardless of the spatial point of origin.
However, this is not a gauge invariant mechanism.
Under a gauge transformation
$D(x,y)=U^\dagger(x)\tilde{D}(x,y)U(y)$, etc, the wall source
becomes $\openone U(\vec{x}{'}t')\,\delta_{t',t}$. Thus, in contrast to
$G(\vec{x}{''}t'',\vec{x}t)$, the
(numerical) solution $G_t(\vec{x}{''}t'')$ of (\ref{wallsrc}) does not
transform covariantly.
Gauge dependent noise in hadronic correlation functions
caused by the wall sources is specifically created
at time slices where two wall sources or one wall source and one sink are placed.
This can be seen from Fig.~\ref{fig123} which shows a diagramatic analysis
of the correlators for the $\pi$--$\pi$, $\pi$--N, and N--N systems in terms
of quark propagators and the corresponding wall sources and sinks.
Wall sources are depicted as short lines and sinks as circles. The time slices
refer to $t_1=0, t_2=1, t_3=t, t_4=t+1$.
For example the $\pi$--$\pi$ rectangular diagram in Fig.~\ref{fig123}(i)
corresponds to
\begin{equation}
C^R(t_4,t_3,t_2,t_1)=\sum_{\vec{x}_2,\vec{x}_3}\langle{\rm Re}{\rm Tr}[
G^\dagger_{t_1}(\vec{x}_2t_2) G_{t_4}(\vec{x}_2t_2)
G^\dagger_{t_4}(\vec{x}_3t_3) G_{t_1}(\vec{x}_3t_3)]\rangle\,.
\label{CR}\end{equation}
The paths involved in the computation of $C^R$ are not all closed, they are
interupted at time slices where a (nonlocal) wall source meets a (local) sink,
or two wall sources meet. This happens, for example, at $t_1$ and $t_4$,
see (\ref{CR}) and Fig.~\ref{fig123}.

In order to deal with the gauge noise problem two strategies may be employed:
The first one is gauge fixing, the other one is to rely on gauge fluctuations
to cancel out in the final correlation function. For the latter the gauge noise
is expected to decrease as $\sim L^{3/2}$ for sufficiently
large $L$ \cite{Fukugita:1995ve}.

Correlator ratios (\ref{RCCCC}) for various channels can be expressed in terms
of linear combinations of certain diagrams which are referred to by labels
D, C, R, V, etc, in Fig.~\ref{fig123}. For example, in the Wilson fermion scheme
\begin{eqnarray}
R^{\pi\pi}_{I=0}(t)&=&R^D(t)+\frac{1}{2}R^C(t)-3R^R(t)+\frac{3}{2}R^V(t)\label{RI0}\\
R^{\pi\pi}_{I=2}(t)&=&R^D(t)-R^C(t)\,.\label{RI2}
\end{eqnarray}
For staggered fermions the corresponding decomposition is slightly different.

\begin{figure}[htb]
\centerline{\includegraphics[width=56mm,angle=-90]{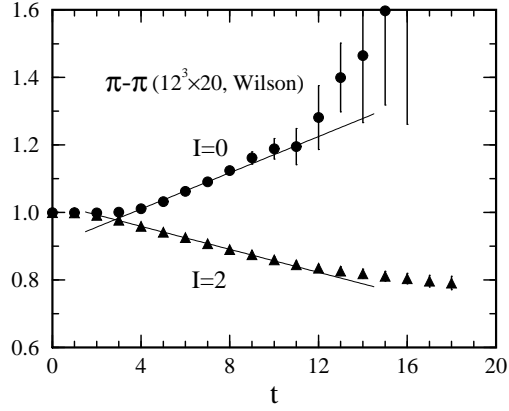}}
\vspace*{8pt}
\caption{Ratios (\protect\ref{RI0}) and (\protect\ref{RI2})
for the $\pi$--$\pi$ four-point function, from Ref.~\protect\cite{Fukugita:1995ve}.
This example is for Wilson fermions at $\beta=5.7$, $\kappa=0.164$, and
wall sources without gauge fixing. The lattice size is $12^3\times 20$.\label{fig10}}
\end{figure}

\begin{figure}[htb]
\centerline{\includegraphics[,width=56mm,angle=-90]{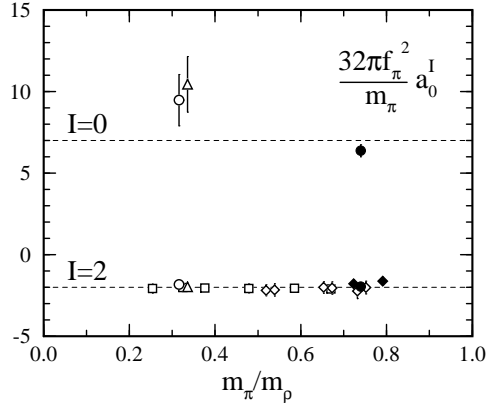}}
\vspace*{8pt}
\caption{Comparison of various scattering length results in the $I=0$ and $I=2$
channels for the $\pi$--$\pi$ system, with units as given in the picture.
Solid and open plot symbols refer to Wilson and Kogut-Sussking fermions,
respectively. Results are from Ref.~\protect\cite{Fukugita:1995ve},
except for the squares and diamonds, which are from
Refs.~\protect\cite{Sharpe:1992pp,Gupta:1993rn}.
Dotted lines are current algebra predictions.\label{fig12}}
\end{figure}

\begin{figure}[htb]
\centerline{\includegraphics[width=56mm,angle=-90]{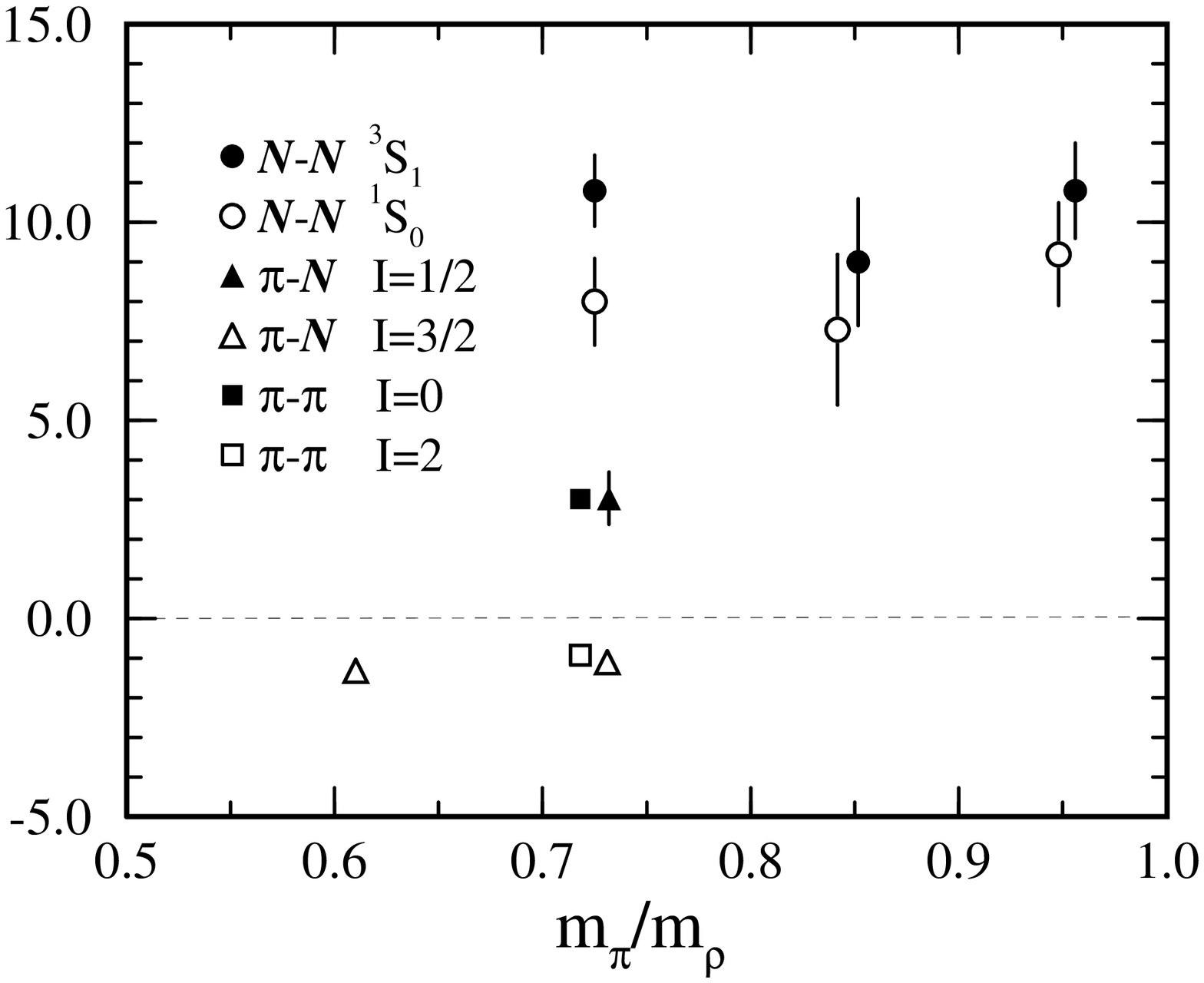}}
\vspace*{8pt}
\caption{Comparison of scattering lengths for the $\pi$--$\pi$,
$\pi$--N, and N--N systems in physical units $a_0$~[fm] versus the
$\pi$--$\rho$ mass ratio.
These are (quenched) results for Wilson fermions
from Ref.~\protect\cite{Fukugita:1995ve}
at lattice constant $a=0.137(2)~{\rm fm}$.\label{fig21}}
\end{figure}

Typical results for a ratio function (\ref{RCCCC}) are shown in 
Fig.~\ref{fig10}
for the $I=0$ and $I=2$ channels of the $\pi$--$\pi$ system.
A selection of scattering lengths, again for the $\pi$--$\pi$ system,
is displayed in Fig.~\ref{fig12}.
In Fig.~\ref{fig21} a comparison of scattering lengths for the systems
$\pi$--$\pi$, $\pi$--N, and N--N is made.
Table~\ref{tab1} contains the main results of the lattice simulation and current
algebra predictions.
A chiral extrapolation has not been made in \cite{Fukugita:1995ve}.
Thus, perhaps not surprisingly, the lattice results agree best with the
current algebra predictions that use the hadron masses supplied by the actual
lattice simulation. The N--N system is special because of its large size.
Realistic lattice results for the scattering length cannot be expected
with contemporary resources.

\begin{table}[htb]
\tbl{Selected results from Ref.~\protect\cite{Fukugita:1995ve}.
Scattering lengths $a_0$ for the s-wave in various channels are shown
in physical units [fm], for the lattice ({\sc lat}), experiment ({\sc exp}),
and current algebra ({\sc cua}).
The lattice results are
for Wilson fermions on a $12^3\times 20$ lattice with $\beta=5.7$.
Light-quark masses are large as indicated by $m_\pi/m_\rho=0.74$ and
$m_N/m_\rho=1.57$.
Experimental and current algebra results are compiled from
Refs.~\protect\cite{Nagels:1979xh,Dumbrajs:1983jd}
and \protect\cite{Weinberg:1966kf}, respectively.
In the last column the masses from the lattice simulation enter the
current algebra predictions.\vspace{4pt}}
{\tabcolsep7pt
\begin{tabular}{llllll} \hline
& & $\phantom{-}${\sc lat} & $\phantom{-}${\sc exp} &
$\phantom{-}${\sc cua} & $\phantom{-}\mbox{\sc cua(lat)}$ \\ \hline
$\pi$--$\pi$ & $I=0$   & $+3.02(17)$  & $+0.37(7)$   & $+0.222$  & $+3.47(5)$   \\
             & $I=2$   & $-0.924(40)$ & $-0.040(17)$ & $-0.0635$ & $-0.993(16)$ \\
$\pi$--N     & $I=1/2$ & $+3.04(66)$  & $+0.245(4)$  & $+0.221$  & $+2.701(41)$ \\
             & $I=3/2$ & $-1.10(20)$  & $-0.143(6)$  & $-0.111$  & $-1.350(20)$ \vspace{0.5ex}\\
N--N & ${}^3S_1$ & $+10.8(9)$ & $-5.432(5)$ & & \\
     & ${}^1S_0$ & $+8.0(1.1)$ & $+20.1(4)$  & & \vspace{0.5ex}\\
K--N                & $I=0$ & $+0.55(47)$   & $-0.0075$       & $\phantom{-}0$ & $\phantom{-}0$\\
                    & $I=1$ & $-1.56(13)$ & $-0.225$        & $-0.399$ & $-2.701(41)$ \\
$\bar{\mbox{K}}$--N & $I=0$ & $+4.64(37)$ & $-1.16+i\,0.49$ & $+0.598$ & $+4.051(61)$ \\
                    & $I=1$ & $+2.63(64)$ & $+0.17+i\,0.41$   & $+0.199$ & $+1.350(20)$ \\ \hline
\end{tabular}}
\label{tab1}
\end{table}

\clearpage
\subsection{Static $N-N$ and $N-{\bar N}$ Potentials}

Beyond the zero-momentum limit (scattering lengths) systems of hadrons
with at least one heavy quark each pose the least technical difficulties.
Going to the extreme, an infinitely heavy quark may be interpreted
as a static color source. This idea goes back to Wilson's original
work \cite{Wilson:1974sk} and played a role in demonstrating confinement.

\begin{figure}[ht]
\centering
\begin{tabular}{cc}
\includegraphics[width=56mm]{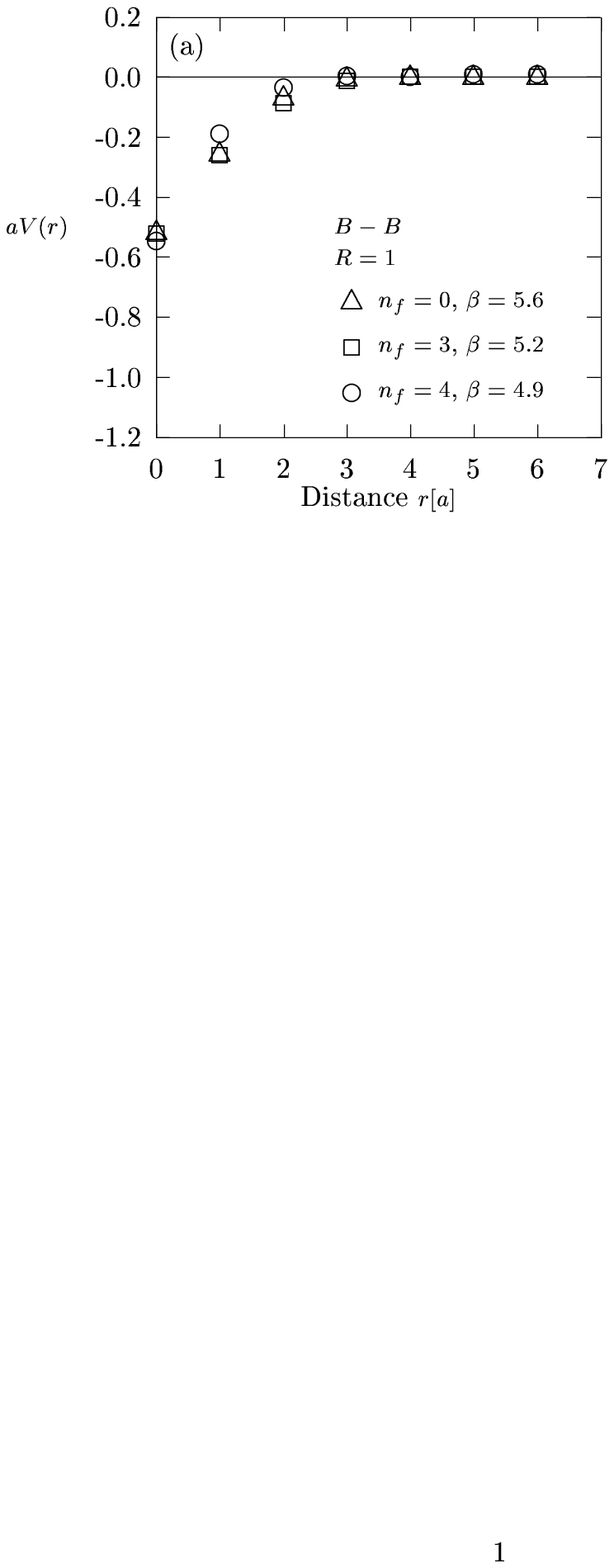}
\includegraphics[width=56mm]{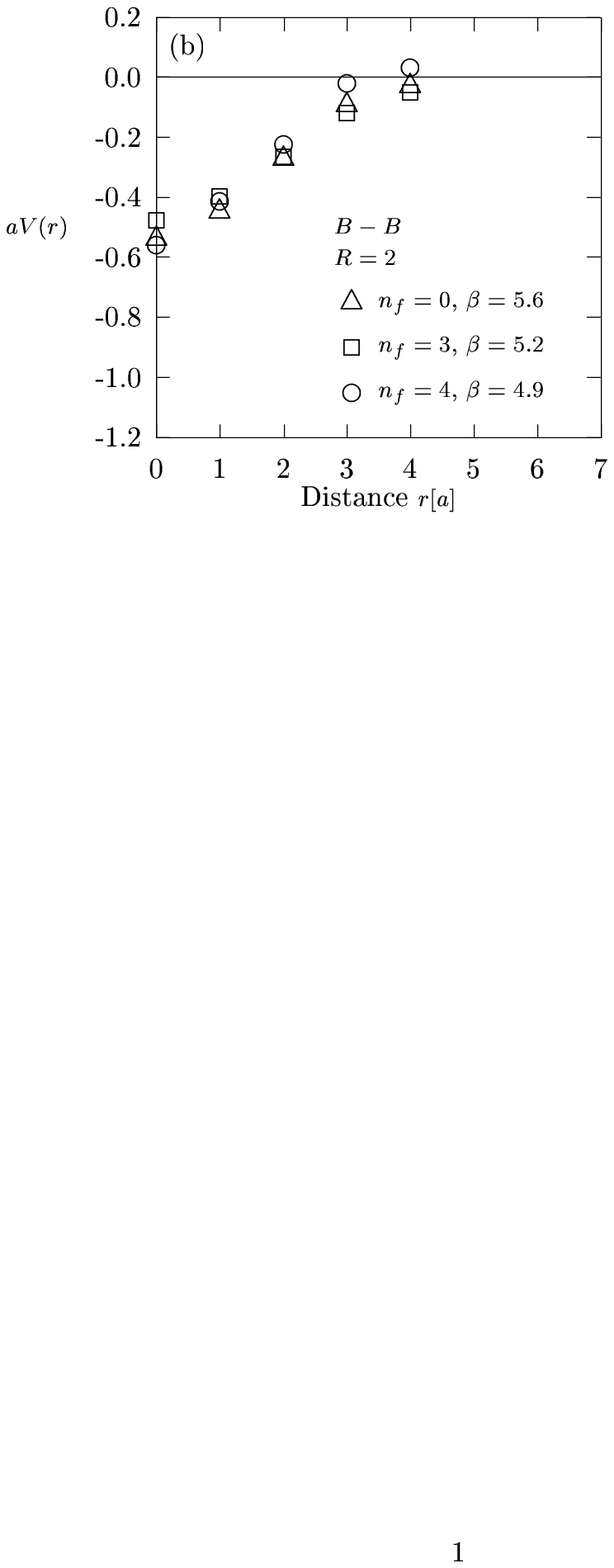}\\
\includegraphics[width=56mm]{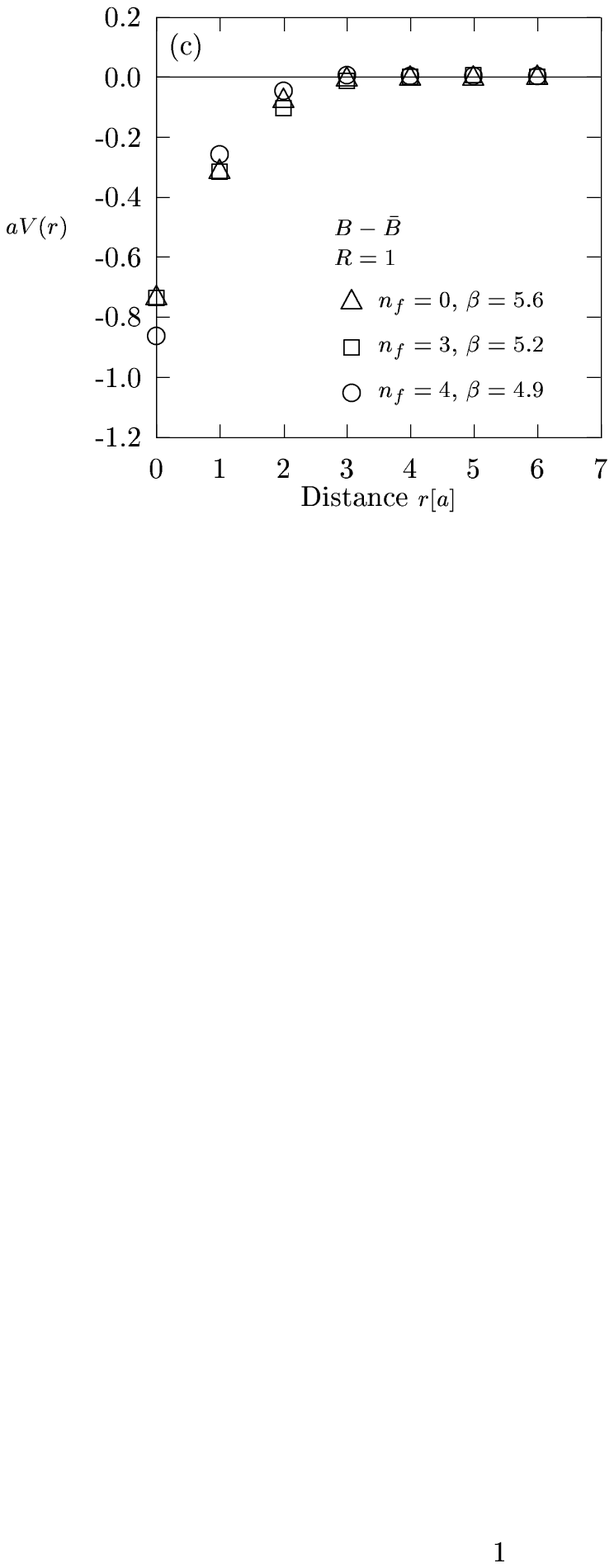}
\includegraphics[width=56mm]{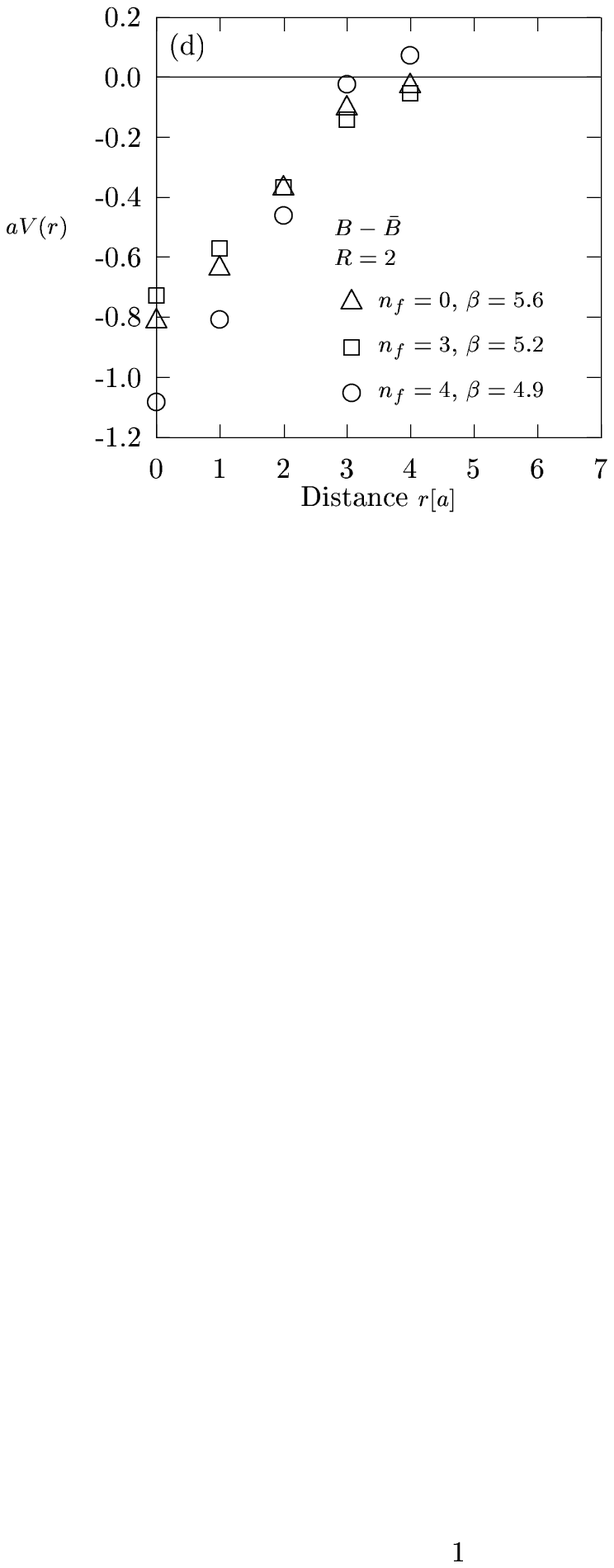}
\end{tabular}
\vspace*{8pt}
\caption{Interaction energies of hadron-hadron systems versus the
relative distance \protect\cite{Rab93a,Buerger:1994ma} in the finite-temperature
formalism. Hadronic structure is represented by a static cluster. The radius
of the cluster is $R=2$ in lattice units.\label{figStat}}
\end{figure}

The time evolution operator of a static color source
(a quark) located at fixed $\vec{x}$ is given by the product of link
variables $U_{4}(\vec{x}t)\in SU(3)$ along a line
$t_i\rightarrow t_i+1 \rightarrow \ldots \rightarrow t_f$ in
the time direction, $\mu=4$. For a lattice $L^3\times T$ with periodic boundary
conditions in the time direction the line is closed, $t_i=t_f$. The
trace of such an operator
\begin{equation}
Q(\vec{x}) = \frac13 \mbox{Tr} \prod_{t=1}^{T} U_{4}(\vec{x}t)
\label{Polyakov}\end{equation}
is gauge invariant and
is known as a Polyakov loop. It represents a static quark at $\vec{x}$,
and $Q^{\dagger}(\vec{x})$ represents a static antiquark at $\vec{x}$.
In the finite-temperature formalism\footnote
{The temperature is the reciprocal of the time extent of the lattice,
$1/Ta$. The path integral over fields becomes periodic in euclidean
time \protect\cite{Mon94}.}
the expectation value \cite{McLerran:1981pk,Kuti:1981gh,Fiebig:1996mx}
\begin{equation}
L^{-3}\sum_{\vec{x}}\langle Q(\vec{r}+\vec{x})Q^{\dagger}(\vec{x})\rangle
=e^{-T\,F_{Q\overline{Q}}(\vec{r}\,)}
\end{equation}
is related to the free energy $F_{Q\overline{Q}}(\vec{r}\,)$, loosely (but
imprecisely) referred to as the heavy quark-antiquark potential.
Adopting this interpretation it is interesting to compute
the free energies of two clusters of two or three
quarks \cite{Rab93a,Buerger:1994ma,Rab97}. Consider
\begin{equation}
O(\vec{x}) = \sum_{\vec{y}_1\vec{y}_2\vec{y}_3}
\rho(\vec{y}_1\vec{y}_2\vec{y}_3)
Q(\vec{x}+\vec{y}_1) Q(\vec{x}+\vec{y}_2) Q(\vec{x}+\vec{y}_3)\,,
\end{equation}
where $\rho$ is some function symmetric in $\vec{y}_1\vec{y}_2\vec{y}_3$,
for example a Gaussian or the characteristic function of a sphere with
radius $R$, which serves as an ad-hoc model for the three-quark cluster.
The free energy of two three-quark clusters at separation $\vec{r}$ then is
\begin{equation}
F(\vec{r}\,)=-\frac{1}{T}\ln L^{-3}\sum_{\vec{x}}
\langle O(\vec{r}+\vec{x})O(\vec{x})\rangle \,.
\end{equation}
An example of $F(\vec{r}\,)$ can be seen in Fig.~\ref{figStat}.
There is attraction
in the overlap region of the clusters, somewhat reflecting the shape
of $\rho$. The cluster interaction possesses no dynamics in this construction.
%

\subsection{Heavy-Light Meson-Meson Systems}

A step towards a more realistic calculation is to make some of the quarks dynamical.
In a system of two mesons which consist of a heavy and a light quark each,
we may still approximate the heavy quark by a Polyakov loop, as above. The role
of the static quarks then is to localize the mesons. This situation suggests
using the coordinate-space approach of Sec.~\ref{secAdi}.

\begin{figure}[htb]
\centerline{\includegraphics[width=56mm]{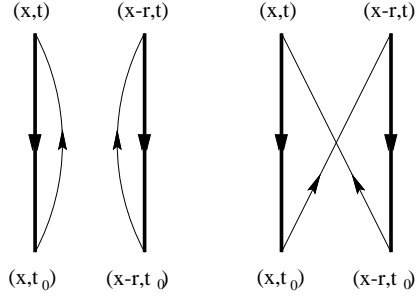}}
\vspace*{8pt}
\caption{Diagrams for the heavy-light meson-meson system. The thick
and thin lines represent propagators for the heavy (static) and the light
(dynamical) quark, respectively.\label{figHlAB}}
\end{figure}

Thus we take, say in the staggered scheme,
\begin{equation}
\phi_{\vec{x}}(t)=\bar{\chi}_{h}(\vec{x},t)\,\chi_{u}(\vec{x},t) \,,
\end{equation}
where $h$ is the heavy flavor ($h=s,c,\ldots$). The two-meson interpolating
field is defined just like (\ref{Phir}).  However, 
the heavy-flavor propagator, $G^{(h)}$, is taken in the limit
$m_h\rightarrow\infty$. This is achieved by using the hopping parameter
expansion \cite{Rot92,Mon94} and keeping the leading term only. We thus use
\begin{equation}
G^{(h)}(\vec{x}t,\vec{y}t_0) = \delta_{\vec{x},\vec{y}} \,
[\frac{1}{2m_h} (-1)^{x_1+x_2+x_3}]^{t-t_0}
\prod_{\tau=1}^{t-t_0}\,U_{4}(\vec{x}\tau) \,.
\end{equation}
The product of link variables becomes proportional to the Polyakov loop
(\ref{Polyakov}) for $t-t_0=T$.
With $G^{(h)}$ diagonal in the space sites the 2-point and 4-point
correlators become
\begin{eqnarray}
C^{(2)}(t,t_0)&=&\langle L^{-6}\sum_{\vec{x}} 
G^{(h)\ast}_{BA}(\vec{x}t,\vec{x}t_0)\,G_{BA}(\vec{x}t,\vec{x}t_0)\rangle\label{C2h}\\
C^{(4)}_{\vec{r}}(t,t_0)&=&\langle L^{-12}\sum_{\vec{x}}[
G^{(h)\ast}_{BA}(\vec{x}t,\vec{x}t_0)\,G_{BA}(\vec{x}t,\vec{x}t_0)\label{C4h}\\
&&\rule{7.3ex}{0mm}\times G^{(h)\ast}_{DC}(\vec{x}-\vec{r}\,t,\vec{x}-\vec{r}\,t_0)\,
G_{DC}(\vec{x}-\vec{r}\,t,\vec{x}-\vec{r}\,t_0)\nonumber\\
& &\rule{7.3ex}{0mm}-G^{(h)\ast}_{BA}(\vec{x}t,\vec{x}t_0)\,G_{BC}(\vec{x}-\vec{r}\,t,\vec{x}t_0)
\nonumber\\
&&\rule{7.3ex}{0mm}\times G^{(h)\ast}_{DC}(\vec{x}-\vec{r}\,t,\vec{x}-\vec{r}\,t_0)\,
G_{DA}(\vec{x}t,\vec{x}-\vec{r}\,t_0)]\rangle\,.\nonumber
\end{eqnarray}
Sums over color indices $A,B,C,D$ are understood. The 4-point correlator
(\ref{C4h}) is a sum of two terms which are illustrated in Fig.~\ref{figHlAB}.
The light quarks can be exchanged between the mesons.
Thus, besides gluon effects, also quark dynamics is included
in the \ERI.

\begin{figure}[htb]
\centerline{\includegraphics[width=112mm]{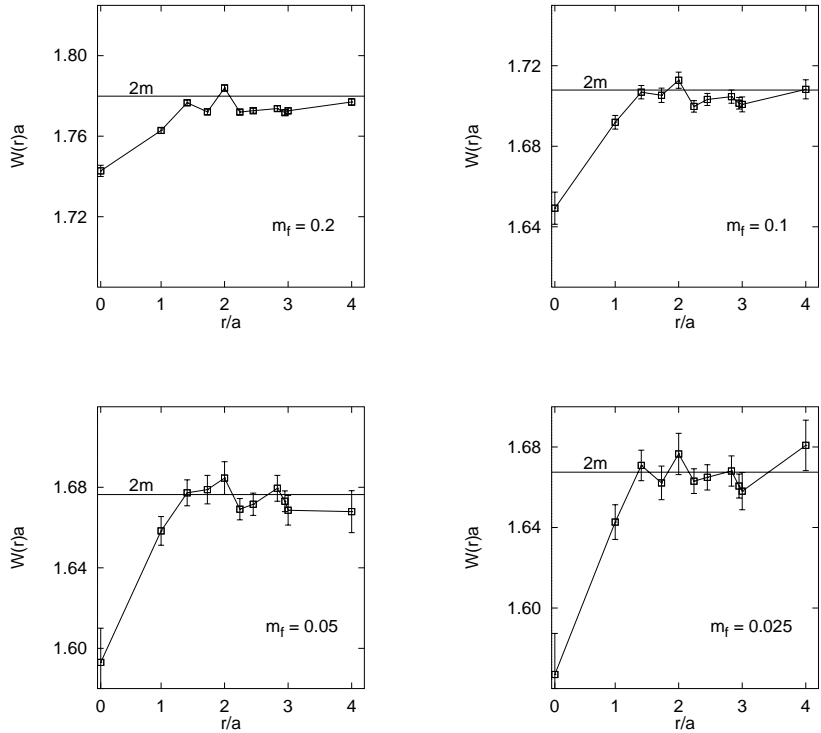}}
\vspace*{8pt}
\caption{Meson-meson energies $W(r)=V^{(A_1)}(r)$, see (\ref{calCr}), for
several light-quark masses of staggered fermions.
Energies at distance $r/a=3$ are degenerate,
$|\vec{r}|=|(2,2,1)|=(3,0,0)|$. Those data points are slightly
shifted \protect\cite{Mih98}.
\label{figinter-fig4}}
\end{figure}

For distances $\vec{r}\neq 0$ parallel to the coordinate axes, the $A_1$ sector 
projection, see Sec.~\ref{secLatSym}, is similar to (\ref{A1p}), thus
\begin{equation}
C^{(4;A_1)}_{r}(t,t_0) = \frac14\sum_{\vec{r},|\vec{r}|=r} 
C^{(4)}_{\vec{r}}(t,t_0) \,.
\end{equation}
According to (\ref{calCt}) the effective correlator
\begin{equation}
{\mathfrak C}^{(4;A_1)}_{r\,r}(t,t_0) \approx
\frac{C^{(4;A_1)}_{r}(t,t_0)}{|C^{(2)}(t,t_0)|^2}\simeq 
c_r\, e^{-V^{(A_1)}(r)(t-t_0)}
\label{calCr}\end{equation}
serves to extract the (adiabatic) potential $V^{(A_1)}(r)$.
Examples from an actual
simulation \cite{Mihaly:1997ue} are shown in Fig.~\ref{figinter-fig4}.

In the simple case of a one-hadron local operator it is possible to
express the corresponding time correlation function in terms of propagator
matrix elements from a source located at only one fixed lattice site,
say $x=(\vec{x}_0t_0)$.
This is done by using translational invariance of the lattice action in the
manner outlined in Sec.~\ref{secMesF} leading to (\ref{eq44}), for example.
The cost of computing this is greatly reduced because only one column
$G(\vec{x}t,\vec{x}_0t_0)$ of the inverse fermion matrix (per color-Dirac
index) is needed.
In the case of a two-hadron system translational invariance might still be
utilized to replace some propagator elements with $G(\vec{x}t,\vec{x}_0t_0)$,
for example via $\vec{x}_1\rightarrow\vec{x}_1+\vec{y}_1-\vec{x}_0$
in (\ref{eq46}). In the latter case elements $G(\vec{x}t,\vec{y}t_0)$
where $\vec{x}$ and $\vec{y}$ run separately over all spatial lattice
sites still remain. Those are known as all-to-all propagators.
Their computation poses an almost unavoidable problem in multi-hadron systems.
The wall sources defined through (\ref{wallsrc}) seemingly circumvent this
difficulty, however, only at the cost of violating gauge invariance of the
correlation functions, see Sec.~\ref{secScatLen}.
In a heavy-light hadron-hadron system,
like (\ref{C4h}) for example, the needed minimum number of
sources is the number of relative spatial distances distances $\vec{r}$ that are
probed (including those related by $O(3,\mathbb Z)$ symmetry).

The use of stochastic methods for dealing with the all-to-all propagator problem
was already discussed in some detail. Even in cases where translational
invariance can be utilized it may be advantageous to employ random sources instead.
The reason is that the additional (large) sums over lattice sites work to
reduce statistical errors in the correlation functions.
Besides, it may also be the only computationally viable strategy.
In a typical application a random source is chosen to be nonzero on one time slice
only, say $t_0$, as in (\ref{Rt0}) and (\ref{Rt0p}).
In this case propagation from any lattice site to any other lattice site within
the same time slice is computed using the stochastic estimator, whereas
propagation between lattice sites with different times, such as $t$ and $t_0$,
is computed with machine precision.
An important consequence is that the errors for propagation in space and time
directions are different. While the latter is as small as the matrix
inversion algorithm permits (e.g. a conjugate gradient) the former is limited
by the variance of the random source probability distribution and, in particular,
is independent of the distance between the lattice sites.
This is a serious drawback, it means that stochastic estimation is futile
in computing the usual (hadronic) time correlation functions since their
exponentially decreasing signal would be concealed by the essentially constant
error of the stochastic estimator.

This problem can be dealt with to some extent by a method called maximum
variance reduction \cite{Michael:1998sg},
see also \cite{Michael:1998rk,Peisa:1998ek}. 
The idea is to enclose disjoint regions of lattice sites by fixed-source
boundaries in order to increase sample sizes.
One such region may consist of a single
point \cite{deDivitiis:1996qx}. Another, somewhat related, example is
$R=\{(\vec{x},t_0)|\vec{x}\in L^3\}$, i.e.
the spatial volume at time slice $t_0$.
An estimator for propagation between any
two points in disjoint regions is then variance reduced.
Specifically, starting from
\begin{equation}
Z = \int [d\phi\,d\phi^\ast] \exp(-\frac{1}{2}\phi^\dagger A\phi)\,,
\label{var1}\end{equation}
where $\phi$ is an auxiliary scalar complex lattice field
and $A$ is a hermitean positive definite matrix,
one may generate stochastically independent
ensembles of fields $\{\phi\}$ by way of common Monte Carlo methods
with the gaussian probability distribution implied in (\ref{var1}).
Writing $i=C\mu x$, etc, for the color-Dirac-site indices it is
then straightforward to show that
\begin{equation}
A_{ij}^{-1}=\frac{1}{Z}\int [d\phi\,d\phi^\ast] \exp(-\frac{1}{2}\phi^\dagger A\phi)\,
\phi_j^\ast \phi_i \cong \langle \phi_j^\ast \phi_i \rangle\,,
\label{var2}\end{equation}
where in this context $\langle\ldots\rangle$ means the average over the
ensemble $\{\phi\}$. The above equations become useful for the choice
\begin{equation}
A=Q^\dagger Q\,,
\label{var3}\end{equation}
with $Q=Q[U]$ being the fermion matrix for a given gauge field configuration $U$.
In the Wilson scheme, for example, it has the form $Q[U]=\openone-\kappa D[U]$.
Quark propagator elements $G_{ij}$ can then be estimated from the ensemble averages
\begin{equation}
G_{ij} \stackrel{\rm def}{=} Q_{ij}^{-1} \cong \langle (Q\phi)^\ast_j\, \phi_i \rangle\,.
\label{var4}\end{equation}
If the variance of the above estimator is $\sigma_{ij}$ the statistical error is
$\sigma_{ij}/\sqrt{N_\phi}$ for an ensemble of size $N_\phi$.\footnote{In fact,
since a new estimator is computed for each of the gauge field
configurations in the lattice simulation, the resulting error of a hadronic
time correlation function is much smaller. For this reason a value of $N_\phi$
that is about 1/10 -- 1/20 of the number of gauge field configurations is usually
sufficient.}
However, the crucial point here is that the variance $\sigma_{ij}$,
for $i=C\mu x$ and $j=B\nu y$,
is essentially independent of the space-time distance $|x-y|$, say
$\sigma_{ij}\approx\sigma$ while $\sigma$ is of the order of one
inherited from its gaussian ancestry.
This effect spoils using the stochastic estimator for the purpose of
computing exponentially decreasing time correlation functions.

To alleviate the problem the authors of Ref.~\cite{Michael:1998sg}
proceed to consider a subset, or region, of contiguous lattice sites.
Describe this region by sets of sites $R$ and $S$, with $R\cap S=\emptyset$,
where $R$ contains all entirely interior points (having 8 neighbors $\in R\cup S$)
and $S$ is the boundary (having 1 through 7 neighbors $\in R\cup S$).
In the process of generating ensembles $\{\phi\}$ from
(\ref{var1}), consider a certain field $\phi$. Then, the field conponents
on the boundary $S$ are kept fixed
\begin{equation}
s_l\stackrel{\rm def}{=}\phi_l\quad{\rm for}\quad l\in S\,,
\label{var5}\end{equation}
and instead of (\ref{var1}) one uses
\begin{equation}
{\cal Z} = \int_R [d\phi\,d\phi^\ast] \exp\left(
-\frac{1}{2}( \phi^\dagger_i \bar{A}_{ij} \phi_j
+ \phi^\dagger_i \tilde{A}_{il} s_l + s^\dagger_l \tilde{A}_{lj} \phi_j )\right)\,, 
\label{var6}\end{equation}
where $i,j\in R$ and $l\in S$ is understood for the index sum ranges
and the notations $\bar{A}$ and $\tilde{A}$ are introduced for the
corresponding blocks of $A$.
The integral runs over the field components located on $R$.
Now the $R$-averaged random source field
\begin{equation}
v_k=\frac{1}{\cal Z}\int_R [d\phi\,d\phi^\ast] \exp\left(
-\frac{1}{2}( \phi^\dagger_i \bar{A}_{ij} \phi_j
+ \phi^\dagger_i \tilde{A}_{il} s_l + s^\dagger_l \tilde{A}_{lj} \phi_j )\right) \phi_k\,, 
\label{var7}\end{equation}
with $k\in R$, is called the variance reduced estimator for $\phi$.
Evaluation of the gaussian integral in (\ref{var7}) gives
\begin{equation}
v_k = - \bar{A}^{-1}_{ki} \tilde{A}_{il} s_l. 
\label{var8}\end{equation}

Recall that in (\ref{var4}) single-site components of the random
source field $\phi$ are employed to estimate the quark propagator.
Replacing those with site-averaged variance reduced estimators $v$
essentially amounts to using extended sources which of course reduce fluctuations.
This is the main idea.
To obtain a variance reduced estimator for quark propagator elements
two disjoint regions $R$ and $R'$ are needed. Then, within the above
framework, one may use (\ref{var4}) to show that
\begin{equation}
G_{k'k} \simeq \langle (Qv)^\ast_k\, v'_{k'} \rangle
\quad{\rm for}\quad k\in R,\mbox{\ }k'\in R'\,,
\label{var9}\end{equation}
where $v$ and $v'$ are the variance reduced sources corresponding to the
regions $R$ and $R'$ respectively \cite{Peisa:1998ek}.
Clearly, only propagator elements between any site in $R$ and any site in $R'$
can obtained in this way. Thus the all-to-all propagator problem is only
partially solved.
In order to maximize the variance reduction effect, $R$ and $R'$ should be equal
in size and together with $S$ and $S'$ cover the entire lattice. This suggests the choices
of all lattice sites with $0 < t < T/2$ and $T/2 < t < T$ for $R$ and $R'$,
respectively. Further computational and technical details for the practitioner
are given in Ref.~\cite{Michael:1998sg}.

Michael and Pennanen have utilized maximal variance reduced estimators in their
work on heavy-light meson-meson systems \cite{Michael:1999nq}
using improved Wilson fermions with a Sheikholeslami-Wohlert
action \cite{Sheikholeslami:1985ij}.
Their main motivation was to explore the possibility of binding heavy-light
systems. Since the kinetic energy of the quarks has a repulsive effect at shorter
relative distances two hadrons containing a heavy quark each should have a
better chance of forming a bound state.
Concerning experiment, the simulation is expected to describe mesons with one (heavy)
b-quark flavor. Using the static limit for the heavy flavor has the
consequence that heavy-light pseudoscalar and vector mesons (B and B$^\ast$
respectively) are degenerate.
However, experimentally their mass difference is less that 0.9\% of the
ground-state B-meson mass \cite{Groom:2000in}, so the static approximation is
quite reasonable. The notation ${\cal B}$ is used in in Ref.~\cite{Michael:1999nq}
to refer to the degenerate heavy-light mesons. For the light quarks u and d flavors are
employed, the corresponding Wilson hopping parameter $\kappa$ puts their masses
in the neighborhood of the strange quark mass.   

Thus a ${\cal B}$--${\cal B}$ system with relative distance $r$ between the (static)
quarks is classified in terms of the total spin $S_q$ and isospin $I_q$ of
the light quarks, while imposing the usual particle exchange symmetries through
appropriate heavy quark total spin $S_b$ and spatial symmetries. 
The latter states are degenerate in the static quark limit.
In the limit of an isotropic spatial wavefunction (L=0) there are only
four degenerate ground state levels of the ${\cal B}$--${\cal B}$ system characterized
by $I_q=0,1$ and $S_q=0,1$. The situation is illustrated in Table ~\ref{tab2}.
\begin{table}[htb]
\tbl{Allowed ${\cal B}$--${\cal B}$ states with $L=0$ according to \protect\cite{Michael:1999nq}.
Total isospin and spin of the light-quark subsystem are denoted by $I_q$ and $S_q$,
respectively. The total spin $S_b$ of the heavy quarks only distinguishes
degenerate states. Also shown are the physical heavy-light systems and their $J^P$
coupling to these states.\vspace{4pt}}
{\tabcolsep8pt
\begin{tabular}{cccclccc} \hline
 $I_q$ & $S_q$ & & $S_b$ & $J^{P}$\rule{0pt}{9pt} & BB & BB$^*$ & B$^*$B$^*$ \\
\hline
 1 & 1 & & 1 & 0$^+$ & $\checkmark$ &              & $\checkmark$ \\
 1 & 1 & & 1 & 1$^+$ &              & $\checkmark$ &              \\
 1 & 1 & & 1 & 2$^+$ &              &              & $\checkmark$ \\
\hline
 1 & 0 & & 0 & 0$^+$ & $\checkmark$ &              & $\checkmark$ \\
\hline
 0 & 1 & & 0 & 1$^+$ &              & $\checkmark$ & $\checkmark$ \\
\hline
 0 & 0 & & 1 & 1$^+$ &              & $\checkmark$ & $\checkmark$ \\
\hline
\end{tabular}}
\label{tab2}
\end{table}

\begin{figure}[htb]
\centerline{\includegraphics[width=50mm]{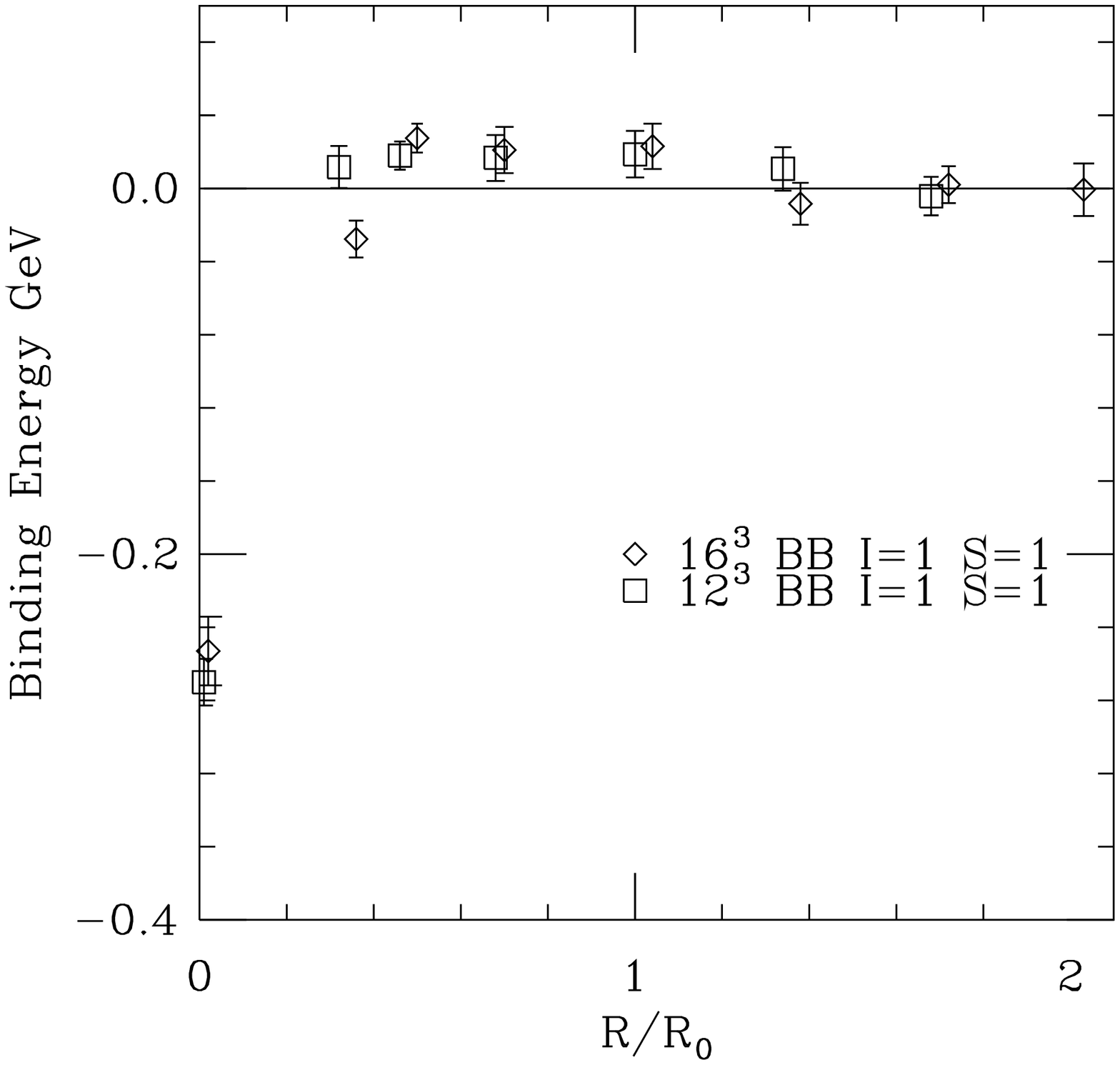}\hspace{2mm}
            \includegraphics[width=50mm]{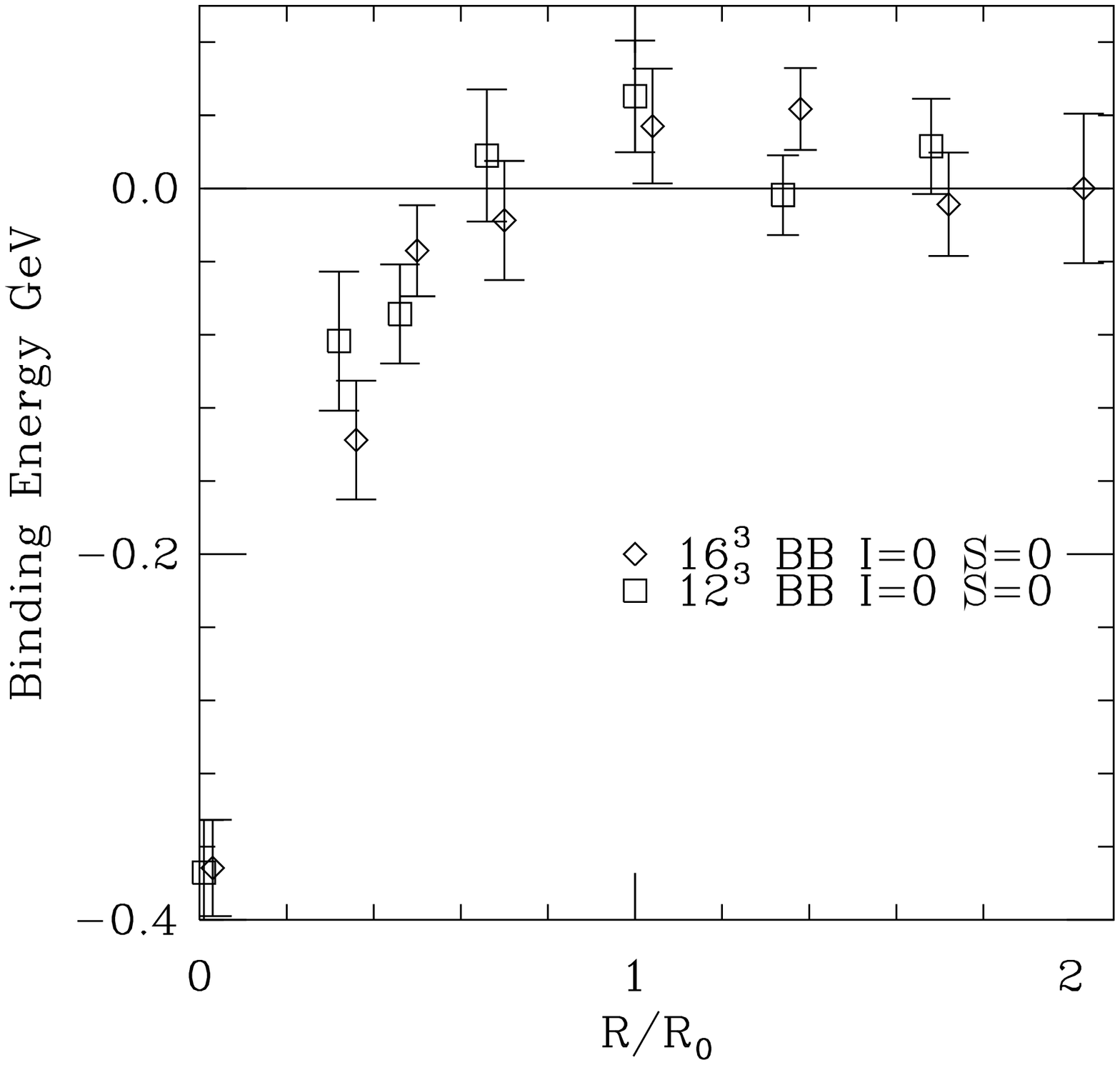}}
\centerline{\includegraphics[width=50mm]{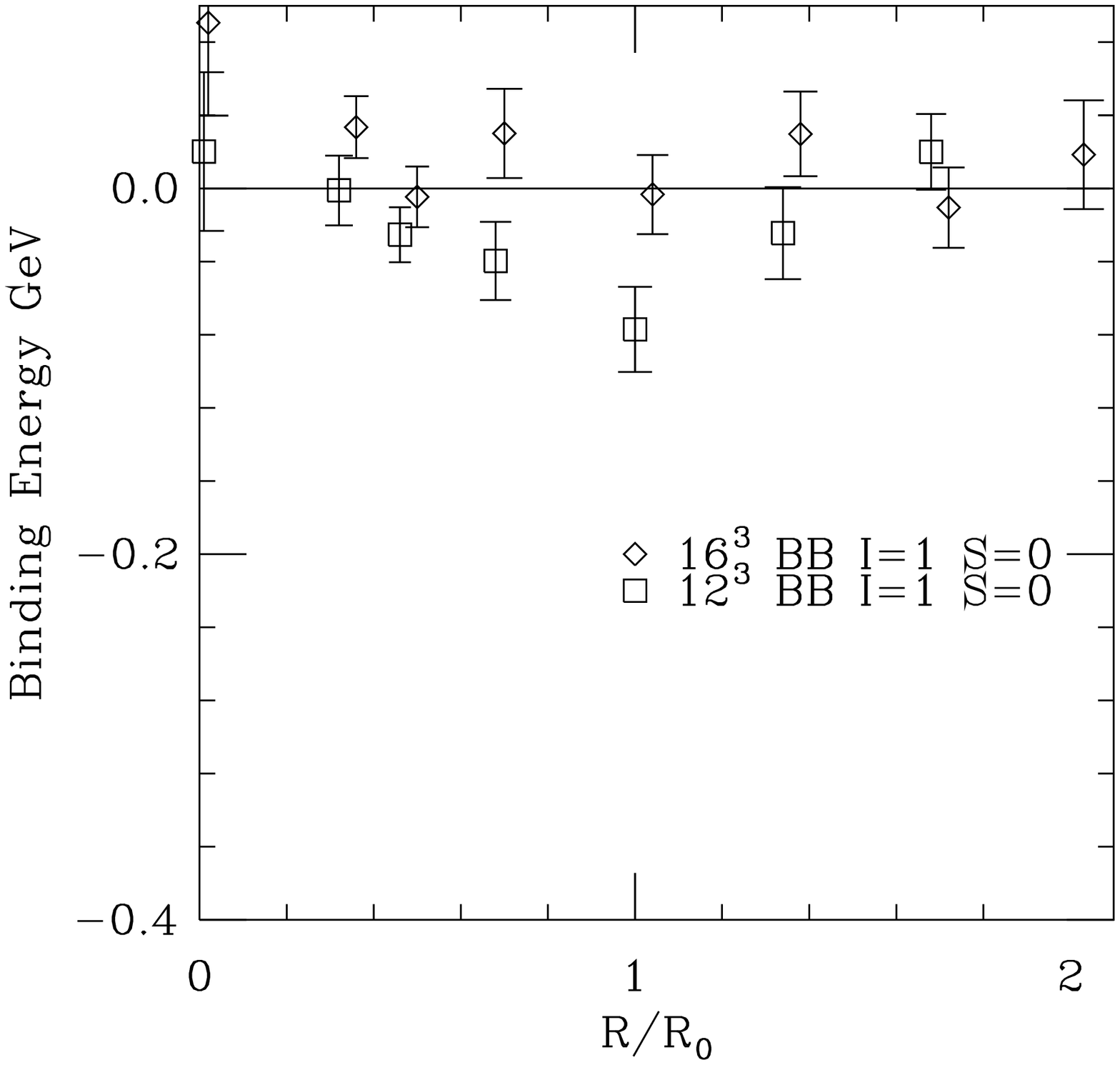}\hspace{2mm}
            \includegraphics[width=50mm]{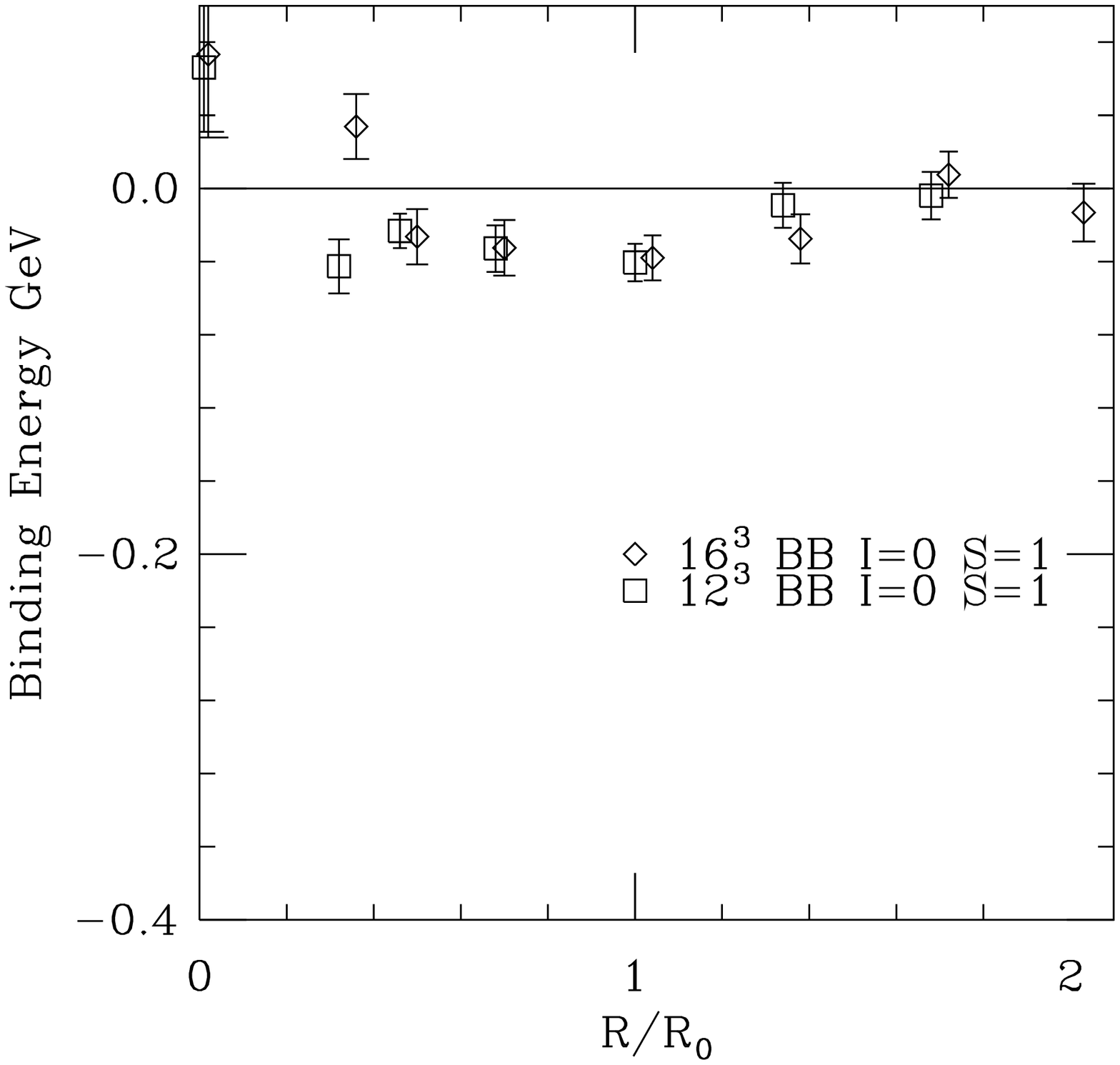}}
\vspace*{8pt}
\caption{Results for the binding energy $E({\cal B}$--${\cal B})-2E({\cal B})$ between
two ${\cal B}$ mesons with 
light quarks in $(I_q,S_q)=(1,1), (0,0), (1,0), (0,1)$  at separation $R$
(called $r$ throughout this chapter)
in units of $R_0 \approx 0.5$~fm. The light quark mass used corresponds
to strange quarks. Results at different spatial lattice sizes are displaced
in $R$ for legibility \protect\cite{Michael:1999nq}.\label{bb4}}
\end{figure}

Selected results of the simulation \cite{Michael:1999nq} are shown
in Fig.~\ref{bb4} exhibiting the energies of the ${\cal B}$--${\cal B}$ system
at fixed separation $r$ of the static $b$-quarks.
The authors suggest that there is evidence for deep binding at small $r$ with
a light di-quark configuration having $(I_q,S_q)=(0,0)$ and $(1,1)$ respectively.
This  binding energy is 400 -- 200~MeV at $r=0$ but is
very short-ranged. It is essentially a gluonic effect and is
rather insensitive to the light quark mass as shown by studies of
static baryons with varying light quark masses \cite{Michael:1998sg}.
Possibly, a configuration where the color of two heavy quarks
combines to a color state symmetric under particle exchange (sextet)
provides the interaction mechanism here. This situation is particular
to relative distance $r=0$ since two-quark color states other than local
singlets are not suppressed by confinement dynamics.

At larger
$r$, around 0.5~fm, evidence is seen for weak binding when the light
quarks are in the $(I_q,S_q)=(0,1)$ and $(1,0)$  states. This can be
related to meson exchange and one finds evidence of an interaction in the
spin-dependent quark-exchange (cross) diagram which is compatible with
the theoretical contribution from pion exchange in the study. Using
lighter, more physical, light quark masses below the strange
quark mass and, most desirably a chiral extrapolation, will be necessary
to confirm this interaction mechanism.

\subsection{Momentum-Space Work on the $\pi$--$\pi$ System\label{secMomSpcWrk}}

\begin{figure}[htb]
\centerline{\includegraphics[width=84mm]{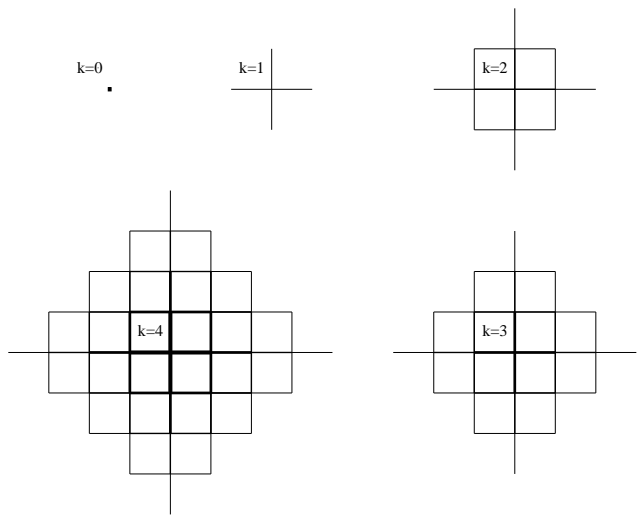}}
\vspace*{8pt}
\caption{Illustration of the `smearing' technique for fermion operators.
Various iterations ($k=0$ local, etc) are shown.
The thickness of the lines indicates the `multiplicity of use' of a
particular link.\label{figSmear}}
\end{figure}

The calculation of euclidean correlation functions between different hadrons
in the incoming and outgoing channel, like K$\rightarrow\pi\pi$ decays,
has to be treated with caution \cite{Maiani:1990ca}.
The program laid out in Secs.~\ref{secMesF}--\ref{secRanS} and
\ref{secTmb} has been put to the test in \QCD\
for a $\pi$--$\pi$ system of Wilson fermions \cite{Rab97}.
The formalism is essentially unchanged, except for the appearance of
color indices on the link variables $U_{\mu}(x)\in\mbox{SU(3)}$ and, replacing
staggered with Wilson fermions, the appearance of both a color
$A,B\ldots=1,2,3$ and a Dirac $\mu,\nu\ldots=1,2,3,4$ index on the
Grassmann quark fields $\psi_{fA\mu}(x)$.
In addition, there is a rather technical, but important point: 

A technique known as `smearing' uses spatially extended
field operators. For example, omitting
flavor and Dirac indices for the moment, define the quark field recursively
\begin{equation}
\psi^{\{ 0\}}_{A}(\vec{x}t) = \psi_{A}(x) \quad\quad
\psi^{\{ k\}}_{A}(\vec{x}t) = \sum_{\vec{y}} \sum_{B}
K_{AB}(\vec{x},\vec{y}\,)\;\psi^{\{ k-1\}}_{B}(\vec{y}t)\,,
\label{Kpsi}\end{equation}
with $k\in{\mathbb N}$, and the matrix
\begin{eqnarray}
K_{AB}(\vec{x},\vec{y}\,) &=& \delta_{AB}\,\delta_{\vec{x},\vec{y}}
+\alpha H_{AB}(\vec{x},\vec{y}\,) \\
H_{AB}(\vec{x},\vec{y}\,)&=&\sum_{m=1}^{3}
\left[ U_{m,AB}(\vec{x}\,t)\delta_{\vec{x},\vec{y}-\hat{m}}
+U^\dagger_{m,AB}(\vec{y}\,t)\delta_{\vec{x},\vec{y}+\hat{m}} \right]
\end{eqnarray}
connecting (hopping) to next-neighbor sites.
With this particular choice of the matrix $K$ the procedure is known as 
Gaussian smearing \cite{Alexandrou:1994ti}. The real number $\alpha$
and the maximal value for $k$ are considered parameters. 
The advantage of smearing lies in the fact that
interpolating operators constructed with smeared fields often have a larger
overlap with the ground state of the hadron in question,
see Fig.~\ref{figSmear}.
The signal from the corresponding correlator is less contaminated
by fluctuations from excited states and thus masses can be extracted
from shorter euclidean time extents. This leads to better statistics and 
is often crucial for the numerical work. A similar technique, known as
`fuzzing' applies to the link variables \cite{Rot92} but will not be detailed
here.

Thus, the equivalent of the one-meson field (\ref{onemes}) is replaced by
\begin{equation}
\phi_{\vec{p}}(t)=L^{-3}\sum_{\vec{x}}\,e^{i\vec{p}\cdot\vec{x}}
\bar{\psi}^{\{ k\}}_{dA}(\vec{x},t)\gamma_5\psi^{\{ k\}}_{uA}(\vec{x},t)
\label{oneWmes}\end{equation}
for a $\pi^+$ pseudoscalar meson. As usual, the quark propagator is obtained
from contractions between the (unsmeared) quark fields
\begin{eqnarray}
\ldots \stackrel{n}{\psi}_{fA\mu}(x)
\stackrel{n}{\bar{\psi}}_{gB\nu}(y)\ldots &=&
\ldots G_{fA\mu,gB\nu}(x,y)\ldots \\
\ldots \stackrel{n}{\bar{\psi}}_{fA\mu}(x)
\stackrel{n}{\psi}_{gB\nu}(y)\ldots &=&
\ldots - \gamma_{5,\mu'\mu} G^{\ast}_{fA\mu',\,gB\nu'}(x,y)
\gamma_{5,\nu\nu'}\ldots \,.\rule{26pt}{0pt}
\label{con1122W}\end{eqnarray}
The latter equation is a consequence of the identity
\begin{equation}
G(y,x)=\gamma_5 G^\dagger(x,y)\gamma_5 \,.
\label{CPT}\end{equation}
The $\dagger$ refers to flavor-color-Dirac indices. The identity is valid
if the lattice action is CPT invariant \cite{Weingarten:1982jy}.
Further, if the mass matrix in the Wilson fermion action is flavor diagonal,
then also
\begin{equation}
G_{fA\mu,gB\nu}(x,y)=\delta_{f\,g} G^{(f)}_{A\mu,B\nu}(x,y) \,.
\end{equation}
Typically, one would choose equal Wilson hopping parameters $\kappa$
(those determine the quark masses) for $u$ and $d$ quarks and different
ones for the strange quark.

We now give a list of essential things to do, according to the plan of
Secs.~\ref{secERI} and \ref{secU1}.

Consider the linear equation
\begin{equation}
\sum_{\vec{y}y_4} \sum_{B\nu}\,G^{-1(f)}_{A\mu,B\nu}(\vec{x}x_4,\vec{y}y_4)\,
X^{(f;\,A'\mu'r\,x_4')}_{B\nu}(\vec{y}y_4)\:=\:
\delta_{AA'}\,\delta_{\mu\mu'}\,R^{(A'\mu'r\,x_4')}(\vec{x})\,
\delta_{x_4 x_4'}.
\label{DXR}\end{equation}
The meaning of the indices are $A,B\ldots=1,2,3$ {\rm color}, 
$\mu,\nu\ldots=1,2,3,4$ {\rm Dirac}, $\vec{x},\vec{y}\ldots$ {\rm space}
($d=3$), $x_4,y_4$ {\rm time}, and $r=1\dots N_R$ labels the random sources
$R$ for each source point. A prime $'$ denotes a source point. There is some 
freedom in choosing the latter. In (\ref{DXR}) the sources are nonzero on one 
time slice only. The same set of sources is used for different flavors 
(Wilson hopping parameters $\kappa$), but a separate source is chosen for each 
color, Dirac and time index.
The features of the random sources are somewhat linked to the choice 
of the solution algorithm \cite{Glassner:1996gz}.
Also, Gaussian, ${\mathbb Z}_2$, or other random sources may be employed.
In any case, the random-source average, which we approximate numerically as
\begin{equation}
\sum_{\langle r\rangle}\ldots \approx \frac{1}{N_R}\sum_{r=1}^{N_R}\ldots \,,
\end{equation}
must satisfy
\begin{equation}
\sum_{\langle r\rangle} R^{(A'\mu'r\,x_4')}(\vec{x})
R^{(B'\nu'r\,y_4')\ast}(\vec{y}\,) = \delta_{A'B'}\,\delta_{\mu'\nu'}\,
\delta_{\vec{x}\vec{y}}\,\delta_{x_4' y_4'} \,.
\end{equation}
A conjugate-gradient algorithm \cite{Bec60}, or a variant, is
suitable for solving the linear Eq.~(\ref{DXR}).
An estimator for the propagator matrix elements then is
\begin{equation}
G^{(f)}_{B\nu,A\mu}(\vec{y}y_4,\vec{x}x_4) =
\sum_{\langle r\rangle} X^{(f;\,A\mu\,r\,x_4)}_{B\nu}(\vec{y}y_4)
R^{(A\mu\,r\,x_4)\ast}(\vec{x}) \,.
\end{equation}
For simple hadron-hadron systems it may be sufficient to place sources on 
one time slice $x_4$ only. This is the situation for the $\pi^+$--$\pi^+$ 
system ($I=2$ channel). Matching the notation in the examples 
discussed in previous sections we set $x_4=t_0$. 
Putting things together, the correlation matrices are then computed through 
the following sequence of steps:
\begin{eqnarray}
&&R^{\{ 0\}(B\nu\,r\,t')}_{C}(\vec{x}\,) =
\delta_{CB} R^{(B\nu\,r\,t')}(\vec{x}\,)\quad\mbox{(\rm no sum over $B$)} \\
&&R^{\{ k\}(B\nu\,r\,t')}_{C}(\vec{x}\,) =
\sum_{\vec{y}\,}  \sum_{A}  K_{CA}(\vec{x}\,,\vec{y}\,)R^{\{ k-1\}(B\nu\,r\,t')}_{A}(\vec{y}\,) \\
&&X^{\{ 0\}(f;\,B\nu\,r\,t')}_{C\mu}(\vec{x}t) =
X^{(f;\,B\nu\,r\,t')}_{C\mu}(\vec{x}t) \\
&&X^{\{ k\}(f;\,B\nu\,r\,t')}_{C\mu}(\vec{x}t) =
\sum_{\vec{y}} \sum_{A} K_{CA}(\vec{x},\vec{y})X^{\{ k-1\}(f;\,B\nu\,r\,t')}_{A\mu}(\vec{y}t) \\
&&{\cal R}^{\{ k\}(AB\,\mu\nu\,r'r\,t')}(\vec{p}\,) =
\sum_{\vec{x}} L^{-3} e^{-i\vec{p}\,\cdot\vec{x}\,}\sum_{C}\nonumber\\
&&\rule{20ex}{0ex}\times R^{\{ k\}(A\mu\,r'\,t')\ast}_{C}(\vec{x}\,) R^{\{ k\}(B\nu\,r\,t')}_{C}(\vec{x}\,) \\
&&{\cal X}^{\{ k\}(gf;\, AB\,\mu'\nu'\,r'r\,t')}_{\nu\mu}(\vec{p}t)=
\sum_{\vec{x}} L^{-3} e^{-i\vec{p}\cdot\vec{x}}\sum_{C}\nonumber\\
&&\rule{20ex}{0ex}\times X^{\{ k\}(g;\,A\mu'\,r'\,t')\ast}_{C\nu}(\vec{x}t)
X^{\{ k\}(f;\,B\nu'\,r\,t')}_{C\mu}(\vec{x}t)\rule{5ex}{0ex}
\end{eqnarray}
\begin{eqnarray}
&&C^{(2)}_{\vec{p}\vec{p}}(t,t_0) =
-\gamma_{5,\nu\mu}^\ast{\cal X}^{\{ k\}(ud;\, AB\,\mu'\nu'\,r'r\,t_0)\ast}_{\nu\mu}(\vec{p}t)\nonumber\\
&&\rule{20ex}{0ex}\times \gamma_{5,\mu'\nu'}{\cal R}^{\{ k\}(ud;\, AB\,\mu'\nu'\,r'r\,t_0)} \,.
\end{eqnarray}
Summation over doubly-occuring indices is understood,
except indicated otherwise.
The expressions for the 4-point correlators $\overline{C}^{(4)}$ and $C^{(4)}$ are more 
complicated, but similar.

\begin{figure}[htb]
\centerline{\includegraphics[width=56mm]{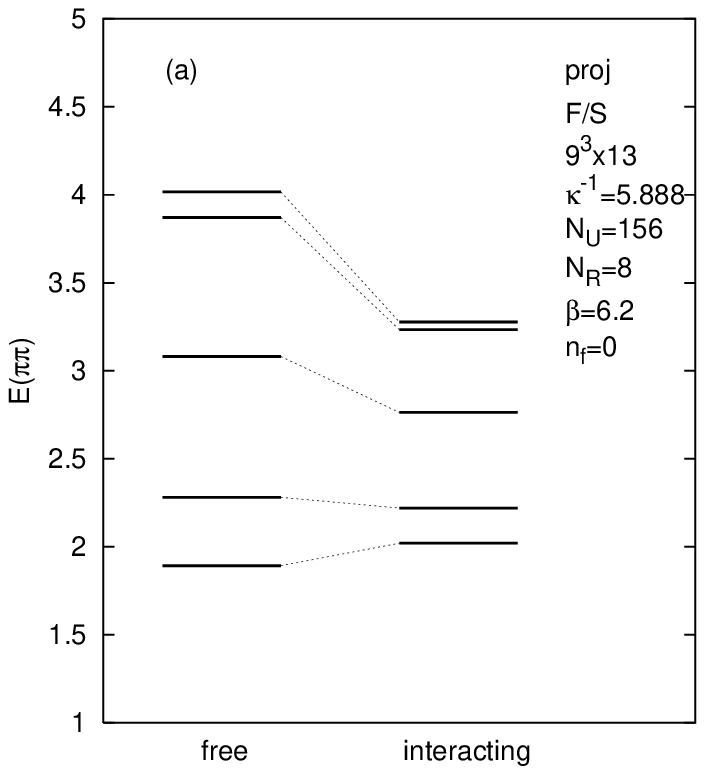}
            \includegraphics[width=56mm]{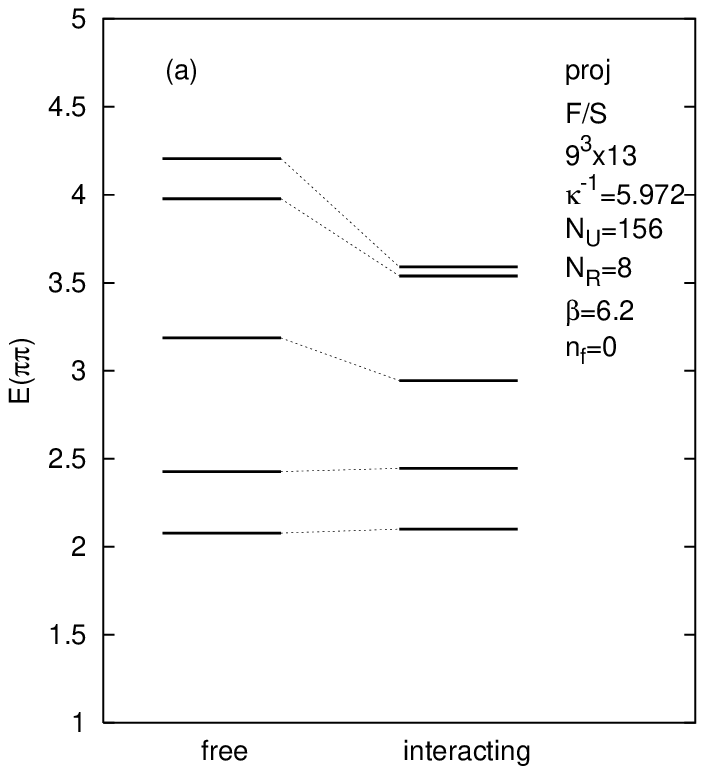}}
\vspace*{8pt}
\caption{Meson-meson energy level splitting due to the residual interaction.
Two different values of the hopping parameter (quark masses) are
shown \protect\cite{Rab97}.\label{figPiPiSpec}}
\end{figure}

Figure~\ref{figPiPiSpec} shows energy-level diagrams of the
non-interacting, from $\overline{C}^{(4)}$, and the interacting, from
$C^{(4)}$, $\pi$--$\pi$ systems for two values of the
hopping parameter, $\kappa^{-1}=5.888$ and $\kappa^{-1}=5.972$.
These are from a set of six values $\kappa^{-1}$ which have been
used to perform the chiral extrapolation $m_\pi^2\rightarrow 0$
(equivalently $\kappa\rightarrow \kappa_c$).
The other lattice parameters \cite{Fiebig:1998va} were $L=9$,
$T=13$, and $\beta=6.2$ with the next-nearest-neighbor $o(a^2)$
tadpole-improved action \cite{Hamber:1983vu,Eguchi:1984xr}. The lattice constant,
matched to the string tension, is $a=0.4$~fm or $a^{-1}=500$~MeV.
An interesting feature of the spectra is that high-momentum levels
are lowered by the interaction indicating attraction at short distances,
and some of the lower levels move up indicating some repulsion in the
system.

\begin{figure}[htb]
\centerline{\includegraphics[width=48mm]{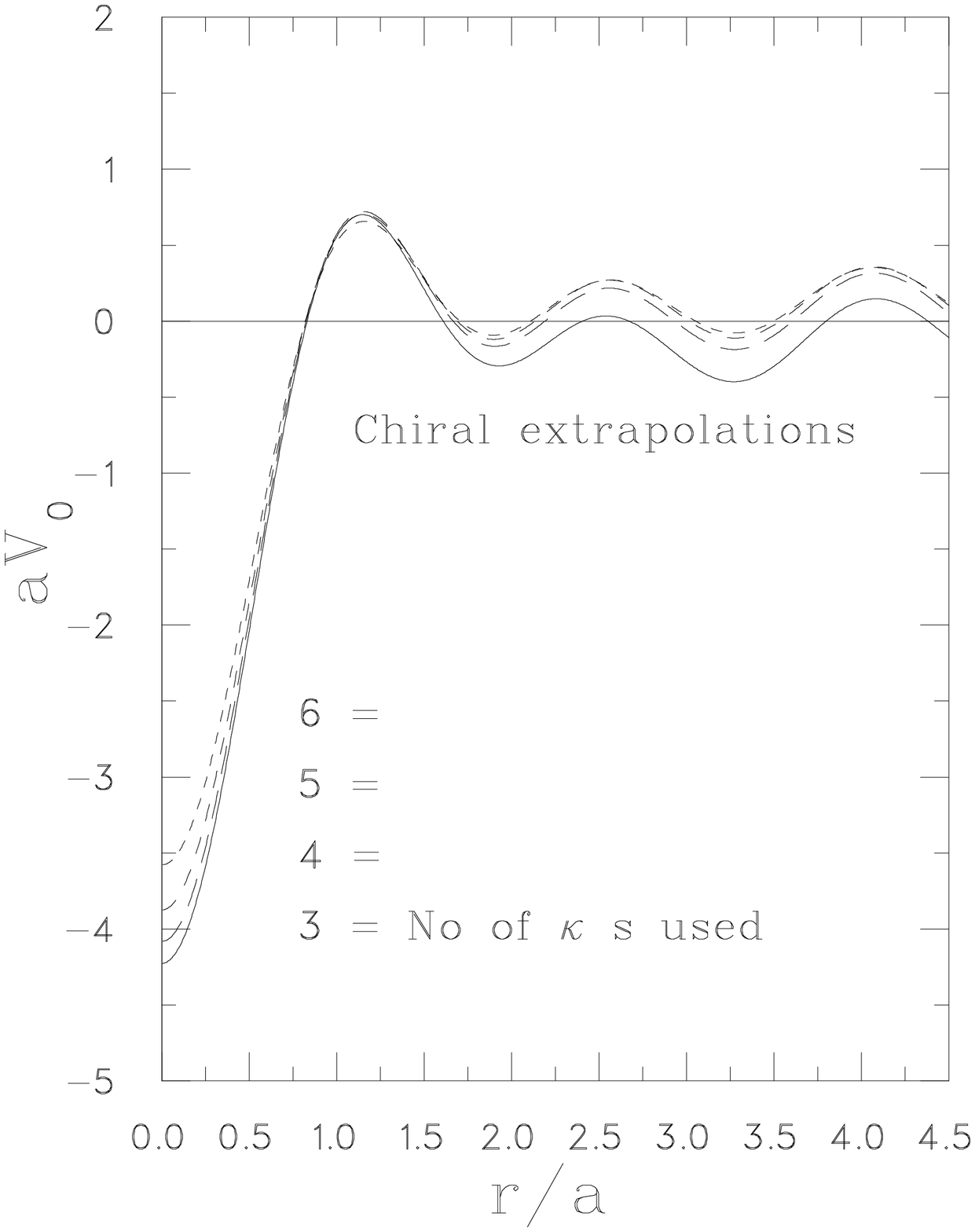}
            \includegraphics[width=48mm]{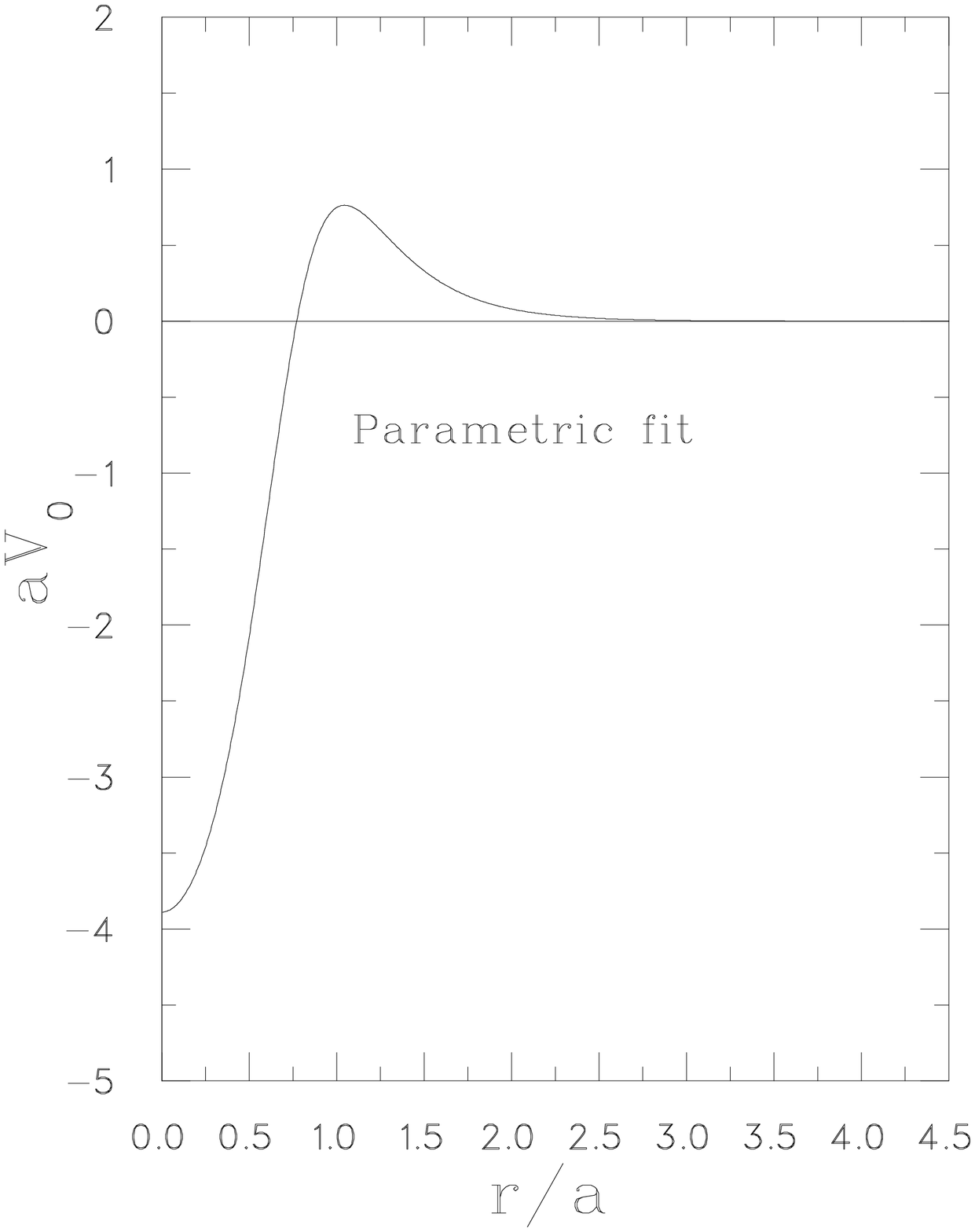}}
\caption{Approximate s-wave projected local meson-meson potentials
 according to
(\protect\ref{Vpot}) and (\protect\ref{Vl0}).
Shown is the result of a chiral extrapolation ($m_\pi^2\rightarrow 0$)
of the potential based on various hopping parameter values; curves resulting
from the use of 3 to 6 $\kappa$-values are compared. A parametric fit
with $V^{(\alpha)}(r)$ based on (\protect\ref{Valpha}), using 3 values
of $\kappa$, is presented \protect\cite{Fiebig:1999hs}.\label{figPiPiVloc}}
\end{figure}

\begin{figure}[htb]
\centerline{\includegraphics[width=48mm]{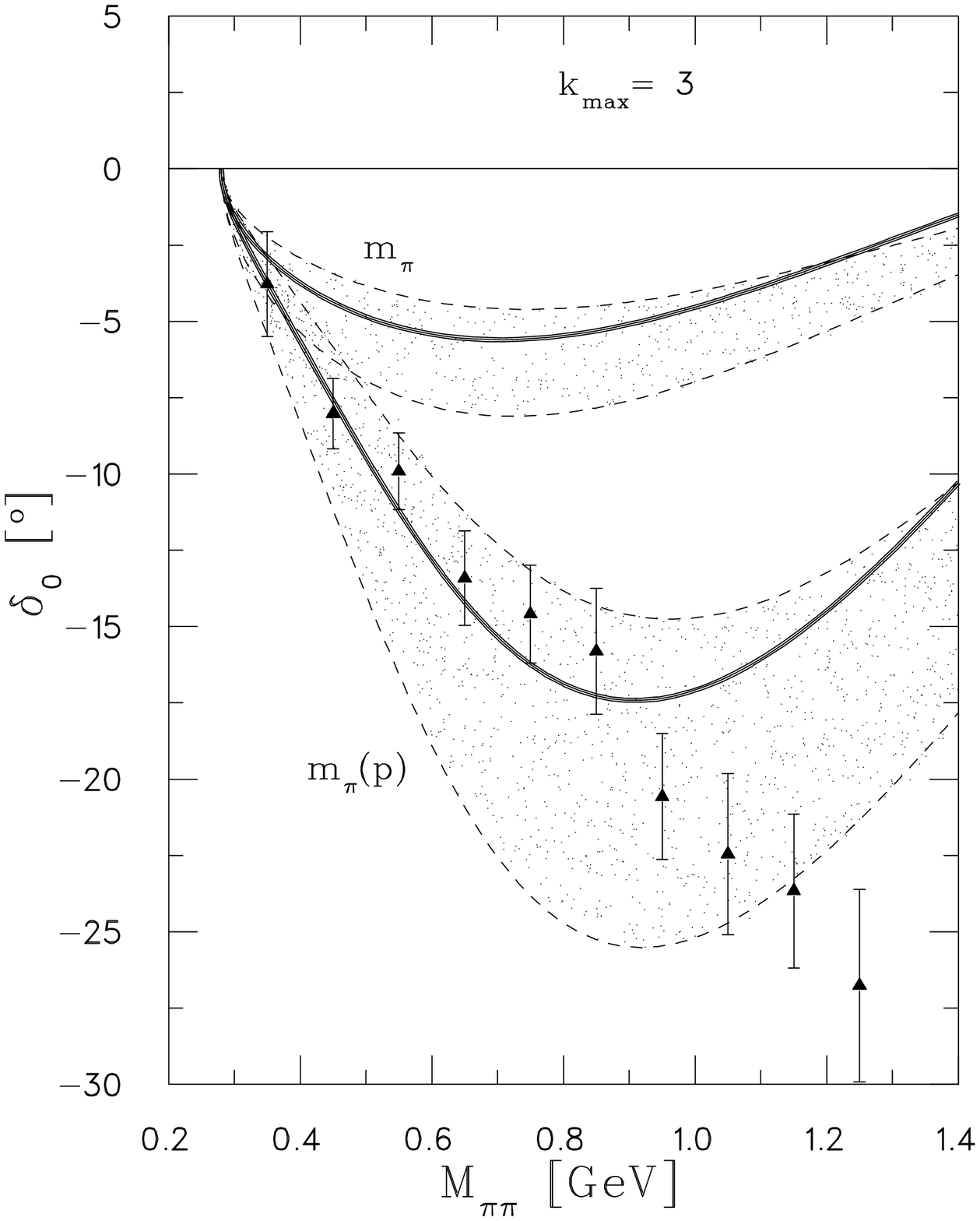}
            \includegraphics[width=48mm]{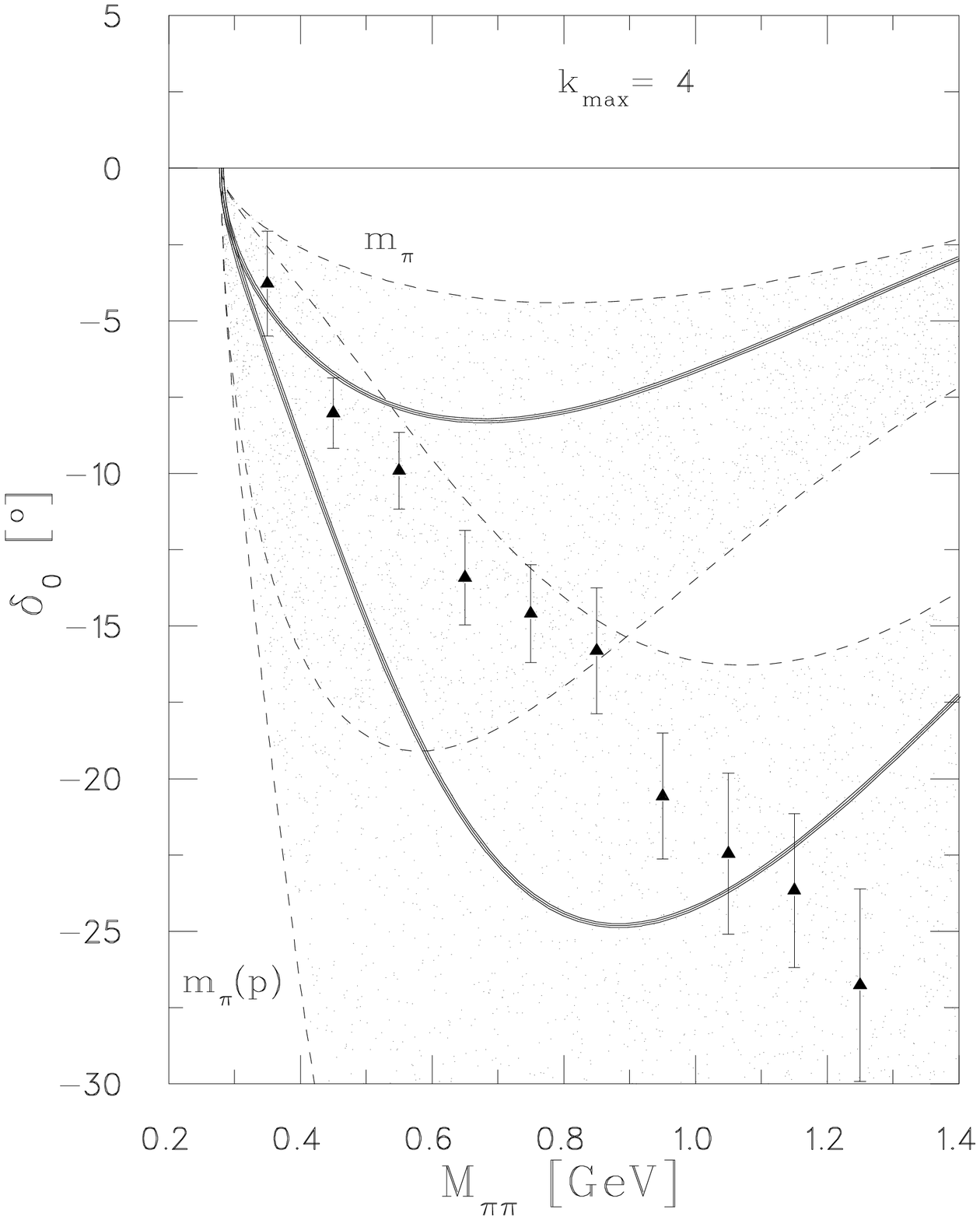}}
\caption{Scattering phase shifts $\delta^{I=2}_{\ell=0}$ calculated from the
$\pi$--$\pi$ potential of the lattice QCD simulation (thick lines)
\protect\cite{Fiebig:1999hs}.
Influence of momentum cutoff $k_{\max}=3$ (left) and $k_{\max}=4$ (right) is shown.
Results using the classical and relativistic dispersion relation are distinguished
by $m_\pi$ and $m_\pi(p)$, respectively.
Errors are represented by the dotted regions. Their boundaries (dashed lines) 
correspond to the phase shifts calculated with extremal (bootstrap)
potentials.
The experimental data are from Ref.~\protect\cite{Hoogland:1977kt}.\label{figPhasPiPi}}
\end{figure}

The corresponding approximate local potentials, according to (\ref{Vpot})
of Sec.~\ref{secTmb}, were projected onto the partial wave $\ell=0$
\begin{equation}
{\cal V}_0(r) = \frac{1}{4\pi}\int d\Omega_{\vec{r}}\,{\cal V}(\vec{r}\,) =
\sum_q\,j_0(2qr)\,\langle q|{\cal H}_I^{(A_1)}|q\rangle\,,
\label{Vl0}\end{equation}
with $q=\frac{2\pi}{L}k, k=0\ldots k_{\max}$.
Results of the chiral
extrapolation plot \cite{Fiebig:1999hs} are shown in Fig.~\ref{figPiPiVloc}.
The Fourier nature of the obtained potentials is apparent.
A parametric fit, using (\ref{Valpha}),(\ref{ValphaL}) and
(\ref{chi2alpha}), is also shown. 
Of course short distances $r$ are beyond the scope of the truncated 
momentum basis and should be interpreted with caution.
Thus only small relative momenta make sense when calculating
scattering phase shifts with this potential.
Results using standard (non-relativistic) scattering theory \cite{Tay72}
with parameterized potentials (\ref{Valpha}) are 
displayed in Fig.~\ref{figPhasPiPi}.

\clearpage

\subsection{Coordinate-Space Work on the $\pi$--$\pi$ System\label{secCorSpcWrk}}

\begin{figure}[htb]
\centerline{\includegraphics[width=40mm]{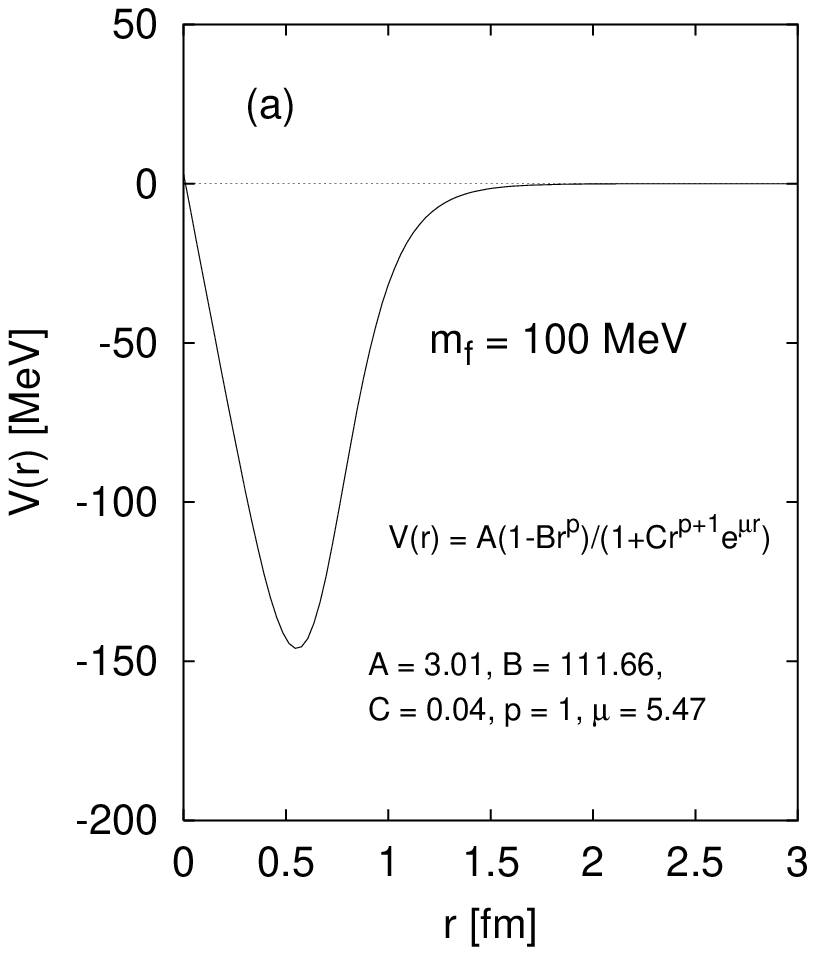}\hspace{12mm}
            \includegraphics[width=40mm]{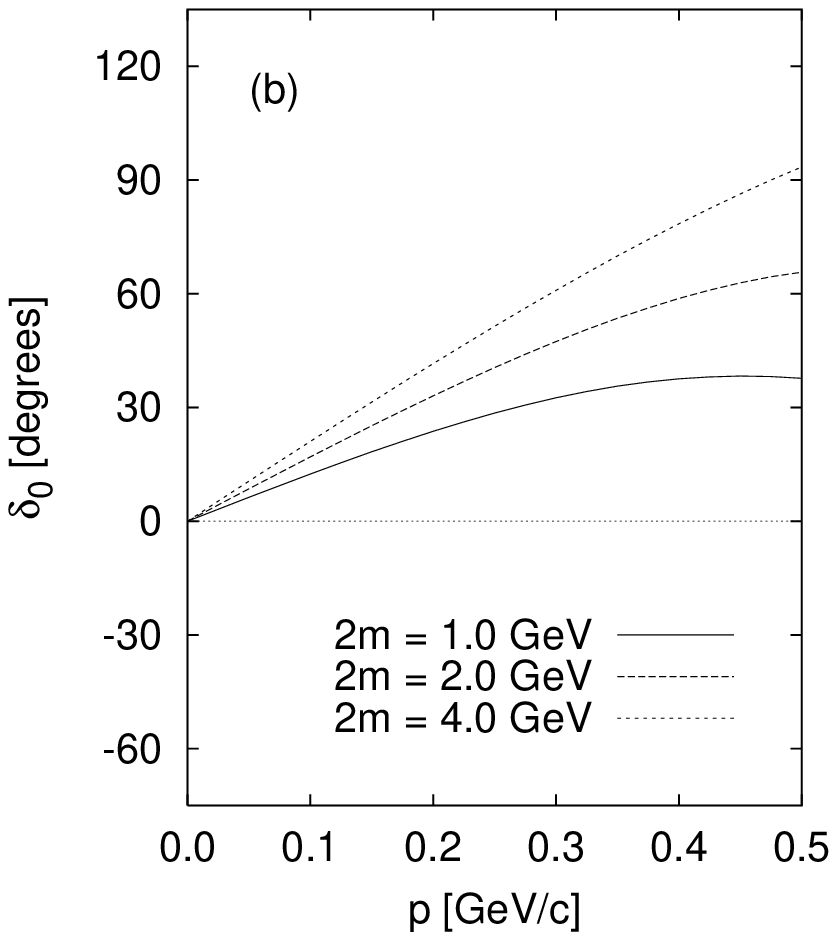}}
\centerline{\includegraphics[width=40mm]{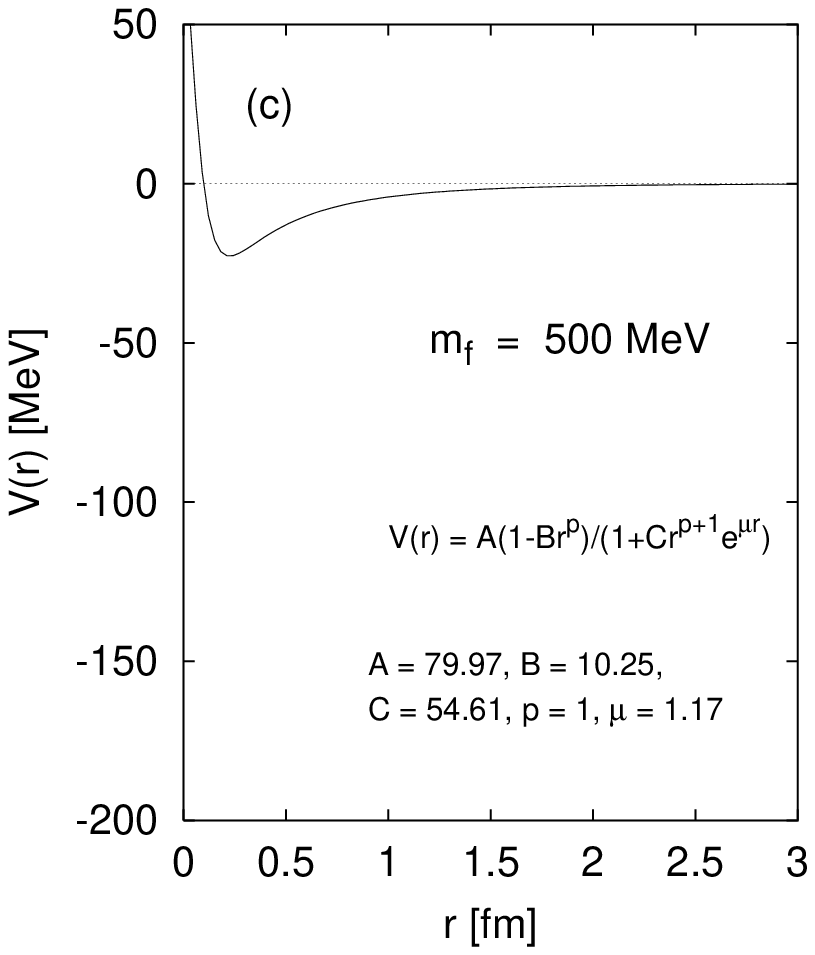}\hspace{12mm}
            \includegraphics[width=40mm]{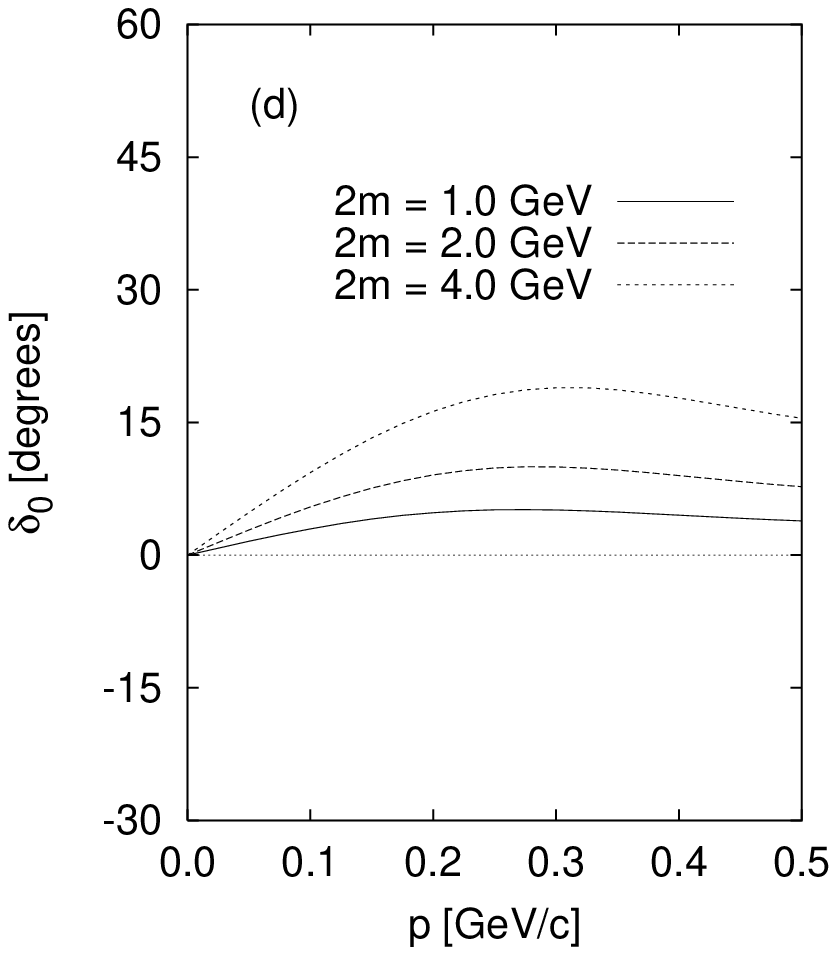}}
\centerline{\includegraphics[width=40mm]{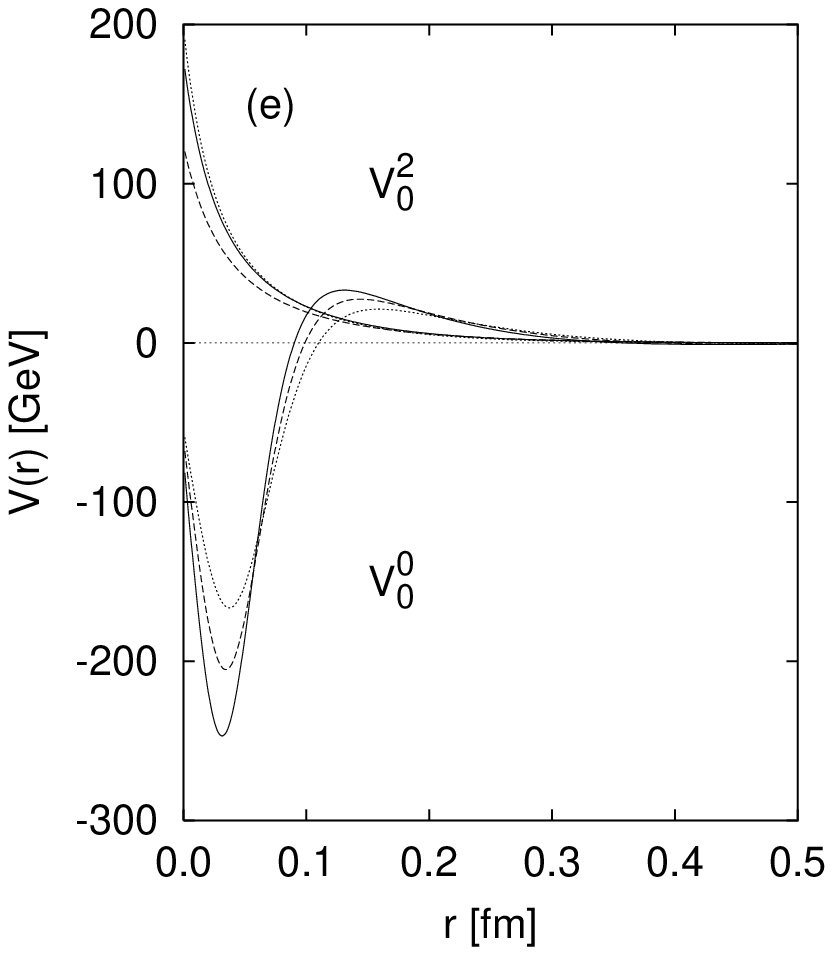}\hspace{12mm}
            \includegraphics[width=40mm]{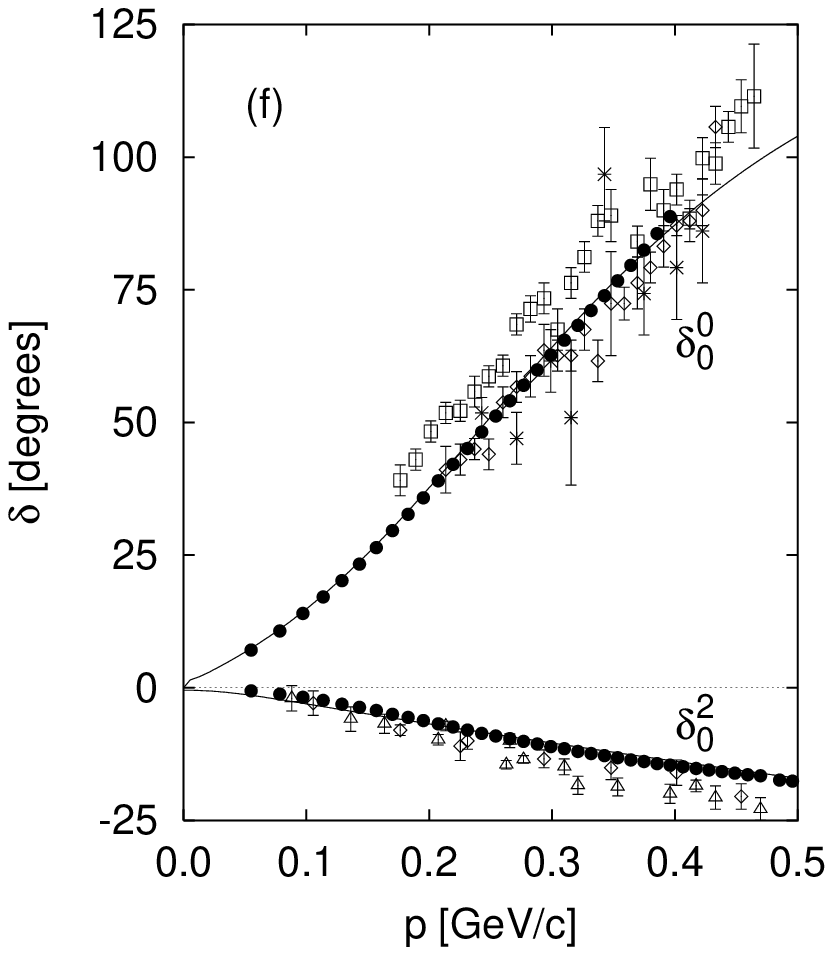}}
*\vspace*{8pt}
\caption{(a)--(d): Lattice results for meson-meson interaction potentials
$V(r)$ after the subtraction of the effects of the mirror
particles, and phase shifts $\delta$, for quark masses
$m_f=100$~MeV and $m_f=500$~MeV. (e):
$\pi\pi$ $L = 0$, $I=0$ and 2 inversion potentials.
Froggatt \protect\cite{Froggatt:1977hu} (solid line),
$\chi$PT \protect\cite{Gasser:1983kx,Gasser:1984yg} (dashed)
and meson exchange \protect\cite{Lohse:1990ew} (dotted). (f): from
$\pi\pi$ $L=0$, $I=0$ and 2 phase shifts of
Froggatt data \protect\cite{Froggatt:1977hu}
and interpolation (dots and solid line),
Estabrooks et al. \protect\cite{Estabrooks:1974vu} (boxes),
Grayer et al. \protect\cite{Grayer:1974cr} (diamonds),
M\"anner \protect\cite{Mae74} (triangles)
and Baillon et al. \protect\cite{Baillon:1972tx} (asterixes).
\label{figcomparepipi}}
\end{figure}

We turn to another exploration in coordinate space with staggered fermions \cite{Rab97}.
The final interaction potential is shown in
Figs.~\ref{figcomparepipi}(a) and (c)
for $m_f=100$~MeV and $m_f=500$~MeV, respectively.
The fits of the periodic potential were stable for the data with
$m_f=100$~MeV. The data for $m_f=500$~MeV suffer from very small
energy differences which makes a fit unreliable.
The potentials exhibit attraction at intermediate range
and may have a repulsive behavior at short distances. The potential at
$m_f=500$~MeV is compatible with zero interaction.
The analytic form (\ref{Valpha}) of the interaction potential
was used as an input to the Schr\"odinger equation for a phase shift
calculation according to scattering theory \cite{Tay72}. The resulting
phase shifts are displayed in Figs.~\ref{figcomparepipi}(b) and (d)
for different values of the meson mass parameter used in the Schr\"odinger
equation.

One aim of this study was the comparison of results obtained from
lattice QCD with results calculated from inverse scattering theory.
There is a wealth of
inversion results from hadron-hadron scattering \cite{Sander:1996jq}.
We here restrict ourselves to cases with partial wave $L=0$ meson-meson scattering.
The basis for any inversion is the existence of experimental data which,
after a suitable interpolation and extrapolation, leads to a solution of
the underlying ill-posed problem.
For the $\pi$--$\pi$ system the experimental situation is acceptable.
The main source of experimental information for the inversion
procedure is the analysis of the CERN--Munich experiment
$ \pi^- p \rightarrow \pi^- \pi^+ n $ \cite{Grayer:1974cr,Froggatt:1977hu,Ochs:1991rw}.
Even though there are ambiguities in
the analysis of the $\pi\pi$ final state, all phase shift analyses reach
comparable results, which can be described by both meson exchange \cite{Lohse:1990ew}
and chiral perturbation theory ($\chi$PT) \cite{Gasser:1983kx,Gasser:1984yg}.
This experimental situation is depicted in Fig.\ \ref{figcomparepipi}(f).
The inversion results for several sources of phase shifts are shown
in Fig.\ \ref{figcomparepipi}(e).
While the isospin $I=2$ channel is purely repulsive,
the isospin $I=0$ channel shows an attractive component.
A resonance is supported in this channel \cite{Sander:1996jq}.

The lattice mesons are coupled to isospin $I = 2$ but
constructed without a specific angular momentum
projection.  Thus there is no one-to-one
correspondence between the quantum numbers of the lattice mesons and the
experimental data.
The lattice potentials in
Figs.~\ref{figcomparepipi}(a) and (c) are attractive with repulsion at
short distances, whereas the potentials obtained from inverse scattering
in Fig.~\ref{figcomparepipi}(e) are only repulsive. This is expressed in
the different form
of the phase shifts in Fig.~\ref{figcomparepipi}(f) compared with
those from the lattice QCD calculations in
Figs.~\ref{figcomparepipi}(b) and (d).
One reason might be the incomplete projection to the correct quantum
numbers. The construction of good quantum numbers is not straightforward
in the Kogut-Susskind scheme. It is easier with the Wilson fermions but
leads to extra terms in the correlator which have to be resolved
numerically (see Appendix~\ref{appA}).

To summarize Secs.~\ref{secMomSpcWrk} and \ref{secCorSpcWrk}, we have performed
two analyses partly in the same
line both aiming at the extraction of a pion--pion potential from lattice QCD.
The computation within staggered quarks suffered from the construction of spin observables
whereas the more sophisticated calculation with Wilson fermions introduced a somewhat
arbitrary cutoff at short distances. As a general conclusion the reader has seen the
current state of these trials which he might judge to be qualitative. Not only the
increase of computer power but the efforts of physicists can make the 
results more
accurate to compare with and predict experiments.

\clearpage

\section{Conclusion and Outlook}

The field of {\sl hadronic physics} has experienced a change of paradigm in
recent years. From a modern perspective its primary goal now is the understanding
of hadronic phenomena in terms of the fundamental degrees of freedom of
the underlying theory, quantum chromodynamics \cite{Capstick:2000dk}.
Historically, phenomenological descriptions of hadron-hadron interactions
have been with us for more than six decades while new questions related to
the quark-gluon structure of hadrons have emerged as our insight has deepened
over time.
Thus hadronic physics now is closely intertwined with QCD.
Dramatic advances in computational technology, lattice field theory, and
algorithms of computational physics, have moved unprecedented opportunities
for fundamental cognition within our reach.
New accelerators and advanced detector design should of course also be
mentioned in this context.
Among the many topics tackled by lattice QCD -- the hadron
mass spectrum, hadronic structure, heavy flavor and exotic hadrons, to
name just a few -- the physics of hadron-hadron interactions is one of the more
challenging subjects.

It is thus not surprising that the topic of this chapter is only in a
nascent state. Concluding words with a quality of closure would be truly
misleading. The current stage of hadron-hadron interactions on the lattice
is driven by the tools of lattice technology with mass calculations being
directly accessible from the euclidean formulation.
All methods discussed in this chapter essentially use the
energy spectra of certain systems as a starting point.
Three ways to utilize those have been visited: finite volume methods and direct
calculation of scattering
phase shifts (L{\"u}scher's method), construction of an effective
interaction from two-hadron states (perturbative potentials), and direct
computation of the interaction in heavy-light systems (adiabatic potential).

Of those methods the first one addresses measurable data directly, so far
its most prominent application being aimed at scattering lengths. The
other variants leave the realm of field theory in the hope of unveiling
physical mechanisms of the interaction. Perhaps, in the latter case the usual
lattice technology constraints (related to chiral symmetry, quark masses,
finite volume, continuum limit, etc) are a lesser impediment to progress
in the sense that qualitative insights into the physics of the interaction
can still be gained with `less than perfect' lattice parameters.   

Precision simulations reproducing experimental scattering data should probably not be
the first priority at this time. To achieve this the field may not be
advanced enough, particularly in the nucleon-nucleon case.
Also, fermion methods dealing properly with chiral
symmetry \cite{Neuberger:2001nb} should certainly be incorporated.

Understanding hadronic interaction mechanisms, for example the importance of
gluon or quark degrees of freedom at certain relative distances, is probably
more desirable at this stage. Emphasizing this aspect in future studies,
for example heavy-light meson-baryon and baryon-baryon systems, seems
a promising line of work that should shed some light on the QCD aspects of
the hadron-hadron interaction. A related aspect is understanding of
the origin of effective field theories \cite{Colangelo:2001df}.

From the viewpoint of lattice technology the computation of multi-quark
correlation functions, which requires all-to-all fermion propagators, is
certainly the biggest problem. Considerable space was devoted to random
source methods. Those are likely to remain pillars of hadron-hadron
systems simulations. In another development, not mentioned before,
Baysian analysis methods for
lattice generated time correlation functions promise access
to excited states \cite{Nakahara:1999vy,Lepage:2001ym,Fiebig:2002sp}, an aspect
crucial for the topic of this chapter.

Finally, although it is true that the euclidean formulation of lattice
QCD in imaginary time currently dominates the field, one should realize that it may not
be the framework best suited to the task. It is not clear how to overcome
the difficulty of describing asymptotic states, in the sense of LSZ
scattering formalism,
other than trying to solve the problem via spectrum computations at multiple
lattice volumes \cite{Luscher:1991cf}. In the context of scattering the alternative
hamiltonian formulation in real time is largely unexplored \cite{Chaara:1994ar}.   

\section*{Acknowledgments}
\addcontentsline{toc}{section}{Acknowledgments}

It is a great pleasure to acknowledge the many
significant contributions of (in alphabetical order)
J. Canosa,
A. Dominguez,
O. Linsuain,
A. Mih\'{a}ly,
K. Rabitsch,
R.M. Woloshyn,
who have at one time or another advanced the subject matter
of this review in various ways.
The support and hospitality of Thomas Jefferson National Laboratory JLab
during three summers and by the European Center for Theoretical
Studies ECT${}^\ast$ are gratefully acknowledged.
This material is based upon work supported by the U.S. National Science Foundation
under Grant No. 0073362.
Resources made available through the Lattice Hadron Physics Collaboration LHPC
were used.
This work was further supported by the Austrian Science Foundation
FWF under Project P10468-PHY.

\section{Appendix~A: Remarks on Staggered Fields\label{appA}}

\setcounter{equation}{0}
\renewcommand{\theequation}{A.\arabic{equation}}

The two most widely used schemes for discretizing fermion fields are
commonly known as `Wilson fermions' \cite{Wil75} and `Kogut-Susskind' or
`staggered fermions' \cite{Kogut:1975ag,Susskind:1977jm,Burden:1987by}.
Wilson fermions are computationally
more demanding than staggered fermions. While Wilson fermions explicitly
break chiral symmetry, though they have a chiral limit which in practice 
can be recovered via a $m_{\pi}^2\rightarrow 0$ extrapolation of observables,
staggered fermions retain a discrete remnant of chiral symmetry \cite{Kluberg-Stern:1983dg}.
In the Wilson scheme construction of 
interpolating fields for hadrons with definite quantum numbers is
transparent, and works more or less like in the continuum theory. 
This is opposite for staggered fermions, which entail complicated
mathematics when constructing
interpolating fields \cite{Kawamoto:1981hw,Kluberg-Stern:1983dg,Kilcup:1987dg}.

A case in point is an ordinary $\pi^+$ interpolating field. First, the
quark fields, $q_f$, are nonlocal operators which live on elementary
hypercubes of the lattice \cite{Kluberg-Stern:1983dg,Burden:1987by}.
For a U(1) gauge model (no colors) we have
\begin{equation} 
q_f(x) = C^{-3/2}\sum_{\eta}\,{\Gamma}_{\eta}  U(\eta,x)
\chi_f(2x+\eta)
\label{A1q}\end{equation}
built from the staggered fields $\chi_f$.
The sum is over the set of the $d$-dimensional vectors $\eta$ that point to
the corners of the hypercube, $\eta_\mu=0,1$ for $\mu=1\ldots d$,
with $d=4$, see Ref.~\cite{Burden:1987by} for an odd number of dimensions. Further
\begin{equation} 
\Gamma_{\eta} = \gamma_1^{\eta_\mu}\ldots\gamma_d^{\eta_d} \,
\quad\mbox{and}\quad
U(\eta,x) = [U_1(2x)]^{\eta_1} \ldots
[U_d(2x+\eta_1\ldots\eta_{d-1})]^{\eta_d} \,,
\label{A1GamU}\end{equation}
and $C$, in (\ref{A1q}), is the dimension of the Clifford algebra.
A  $\pi^+$ interpolating field then is
\begin{equation}
\phi(x)=\bar{q}_d(x)(\gamma_5\otimes\gamma_5^{\ast})q_u(x) \,.
\label{A1pi}\end{equation}
The direct product $\otimes$ refers to row and column indices, respectively,
of Clifford algebra matrices. Now, turning to the corresponding correlation
function it can be shown that the latter reduces to a sum of local terms
\begin{eqnarray}
\langle \sum_x \phi^{\dagger}(\vec{x},t)
\phi(\vec{x}_0,t_0) \rangle &=& 
\langle 2\sum_{\vec x} [
2 G(\vec{x},2t;\vec{x}_0,t_0) G^{\ast}(\vec{x},2t;\vec{x}_0,t_0) 
\phantom{xxxxxxxx}
\nonumber \\ & &
+ G(\vec{x},2t+1;\vec{x}_0,t_0) G^{\ast}(\vec{x},2t+1;\vec{x}_0,t_0) 
\nonumber \\ & &
+ G(\vec{x},2t-1;\vec{x}_0,t_0) G^{\ast}(\vec{x},2t-1;\vec{x}_0,t_0) 
] \rangle \,.
\label{A1picor}\end{eqnarray}
For a concise exposition see Ref.~\cite{Wilcox:1985bv}. This shows that the following
equivalent correlator may be used
\begin{equation}
C^{(2)}(t,t_0) =  \langle\sum_{\vec x}
G(\vec{x},t;\vec{x}_0,t_0) G^{\ast}(\vec{x},t;\vec{x}_0,t_0)
\rangle \,.
\label{A1piC2}\end{equation}
For our purposes a useful observation is that the local field
\begin{equation}
(-1)^{x_1+\ldots +x_{d-1}+t}\bar{\chi}_d(\vec{x},t)\chi_u(\vec{x},t)
\label{A1}\end{equation}
supplemented by a phase factor as shown gives rise to the
correlator (\ref{A1piC2}).
Hence, rather than using (\ref{A1q}) and going through an involved
derivation, we may as well work with the interpolating field
\begin{equation}
\phi_{\vec{p}}(t)=L^{-d/2}\sum_{\vec{x}}\,e^{i\vec{p}\cdot\vec{x}}
(-1)^{x_1+\ldots +x_{d-1}+t} \, \bar{\chi}_{d}(\vec{x},t)\,\chi_{u}(\vec{x},t)
\label{A1onemes}\end{equation}
from the outset. The construction of correlators requires contractions
between the staggered Grassmann fields
\begin{equation} 
\ldots\stackrel{n}{\chi}_{f}(x) \stackrel{n}{\bar{\chi}}_{f'}(x')
\ldots = \ldots\delta_{ff'} {G}(x,x')\ldots\,,
\label{A1eq42}\end{equation}
where ${n}$ indicates the partners of the contraction, and $G$ is the
inverse fermion matrix, see (\ref{FerMat}). We also need
\begin{equation} 
\ldots \stackrel{n}{\bar{\chi}}_{f}(x) \stackrel{n}{\chi}_{f'}(x')\ldots =
\ldots-\stackrel{n}{\chi}_{f'}(x') \stackrel{n}{\bar{\chi}}_{f}(x)\ldots =
\ldots-\delta_{f'f} {G}(x',x)\ldots \,.
\label{A1eq42a}\end{equation}
In our application $x'$ lives on a fixed time slice $t_0$ whereas $x$ runs
over all lattice sites. This would require computing the entire inverse
fermion matrix, as opposed to only one column. Here a symmetry can
be employed.
In the staggered fermion scheme the propagator satisfies \cite{Kilcup:1987dg}
\begin{equation} 
G(x',x) = (-1)^{x'}\,G^{\dagger}(x,x')\,(-1)^{-x} \quad\mbox{with}\quad
(-1)^x=(-1)^{x_1+\ldots+x_d} \,,
\label{A1eq42b}\end{equation}
which is actually obvious from (\ref{FerMat}).
(The Wilson fermion version of this is (\ref{CPT}).)
Thus
\begin{equation} 
\ldots \stackrel{n}{\bar{\chi}}_{f}(x) \stackrel{n}{\chi}_{f'}(x')\ldots =
\ldots-\delta_{ff'}(-1)^{x'}{G}^{\dagger}(x,x')(-1)^{-x}\ldots \,.
\label{A1eq42c}\end{equation}
It is straightforward to check that the correlation matrices $C^{(2)}$
and $C^{(4)}$ of Sec.~\ref{secMesF} constructed with the fields 
(\ref{A1onemes}) give exactly the same results as stated in (\ref{eq43})
and (\ref{eq46}) respectively. The reason is cancellation of the phase
factors appearing in (\ref{A1onemes}) and (\ref{A1eq42c}).

For the purpose of concise presentation we omitted the phase
factors in Sec.~\ref{secMesF}.

\section{Appendix B: Improved Lattice Actions\label{appB}}

\setcounter{equation}{0}
\renewcommand{\theequation}{B.\arabic{equation}}

The history of improved lattice actions originates with work by
Symanzik \cite{Sym82,Symanzik:1983dc,Symanzik:1983gh}. The idea is to cancel discretization errors through
adding additional, carefully tuned, terms to the lattice action.
Quantum effects (radiative corrections) play an important role in their
construction.
Symanzik's improvement program has received considerable
attention \cite{Curci:1983an,Weisz:1983zw,Weisz:1984bn,Luscher:1984zf,Luscher:1985xn}
since its inception, but has only recently evolved into a form
which is finding widespread application through the work of Lepage and
Mackenzie \cite{Lepage:1993xa}. Through a combination of beyond-elementary geometric
structures in the lattice action and accounting for part of the quantum
fluctuations through
numerically self-consistent renormalization lattice simulations now
operate considerably closer to the continuum limit than ever before.
With improved lattice actions results which can be compared to 
experiment are now within reach in many cases. With regard to the
current topic (hadronic interaction) improved actions are important
because they allow us to use large lattices, say $L\approx 4$~fm or so,
with relatively moderate computing facilities.

Improvement of the Lepage-Mackenzie flavor has been well publicized \cite{Lepage:1996ph}.
Educational articles (for beginners) detailing the basics, ramifications, and
physical significance of this development have been written by its
originators \cite{Lepage:1994yd,Lepage:1996jw}. There is no point in duplicating those
writings. However, given the importance of large lattices for hadronic
interactions it is appropriate to devote at least an appendix to the basics 
of this development. We will do so by way of selected examples.

\subsection{Appendix B.1: A Scalar Example}\label{secASX}

In the lattice discretization of the space-time continuum derivatives are 
approximated by finite differences. For example, considering a scalar
field theory,
\begin{equation}
\partial^2 \phi(x) \approx \sum_\mu \frac{\phi(x+a\hat{\mu})+\phi(x-a\hat{\mu})
-2\phi(x)}{a^2} \,.
\label{fdiff}\end{equation}
In the classical limit, this approximation is `good' if $\phi(x)$ is smooth 
on a scale larger than the lattice constant $a$. We may give a more precise
meaning to this statement by saying that the Fourier transform
\begin{equation}
\tilde{\phi}(p) = (2\pi)^{-2}\int d^4x\,e^{ip\cdot x} \phi(x)
\label{FT}\end{equation}
is zero (within some bound $\epsilon$) for momenta $|p|>\pi/a$. The error of
the finite-difference approximation (\ref{fdiff}) is $o(a^2)$, as a
Taylor expansion shows. In fact we may systematically improve upon the
classical discretization by including next-nearest-neighbor finite
differences. Writing
\begin{equation}
\phi(x+a\hat{\mu})=\sum_{n=0}^{\infty}\frac{a^n}{n!} \partial_{\mu}^{n}
\phi(x) = e^{a\partial_\mu} \phi(x)
\end{equation}
we may solve the four equations
\begin{eqnarray}
e^{\pm a\partial_\mu} &=& 1 \pm a\partial_\mu +\frac12 a^2\partial_\mu^2
\pm\frac16 a^3\partial_\mu^3 +\frac{1}{24} a^4\partial_\mu^4 \pm \ldots \\
e^{\pm 2a\partial_\mu} &=& 1 \pm 2a\partial_\mu +2a^2\partial_\mu^2
\pm\frac43 a^3\partial_\mu^3 +\frac23 a^4\partial_\mu^4 \pm \ldots 
\end{eqnarray}
for $\partial_{\mu}^{n}$, $n=1\ldots 4$, in terms of powers of single-step
shift operators
\begin{equation}
\Delta_\mu^{(\pm)} = e^{\pm a\partial_\mu} \,.
\end{equation}
Specifically we find
\begin{eqnarray}
a\partial_\mu &=& \frac23(\Delta_\mu^{(+)}-\Delta_\mu^{(-)})
-\frac1{12}(\Delta_\mu^{(+)2}-\Delta_\mu^{(-)2}) + o(a^5) \\
a^2\partial_\mu^2 &=& \frac43(\Delta_\mu^{(+)}+\Delta_\mu^{(-)}-2)
-\frac1{12}(\Delta_\mu^{(+)2}+\Delta_\mu^{(-)2}-2) + o(a^6) \,.
\end{eqnarray}
Introducing the finite-difference derivatives
\begin{equation}
\Delta_\mu^{(1)} = \frac{\Delta_\mu^{(+)}-\Delta_\mu^{(-)}}{2a}
\quad\mbox{and}\quad
\Delta_\mu^{(2)} = \frac{\Delta_\mu^{(+)}+\Delta_\mu^{(-)}-2}{a^2}
\end{equation}
we may also write
\begin{eqnarray}
\partial_\mu &=& \Delta_\mu^{(1)} - \frac1{12}a^2(
\Delta_\mu^{(1)}\Delta_\mu^{(2)}+\Delta_\mu^{(1)}\Delta_\mu^{(2)})+o(a^4) \\
\partial_\mu^2&=&\frac43\Delta_\mu^{(2)}-\frac13\Delta_\mu^{(1)2}+o(a^4)\,.
\end{eqnarray}
Note that $\Delta_\mu^{(1)}$ and $\Delta_\mu^{(2)}$ commute. Also, it is
straightforward to generalize the expansions to higher
orders in $a$ in a systematic way. Thus the continuum action, say for a scalar
Klein-Gordon field with quartic self interaction
\begin{equation}
S[\phi]=\int d^4x [-\frac12(\partial_\mu\phi)^2+\frac12m^2\phi^2
+\frac{g}{4!}\phi^4]
\end{equation}
may then be discretized, alternatively, as
\begin{eqnarray}
S[\phi]\approx&&\beta a^4\sum_x\left[-\frac12\sum_\mu\left(\Delta_\mu^{(1)}
\phi(x)-\frac16 a^2 \Delta_\mu^{(1)}\Delta_\mu^{(2)}\phi(x)\right)^2
\right.\nonumber\\&&\left.
+\frac12 m^2 \phi^2(x) +\frac{g}{4!}\phi^4(x) \right] \,,
\label{phi1lat}\end{eqnarray}
or
\begin{eqnarray}
S[\phi]\approx&&\beta a^4\sum_x \left[ -\frac12\sum_\mu \phi(x)\left(\frac43
\Delta_\mu^{(2)}-\frac13\Delta_\mu^{(1)2}\right)\phi(x)
\right.\nonumber\\&&\left.
+\frac12 m^2 \phi^2(x) +\frac{g}{4!}\phi^4(x) \right] \,.
\label{phi2lat}\end{eqnarray}
The above lattice actions are equivalent up to $o(a^4)$. Evidently
the advantage of $o(a^4)$ improvement is that one works close to the
continuum limit ($a\rightarrow 0$), or from a numerical point of view,
at the same `closeness' to the continuum theory on can afford a coarser
(larger $a$) lattice which translates into less computational cost.

Since a classical field $\phi(x)$ is smooth on some scale, say $a$,
improvement should be expected to work at some level of accuracy.

\begin{figure}[htb]
\centerline{\includegraphics[width=56mm]{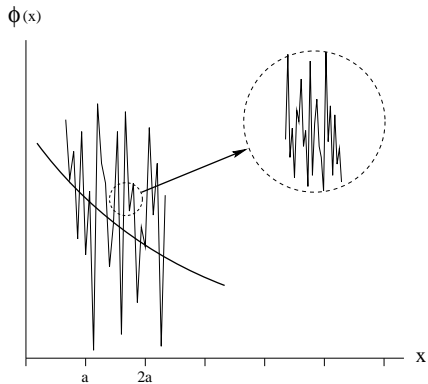}}
\vspace*{8pt}
\caption{Roughness of the quantum field on any scale. The smooth curve
illustrates a Fourier average with a finite momentum
cutoff $\Lambda$.\label{figRough}}
\end{figure}

However, upon quantizing the theory, smoothness of the field is lost.
In the generating functional
\begin{equation}
Z[J]=\int [d\phi]\,e^{-S[\phi]+J\cdot\phi}
\label{GFphi}\end{equation}
$\phi(x)$ and $\phi(y)$ are independent integration variables
for $x\neq y$ even if $x$ and $y$ are infinitesimally close.
At a given scale, say $a$, those fluctuations $|\phi(x)-\phi(y)|$
can in principle be large for $|x-y|\ll a$. Their importance is
determined by the Boltzmann weight in the euclidean path integral.
In a manner of speaking we say that the quantum fields are rough
on any scale, see Fig.~\ref{figRough}.

This means that the Fourier transform (\ref{FT}) is unbounded and,
generally speaking, 
momentum-space integrals diverge. The well-known solution is
renormalization. In a nutshell, introducing a
momentum cutoff $\Lambda$, which makes the momentum-space integrals
finite, then renormalizing the fields
\begin{equation}
\phi(x)= Z_\Lambda^{-1}\phi_\Lambda(x) \,,
\end{equation}
couplings, etc, and, at the level of physical quantities (observables)
removing the cutoff $\Lambda$, becomes an integral part of the quantum
field theory considered.

Clearly, lattice discretization introduces a natural cutoff momentum
\begin{equation}
\Lambda=\frac{\pi}{a}\,.
\end{equation}
In a way, the roughness of the fields at scales less than the lattice 
constant $a$ is rendered harmless at the expense of introducing 
renormalization. 
(One may think of $Z_\Lambda$, for example, as playing a role similar 
to a dielectric constant $\kappa_e$ for a medium of electric
dipoles. The small-scale fluctuations of the electric field on the scale
of the dipoles are lumped into $\kappa_e$.)

The field renormalization constant is without consequence in the present
example because it can be made to disappear by changing  integration 
variables $\phi(x)\rightarrow\phi_\Lambda(x)$ in the generating functional
(\ref{GFphi}). Thus
\begin{equation}
Z[J_\Lambda]=\mbox{const}\,\int [d\phi_\Lambda]\,e^{-S_\Lambda[\phi_\Lambda]
+J_\Lambda\cdot\phi_\Lambda},
\end{equation}
where, $S_\Lambda$ is the same as (\ref{phi1lat}, \ref{phi2lat}) except that
$\beta_\Lambda=\beta Z_\Lambda^{-2}$ and $g_\Lambda=g Z_\Lambda^{-4}$.
Improvement does not change the physics of the quantum theory in this 
particular example of the scalar interacting Klein-Gordon field.

This culture of thinking has, perhaps, impeded a much earlier discovery
of the type of improvement now routinely used in lattice gauge simulations.
The above line of thought does lead to a genuinely different situation
in the case of gauge theories.

\subsection{Appendix B.2: Improvement of a Pure Gauge Theory}\label{secGimp}

\begin{figure}[htb]
\centerline{\includegraphics[width=84mm]{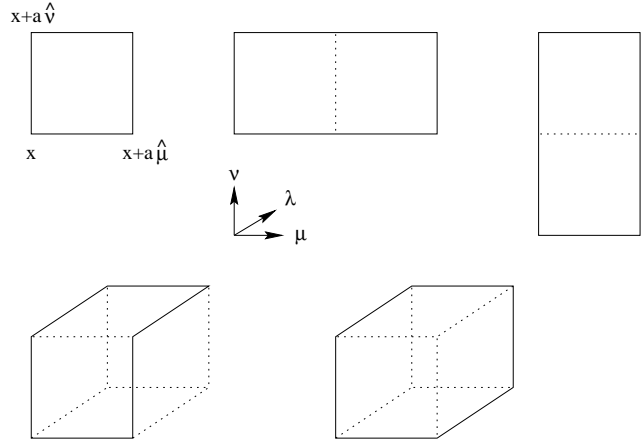}}
\vspace*{8pt}
\caption{Geometry of some loop operators, elementary 4-link plaquette,
6-link planar rectangles, bent rectangle, and parallelogram,
considered for improved gauge field actions.\label{figLoops}}
\end{figure}

\begin{figure}[htb]
\centerline{\includegraphics[height=84mm]{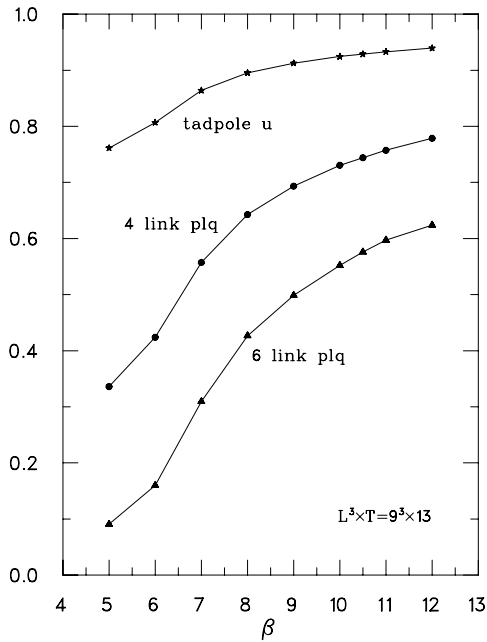}}
\vspace*{8pt}
\caption{Expectation values of the 6-link plaquette (planar rectangle), the
4-link (elementary) plaquette, and the corresponding tadpole renormalization
factor $u$ for versus $\beta$.\label{figPlqRt}}
\end{figure}

We start from the Schwinger representation \cite{Itz80} of the (lattice)
link variables
\begin{equation}
U_{\mu}(x)={\cal P} \exp\left(ig\int_0^a d\xi\, 
A_{\mu}(x+\xi\hat{\mu})\right), \
\label{SchwU}\end{equation}
where ${\cal P}$ denotes path ordering. The latter ensures proper behavior
under gauge transformations. The continuum limit of the classical, smooth,
field is obtained by expanding $U_{\mu}(x)$ into a power series of $a$.
Some formalism proves useful. Motivated by
\begin{equation}
\int_0^a d\xi\, A_{\nu}(x+\xi\hat{\mu}) =
\int_0^a d\xi\, e^{\xi\partial_\mu} A_\nu(x) =
aT_1(a\partial_\mu)A_\nu(x)
\end{equation}
we define the set of (integrated) translators
\begin{equation}
T_n(X)=\sum_{k=0}^\infty\frac{n!}{(k+n)!}X^k \,.
\end{equation}
Those satisfy
\begin{equation}
T_0(X)=e^X \quad\quad  , \ \ T_n(X)={\openone}+\frac{1}{n+1}T_{n+1}(X)X \,.
\label{Trecu}\end{equation}
A straight path along $m$ links in direction $\hat{\mu}$ involves
\begin{equation}
\int_0^{ma}d\xi\,A_{\nu}(x+\xi\hat{\mu}) =
ma T_1(ma\partial_\mu)A_\nu(x).
\end{equation}
These can be used as building blocks for various loop operators, for
example, the planar $m\times n$ plaquettes, see Fig.~\ref{figLoops},
\begin{eqnarray}
P_{\mu\nu}^{(m\times n)}(x)=&&\frac13 \mbox{Re} \mbox{Tr} {\cal P}
\exp\left[
+ig\int_0^{ma}d\xi\,A_{\mu}(x+\xi\hat{\mu})
\right.\nonumber\\&&\left.
+ig\int_0^{na}d\xi\,A_{\nu}(x+ma\hat{\mu}+\xi\hat{\nu})
\right. \nonumber \\ & & \left.
-ig\int_0^{ma}d\xi\,A_{\mu}(x+na\hat{\nu}+\xi\hat{\mu})
\right.\nonumber\\&&\left.
-ig\int_0^{na}d\xi\,A_{\nu}(x+\xi\hat{\nu})
\right] \label{PmnS}\\
=&&\frac13 \mbox{Re} \mbox{Tr} {\cal P} \exp\left[
igmna^2 T_1(na\partial_\nu) T_1(ma\partial_\mu)
\right.\nonumber\\&&\left.
(\partial_\mu A_{\nu}(x)-\partial_\nu A_{\mu}(x))
\right] \,.
\label{Pmn}\end{eqnarray}
The combination of path ordering ${\cal P}$ and taking the color trace  
$\mbox{Tr}$ ensures that $P_{\mu\nu}^{(m\times n)}$ is gauge invariant.
Manifest gauge invariance may be recovered by replacing $\partial_\mu$ 
with the covariant derivative
\begin{equation}
D_\mu=\partial_\mu-igA_{\mu}(x) \,.
\end{equation}
As a consequence, the expansion of (\ref{Pmn}) into a power series of $a$
only contains gauge-invariant terms.
Specifically, for the 4-link elementary plaquette we obtain
\begin{eqnarray}
P_{\mu\nu}^{(1\times 1)}(x) =&& 1 -\frac16 a^4 \mbox{Tr}(g F_{\mu\nu}(x))^2
\nonumber\\
&&-\frac{1}{72} a^6 \mbox{Tr}\left(g F_{\mu\nu}(x)(D_\mu^2+D_\nu^2)
g F_{\mu\nu}(x)\right) + o(a^8)
\label{P4link}\end{eqnarray}
and for the 6-link planar rectangle
\begin{eqnarray}
P_{\mu\nu}^{(2\times 1)}(x) =&& 1 -\frac23 a^4 \mbox{Tr}(g F_{\mu\nu}(x))^2
\nonumber\\&&
-\frac{1}{18} a^6 \mbox{Tr}\left(g F_{\mu\nu}(x)(4D_\mu^2+D_\nu^2)
g F_{\mu\nu}(x)\right) + o(a^8)\,.
\label{P6link}\end{eqnarray}
For $P_{\mu\nu}^{(1\times 2)}$ we have to replace $4D_\mu^2+D_\nu^2$ with 
$D_\mu^2+4D_\nu^2$ in (\ref{P6link}). Thus the classically improved Wilson
action for the pure gauge theory is
\begin{equation}
S_G[U] = \beta\sum_x\sum_{\mu<\nu}\left[ (1-P_{\mu\nu}^{(1\times 1)})
+C_2(1-P_{\mu\nu}^{(2\times 1)})+C_2(1-P_{\mu\nu}^{(1\times 2)})\right]
\label{SGimp}\end{equation}
with $C_2$ chosen such that the $a^6$ terms of the classical limit cancel.
This is achieved by 
\begin{equation}
C_2=-\frac{1}{20} \quad\quad\mbox{(classical)}
\label{C2imp}\end{equation}
in which case $\beta=5/g^2$ and
\begin{eqnarray}
S_G[U]\rightarrow&&\beta a^4\sum_x\sum_{\mu<\nu}\left[\frac{g^2}{10}
\mbox{Tr}(F_{\mu\nu}(x))^2 + o(a^3) \right]
\\
&&= a^4\sum_x\left[\frac14\sum_{\mu,\nu}\mbox{Tr}(F_{\mu\nu}(x))^2
+o(a^3)\right] \,.
\end{eqnarray}
In the quantized theory expansion of the gauge fields into a power series
may no longer be justified. Nevertheless, we expect the expansion to be
`good' if the expectation values of the plaquette operators (\ref{P4link})
and (\ref{P6link}) are `close' to one,
\begin{equation}
\langle \frac{1}{6L^3T}\sum_x\sum_{\mu<\nu}P_{\mu\nu}^{(m\times n)} \rangle
\approx 1 \,.
\end{equation}
This condition can be easily tested in practical situations.
In Fig.~\ref{figPlqRt}
for example, are shown the lattice averages of the 4-link plaquette and
the planar 6-link rectangle. 
The deviations from one are sizeable and increase
if the gauge field coupling ($g^2\propto\beta^{-1}$) is made larger.
This effect is, of course, due to quantum fluctuations. Its significance
is that
\begin{itemize}
\item Quantum fluctuations spoil the classical continuum limit.
\end{itemize}
Again, the standard procedure to deal with short-ranged (ultraviolet)
fluctuations is renormalization. An ultraviolet momentum cutoff is in place
through lattice discretization. The gauge fields are represented by the link
variables
(\ref{SchwU}), thus replace
\begin{equation}
U_{\mu}(x) \rightarrow u^{-1} U_{\mu}(x)
\end{equation}
where $u$ is a renormalization constant. Note that without the 6-link terms
in the action $S_G[U]$, see (\ref{SGimp}), the factor $u$ would just renormalize
the gauge field coupling, and be harmless (and useless) very much like in
the example of Sec.~\ref{secASX}. Link variable renormalization does not
do anything if only the standard Wilson action is used. The crucial point
here is that the 6-link improvement terms in the action carry two more
factors of link variables. Their fluctuations (in the mean-field sense) can 
be divided out by two renormalization factors. Hence, using
\begin{equation}
C_2=-\frac{1}{20u^2}
\label{C2tad}\end{equation}
in the action results in the cancellation of the $o(a^6)$ terms, as above,
between the 4-link and  6-link terms in the action.
\begin{itemize}
\item Classical improvement and gauge link renormalization conspire to
cancel discretization errors  in the quantized lattice field theory.
\end{itemize}
This state of affairs is characteristic for a gauge theory.
In the lingo of perturbation theory the classical limit is known as the
tree-level approximation, because no loop diagrams are taken into account.
The renormalization factor $u$  comprises the effect of so-called tadpole
diagrams, which are singly-attached loops to the gluon propagator.

To be sure, there are radiative corrections to the gluon propagator which are
not contained in the tadpole diagrams. Some of those induce $o(a^2)$
corrections $\propto \alpha_s(\pi/a)$ and, containing loops, are harder to 
come by \cite{Luscher:1985xn,Luscher:1985zq}.

The tadpole factor, on the other hand, can
be monitored during an actual simulation. Starting with an arbitrary
value, say $u=1$, after a number of iterations we simple `measure' the
elementary plaquette (for example) and redefine
\begin{equation}
u = \langle \frac{1}{6L^3T}\sum_x\sum_{\mu<\nu}P_{\mu\nu}^{(1\times 1)}
\rangle^{1/4}  \,.
\label{tadu}\end{equation}
The new value is then used in the action $S_G[U]$ and the iterations continue.
Another $u$ is then `measured' with the new action, and so on. In this way the 
\begin{itemize}
\item tadpole renormalization factor is determined self consistently during the
simulation.
\end{itemize}

Numerical work has shown that tree-level and tadpole improvement (even
without loop corrections) already account for much of the discretization
errors \cite{Alford:1995ui,Alford:1995hw}. The effect of improvement is
quite dramatic. A well-known
example \cite{Lepage:1996jw} is the restoration of rotational invariance of the 
static quark-antiquark potential, usually parametrized as
$V(r)=Kr-\pi/12r+C$. On-axis distances $r/a=1,2,3,4\ldots$ and off-axis
distances $r/a=\sqrt{2},\sqrt{3},\sqrt{5}\ldots$, which are supposed to 
fall on a smooth curve $V(r)$, are off by as much as 38\%.
With a tree-level and tadpole improved action errors are reduced to the
one-percent level!

Improvement schemes are not unique. Various loop geometries (bent, twisted
rectangles, etc) may be used to cancel discretization errors. A corresponding
statement of course holds for improving the fermionic part of the action.

\subsection{Appendix B.3: Improvement of a Fermionic Action}\label{secFimp}

A simple fermion action, which is improved up to the same order as the gauge
field action, can be constructed by employing next-nearest-neighbor
terms \cite{Hamber:1983qa,Eguchi:1984xr,Wetzel:1984hq}. Consider the following
\begin{eqnarray}
S_F^{(0)} &=& a^3\sum_x \bar{\psi}(x)\psi(x) \\
S_F^{(1)} &=& a^3\sum_x\sum_\mu
\left[ \bar{\psi}(x)(r-\gamma_\mu)U_\mu(x)\psi(x+a\hat{\mu})
\right.\nonumber\\&&\left.
+ \bar{\psi}(x+a\hat{\mu})(r+\gamma_\mu)U^\dagger_\mu(x)\psi(x)\right] \\
S_F^{(2)} &=& a^3\sum_x\sum_\mu
\left[ \bar{\psi}(x)(r'-\gamma_\mu)U_\mu(x)
U_\mu(x+a\hat{\mu})\psi(x+2a\hat{\mu})
\right.\nonumber\\&&\left.
+ \bar{\psi}(x+2a\hat{\mu})(r'+\gamma_\mu)U^\dagger_\mu(x+a\hat{\mu})
U^\dagger_\mu(x)\psi(x)\right]
\end{eqnarray}
where $r,r'$ are parameters. Knowing about the essential ingredients for
tree-level and tadpole improvement we construct a fermionic action as a
linear combination
\begin{equation}
S_F = c_0S_F^{(0)}+\frac{c_1}{u}S_F^{(1)}+\frac{c_2}{u^2}S_F^{(2)} \,.
\label{ImpSF}\end{equation}
The factor $u$ is the tadpole renormalization (\ref{tadu}) which reflects the
relative number of gauge links in $S_F^{(0)}$ and $S_F^{(1)}$, $S_F^{(2)}$.
We need the following classical continuum limits
\begin{eqnarray}
S_F^{(1)} &=& 8r\phantom{'} a^3\sum_x \bar{\psi}\psi
-2 a^4\sum_x \bar{\psi} \gamma_\mu D_\mu \psi
\\&&
+r\phantom{'} a^5\sum_x \bar{\psi} D_\mu D_\mu \psi
-\frac13 a^6\sum_x \bar{\psi} \gamma_\mu D_\mu D_\nu D_\nu \psi
+o(a^7) \nonumber\\
S_F^{(2)} &=& 8r' a^3\sum_x \bar{\psi}\psi
-4 a^4\sum_x \bar{\psi} \gamma_\mu D_\mu \psi
\\&&
+4r' a^5\sum_x \bar{\psi} D_\mu D_\mu \psi
-\frac83 a^6\sum_x \bar{\psi} \gamma_\mu D_\mu D_\nu D_\nu \psi
+o(a^7) .\nonumber
\end{eqnarray}
Remarkably it is possible to simultaneously cancel the $o(a^5)$ and
$o(a^6)$ terms. Cancellations must be realized on the tree level, $u=1$.
The requirement is
\begin{equation}
\left(\begin{array}{rr}
r & 4r' \\ & \\
-\frac{1}{3} & -\frac{8}{3} \end{array}\right)
\left(\begin{array}{c} c_1 \\ \\ c_2 \end{array}\right) =
\left(\begin{array}{c} 0 \\ \\ 0 \end{array}\right) \,.
\end{equation}
A nontrivial solution exists if
\begin{equation}
r'=2r \quad\quad \mbox{with} \quad\quad c_1+8c_2=0 \,.
\end{equation}
We solve the last equation writing
\begin{equation}
c_1=-\frac{4}{3}c'
\quad\quad c_2=-\frac{1}{6}c'
\end{equation}
where $c'$ is a normalization. Still on the tree level, we then have
\begin{eqnarray}
S_F & \rightarrow & (c_0-8c'r) a^3\sum_x \bar{\psi}\psi
+2c'a^4\sum_x \bar{\psi} \gamma_\mu D_\mu \psi +o(a^7) \nonumber\\
 & & =  2c'a^4\sum_x \left[ \frac{c_0-8c'r}{2c'a}\bar{\psi}\psi\
+\bar{\psi} \gamma_\mu D_\mu \psi +o(a^3) \right]
\end{eqnarray}
for $a\rightarrow 0$. Thus, in the continuum limit, the bare fermion mass is
$m_f=(c_0-8c'r)/2c'a$.
The choice
\begin{equation}
r=1
\end{equation}
for the Wilson parameter takes care of the doubling problem (violating 
chiral symmetry). In order to meet with the usual conventions we introduce
the `hopping' parameter $\kappa$ through
\begin{equation}
c'=\kappa u c_0
\end{equation}
and choose the normalization
\begin{equation}
c_0=\frac{1}{a^3} \,.
\end{equation}
Thus, finally, the improved lattice action (\ref{ImpSF}) becomes
\begin{eqnarray}
S_F =&& \sum_x \bar{\psi}(x)\psi(x)
 -\frac{4\kappa}{3}\sum_x\sum_\mu \left[
\bar{\psi}(x)(1-\gamma_\mu)U_\mu(x)\psi(x+a\hat{\mu})
\right.\nonumber\\&&\left.
+\bar{\psi}(x+a\hat{\mu})(1+\gamma_\mu)U^\dagger_\mu(x)\psi(x)\right]
\nonumber \\
 & & +\frac{\kappa}{6u}\sum_x\sum_\mu \left[
\bar{\psi}(x)(2-\gamma_\mu)U_\mu(x)U_\mu(x+a\hat{\mu})\psi(x+2a\hat{\mu})
\right.\nonumber\\&&\left.
+\bar{\psi}(x+2a\hat{\mu})(2+\gamma_\mu)U^\dagger_\mu(x+a\hat{\mu})
U^\dagger_\mu(x)\psi(x)\right] \,.
\label{SFnnn}\end{eqnarray}
With this improved action we are hopefully `close' to the continuum limit,
so the fermion mass is `close' to $m_f=(1-8\kappa u)/2\kappa u a$ with
a critical value of $\kappa_c=1/8u$ for $m_f=0$.

\subsection{Appendix B.4: More on Highly Improved Actions}

\begin{figure}[htb]
\centerline{\includegraphics[height=84mm]{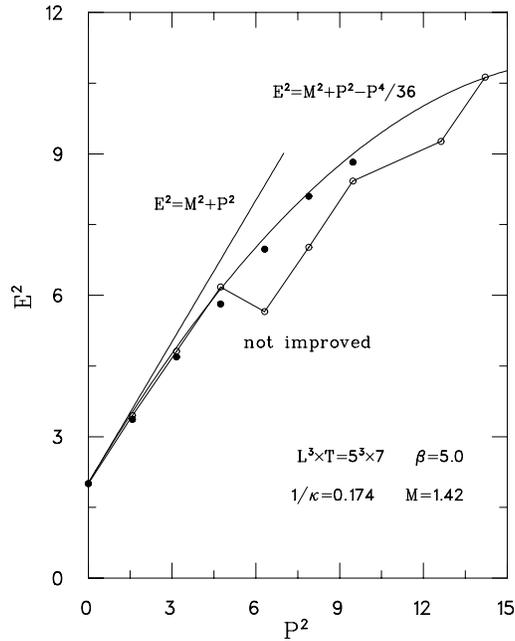}}
\vspace*{8pt}
\caption{Energy-momentum relation for a pseudoscalar meson obtained from
a lattice simulation with the improved action (filled circles), 
and a fitted curve. Also shown are the unimproved lattice dispersion 
relation for a free Bose field \protect\cite{Rot92},
and the continuum relation.\label{figDisp}}
\end{figure}

An example of a dispersion function computed with the $o(a^2)$ improved
action (\ref{SGimp}, \ref{C2tad}) and (\ref{SFnnn}) is shown in 
Fig.~\ref{figDisp}. The filled circles are the energies
of a pseudoscalar meson with momentum $p$ from a $5^3\times7$ lattice
at $\beta=5.0$. The data are close to a smooth (fitted) curve, as opposed to the
open circles, which are from the standard lattice dispersion relation
for bosons \cite{Rot92}.

\begin{figure}[htb]
\centerline{\includegraphics[height=84mm]{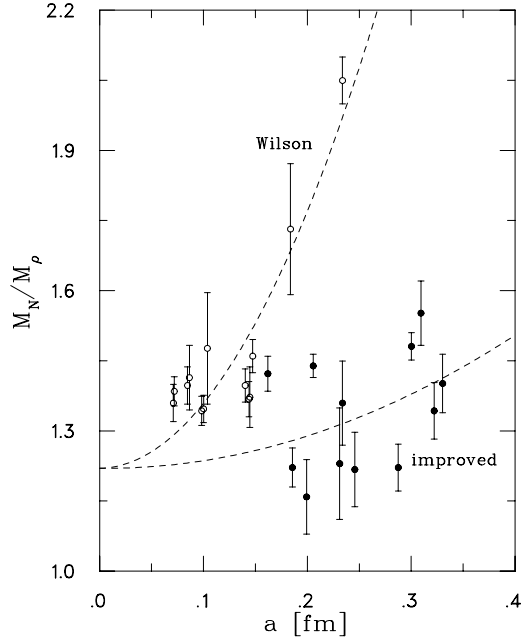}}
\vspace*{8pt}
\caption{Nucleon $\rho$-meson mass ratio versus the lattice constant $a$. The
scale for $a$ was obtained from $M_\rho=770$~MeV.
See Ref.~\protect\cite{Fiebig:1996wg,Collins:1997zc,Lee:1997bq} to trace the references for 
the data points.\label{figMnMrho}}
\end{figure}

Another example of `closeness to the continuum' achievable with
improved lattice actions is shown in Fig.~\ref{figMnMrho}. The nucleon-rho
mass ratio is plotted versus the rho mass. Open plot symbols correspond
to the standard, unimproved, Wilson action (gauge and fermion fields).
The filled symbols refer to various `flavors' of improved actions. The
experimental value of $m_N/m_\rho$ is 1.22. The two curves in 
Fig.~\ref{figMnMrho}, meant to `guide the eye', illustrate dramatically
the `quantum leap' provided to us by improved actions: Physical results,
close to the continuum limit, can be obtained from simulations on
relatively coarse lattices with lattice constants $a$ about a factor
of $\approx 4$ larger compared to the standard Wilson action.  
Systematic errors still are large though, at this time.

The lattice actions discussed in Secs.~\ref{secGimp} and \ref{secFimp}
are improved up to two leading
orders in the lattice constant. Those actions are commonly referred to as
$o(a^2)$ improved, or highly improved actions. It should be mentioned that
highly improved actions are not without problems. The dispersion function
$E(p)$ for free massless quarks, as calculated from the zeros of the
inverse lattice propagator, exhibit unphysical branches that describe
additional particles, or ghosts. Hopefully their masses are sufficiently large,
so as not to interfere with the propagation of the physical quarks on
the lattice. The effect of those particles reveals itself in the
early-time-slice behaviour of hadronic correlation functions. Thus,
in practice, their effect can be isolated and hardly poses a problem.

A class of highly improved actions that has received much attention
recently is known as D234 actions, the nomenclature taken from the way
derivative operators are used \cite{Alford:1996dm}. Here, in addition to the
techniques discussed above, one utilizes a construct known as
the Sheikholeslami-Wohlert (SW), or clover, action \cite{Sheikholeslami:1985ij} as
well as anisotropic lattices. The latter have different lattice constants
$a_s$ and $a_t$ for space and time respectively, typically $a_s>a_t$.
The purpose of combining those features is to push doublers, or ghosts,
into the high-mass region where they can do little harm to
hadronic physics. Actions of this class are well documented, technical
issues are in the foreground at this time \cite{Alford:1996dm,Alford:1997pk,Alford:1997nx,Alford:1998mx}.
We refrain from discussing these further.

\subsection{Appendix B.5: Clover Leaf Fermion Action}

The SW action deserves being mentioned here because
of its widespread use and relatively simple implementation.

\begin{figure}[htb]
\centerline{\includegraphics[width=56mm]{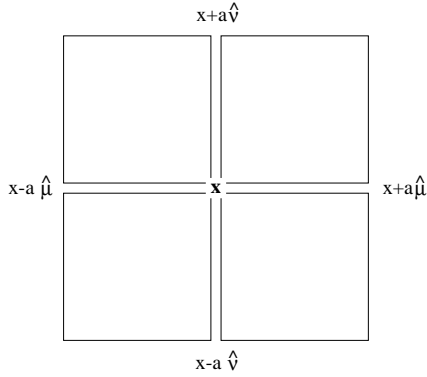}}
\vspace*{8pt}
\caption{Geometry of the paths used in $ P_{\mu\nu}(x)$ for the
clover action.\label{figClover}}
\end{figure}

The idea is to consider (integration) variable transformations of the
fermion fields in the path integral that defines the lattice QFT. This
will of course not alter the physics of the theory but, in general,
change the appearance of the lattice action written in terms of the
transformed fields. The additional terms are called redundant. 
Sheikholeslami and Wohlert \cite{Sheikholeslami:1985ij} considered (among
an extensive list of other possibilities) the transformation
\begin{equation}
\psi = [{\openone}-\frac{ra}{4}(\gamma_\mu\Delta_\mu-m')]\psi' \quad\quad
\bar{\psi} = \bar{\psi}'[{\openone}-\frac{ra}{4}(\Delta_\mu\gamma_\mu-m')]
\label{SWtrans}\end{equation}
where $\psi'$,$\bar{\psi}'$ refer to the original field variables and
mass $m'$. Working out the Dirac action we obtain
\begin{equation}
\bar{\psi}'(\gamma_\mu\Delta_\mu+m')\psi' = 
\bar{\psi}[\gamma_\mu\Delta_\mu+m -\frac{ra}{2}(\Delta_\mu\Delta_\mu
+\frac12\sigma_{\mu\nu}gF_{\mu\nu})]\psi + o(a^2)
\label{SWlim}\end{equation}
with
\begin{equation}
m=m'(1+\frac12ram') \,.
\label{SWlim1}\end{equation}
The $\sigma\cdot F$ term stems from the identity
\begin{equation}
(D_\mu\gamma_\mu)^2=D_\mu D_\mu +\frac12\sigma_{\mu\nu}gF_{\mu\nu}
\quad\mbox{with}\quad \sigma_{\mu\nu}=\frac{1}{2i}[\gamma_\mu,\gamma_\nu] \,.
\end{equation}
This leads to the SW, or clover, action
\begin{equation}
S_F^{(SW)} = \frac{\kappa}{2iu^3} a^4 \sum_x\sum_{\mu,\nu}
\bar{\psi}(x) \sigma_{\mu\nu} P_{\mu\nu}(x) \psi(x),
\end{equation}
which is an additive part to the standard Wilson fermion action.
The name clover action is apparent from Fig.~\ref{figClover} which shows
the geometry of link variables involved in
\begin{eqnarray}
P_{\mu\nu}(x) &=& \frac14 \left( \rule{0mm}{4mm}
\phantom{x}U_\mu(x)U_\nu(x+a\hat{\mu})U^\dagger_\mu(x+a\hat{\nu})
U^\dagger_\nu(x) \right. \nonumber \\
 & & \phantom{xx}-U^\dagger_\nu(x-a\hat{\nu})
U^\dagger_\mu(x-a\hat{\mu}-a\hat{\nu})
U_\nu(x-a\hat{\mu}-a\hat{\nu})U_\mu(x-a\hat{\mu}) \nonumber \\
 & & \phantom{xx}+U_\nu(x)U^\dagger_\mu(x-a\hat{\mu}+a\hat{\nu})
U^\dagger_\nu(x-a\hat{\mu})U_\mu(x-a\hat{\mu}) \nonumber \\
 & & \phantom{xx}\left. -U_\mu(x)U^\dagger_\nu(x+a\hat{\mu}-a\hat{\nu})
U^\dagger_\mu(x-a\hat{\nu})U_\nu(x-a\hat{\nu}) \rule{0mm}{4mm} \right) \,.
\label{Pclover}\end{eqnarray}
The action is $o(a)$ improved on the
tree level, and local, meaning that no `hopping' to next neighbors or 
beyond occurs. This feature is computationally desirable since it leads
to simpler and faster codes.

The clover action and the action discussed in Sec.~\ref{secFimp} are
related \cite{Martinelli:1991ny}.


\end{document}